\documentclass[11pt,a4paper]{article}
\usepackage{jheppub}

\usepackage{amssymb}
\usepackage{amsfonts}
\usepackage{graphicx}
\usepackage{epstopdf}
\usepackage{dcolumn}
\usepackage{amsmath}
\usepackage{latexsym,bm}
\usepackage{amsthm}
\usepackage{slashed}
\usepackage{float}
\usepackage{color}
\usepackage{url}
\usepackage{longtable}
\usepackage{subfig}
\usepackage{tikz}
\usepackage[all,cmtip]{xy}

\usepackage{multirow}
\usepackage{longtable}

\usepackage[titletoc]{appendix}
\makeatletter
\newtheorem*{rep@theorem}{\rep@title}
\newcommand{\newreptheorem}[2]{%
\newenvironment{rep#1}[1]{%
 \def\rep@title{#2 \ref{##1}}%
 \begin{rep@theorem}}%
 {\end{rep@theorem}}}
\makeatother

\newreptheorem{conj}{Conjecture}
\theoremstyle{definition}

\newcommand \xoverline[2][0.75]{
    \sbox{\myboxA}{$\m@th#2$}
    \setbox\myboxB\null
    \ht\myboxB=\ht\myboxA
    \dp\myboxB=\dp\myboxA
    \wd\myboxB=#1\wd\myboxA
    \sbox\myboxB{$\m@th\overline{\copy\myboxB}$}
    \setlength\mylenA{\the\wd\myboxA}
    \addtolength\mylenA{-\the\wd\myboxB}
    \ifdim\wd\myboxB<\wd\myboxA
       \rlap{\hskip 0.5\mylenA\usebox\myboxB}{\usebox\myboxA}%
    \else
        \hskip -0.5\mylenA\rlap{\usebox\myboxA}{\hskip 0.5\mylenA\usebox\myboxB}%
    \fi}
\makeatother

\newcommand{\ba}{\begin{aligned}}
\newcommand{\ea}{\end{aligned}}
\def\be{\begin{equation}}
\def\ee{\end{equation}}
\def\bsp{\begin{split}}
\def\esp{\end{split}}
\def\bea{\begin{eqnarray}}
\def\eea{\end{eqnarray}}

\def \bp{\begin{pmatrix}}
\def\ep{\end{pmatrix}}

\def\R{\mathbb{R}}
\def\N{\mathcal{N}}

\def\C{\mathbb{C}}
\def\Z{\mathbb{Z}}

\def\mk{\mathfrak}

\def\bbD{\mathbb{D}}
\def\bbS{\mathbb{S}}

\usepackage{tikz}
\usepackage{diagbox}
\usetikzlibrary{positioning}
\usetikzlibrary{calc}
\usetikzlibrary{decorations.pathreplacing,calligraphy}

\usepackage{xstring}
\usetikzlibrary{decorations.pathmorphing} 
\usetikzlibrary{decorations.markings} 
\usetikzlibrary{snakes}
\usetikzlibrary{arrows} 
\usetikzlibrary{shapes} 
\usetikzlibrary{matrix} 
\usetikzlibrary{positioning} 
\usepackage[english]{babel} 
\usepackage[autostyle]{csquotes}

\usepackage{tikz}
\usetikzlibrary{arrows}
\usetikzlibrary{arrows.meta}
\usetikzlibrary{shapes.geometric,calc,arrows, positioning,shapes.misc,decorations.markings}
\tikzset{
  big arrow/.style={
    decoration={markings,mark=at position 1 with {\arrow[scale=2,#1]{>}}},
    postaction={decorate},
    shorten >=0.4pt},
  big arrow/.default=black}
  
\pgfdeclarelayer{edgelayer}
\pgfdeclarelayer{nodelayer}
\pgfsetlayers{edgelayer,nodelayer,main} 
\tikzstyle{none}=[inner sep=0pt] 

\tikzstyle{NodeCross}=[draw, shape=circle, cross out, inner sep=0pt, minimum size=6pt,line width=0.25mm]
\tikzstyle{Circle}=[draw, shape=circle, black,  fill=black, inner sep=0pt, minimum size=6pt]
\tikzstyle{Star}=[draw, shape=star, fill=black, star points=8, inner sep=0pt, minimum size=8pt]

\tikzstyle{DashedLine}=[-, densely dashed, line width=0.25mm]
\tikzstyle{DottedLine}=[-, dotted, line width=0.25mm]
\tikzstyle{ThickLine}=[-, line width=0.25mm]
\tikzstyle{ArrowLineRight}=[-, -{Stealth[scale=1.75]}, line width=0.1mm, scale=5]
\tikzstyle{RedLine}=[-, draw={rgb,255: red,191; green,0; blue,0}, fill=none, line width=0.25mm]
\tikzstyle{DottedRed}=[-, dotted, draw={rgb,255: red,191; green,0; blue,0}, fill=none, line width=0.25mm]
\tikzstyle{DashedLineThin}=[-, densely dashed, line width=0.125mm, fill=none, draw=black]
\tikzstyle{ArrowLineRed}=[-, -{Stealth[scale=1.75]}, draw={rgb,255: red,191; green,0; blue,0}, line width=0.1mm, scale=5]
\tikzstyle{brane}=[draw]
\tikzset{D7/.style={circle, draw=black, inner sep=0pt, fill=white, minimum size=3mm}}
\tikzset{hasse/.style={circle, fill,inner sep=2pt}}
\tikzset{flavor/.style={regular polygon,fill=white,regular polygon sides=4,inner sep=2.5pt, draw}}
\tikzset{gauge/.style={circle, draw,inner sep=2.5pt}}
\tikzset{gaugeb/.style={circle, draw,fill=black,inner sep=2.5pt}}
\tikzset{gauger/.style={circle, draw,fill=cyan,inner sep=2.5pt}}
\tikzset{gaugeg/.style={circle, draw,fill=red,inner sep=2.5pt}}
\tikzset{SUd/.style={circle, draw=black, inner sep=0pt, fill=yellow, minimum size=2mm}}
\tikzset{bd/.style={circle, draw=black, inner sep=0pt, fill=black, minimum size=2mm}}
\tikzset{wd/.style={circle, draw=black, inner sep=0pt, fill=white, minimum size=2mm}}
\tikzset{Dynkin/.style={circle, draw=black, inner sep=0pt, fill=white, minimum size=2mm}}
\tikzstyle{ligne}=[draw, thick] 
\tikzset{doublearrow/.style={ draw=black!75, color=black!75, thick, double distance=3pt, }} 

\usepackage{float}
\restylefloat{table}

\DeclareSymbolFont{extraup}{U}{zavm}{m}{n}
\DeclareMathSymbol{\varheart}{\mathalpha}{extraup}{86}
\DeclareMathSymbol{\vardiamond}{\mathalpha}{extraup}{87}

\makeatletter
\patchcmd{\maketitle}{\@fpheader}{}{}{} % {Link: This line https://tex.stackexchange.com/questions/531828/how-to-remove-the-line-prepared-for-submission-to-jcap}
\makeatother

\usepackage{dsfont}

\newcommand\restr[2]{{% we make the whole thing an ordinary symbol
  \left.\kern-\nulldelimiterspace % automatically resize the bar with \right
  #1 % the function
  \vphantom{\big|} % pretend it's a little taller at normal size
  \right|_{#2} % this is the delimiter
  }}

\usepackage{physics}

\usepackage{tikz-cd}

\usepackage{bclogo}%  \bcpanchant

\usepackage{slashed}

\makeatletter
\def\widebreve#1{\mathop{\vbox{\m@th\ialign{##\crcr\noalign{\kern\p@}%
  \brevefill\crcr\noalign{\kern0.1\p@\nointerlineskip}%
  $\hfil\displaystyle{#1}\hfil$\crcr}}}\limits}

\def\brevefill{$\m@th \setbox\z@\hbox{}%
 \hfill\scalebox{0.7}{\rotatebox[origin=c]{90}{(}} \kern4pt $}
\makeatletter

\usepackage{amsmath}

\usepackage{bbm}

\usepackage{enumitem} 

\usepackage{physics}

\usepackage{chemfig}
\usepackage{import}

\makeatletter
\newcommand\xleftrightarrow[2][]{%
  \ext@arrow 9999{\longleftrightarrowfill@}{#1}{#2}}
\newcommand\longleftrightarrowfill@{%
  \arrowfill@\leftarrow\relbar\rightarrow}
\makeatother

\usepackage{booktabs}

\newcommand{\PROVE}[1]{{\textcolor{red}{PROVE}}}

\usepackage{braket}

\usepackage{mathrsfs}

\newcommand{\vol}{\mathrm{vol}}

\usepackage{amsthm}

	\theoremstyle{definition}

\graphicspath{ {figs/} }

\title{\centering{M-theory and T-geometry:\\
Higgs branch moduli and charged matter}}

\author[\clubsuit]{Marwan Najjar\,}
\emailAdd{marwannajjar@bimsa.cn}

\affiliation[\clubsuit]{Beijing Institute of Mathematical Sciences and Applications (BIMSA),\\
Huaibei Town, Huairou
District, Beijing 101408, China}

\abstract{\quad M-theory geometric engineering on manifolds of special holonomy yields a rich class of novel field theories. In this paper, we construct new 3d $\N=2^{\ast}$ and $\N=4^{\ast}$ gauge theories, realized as mass-deformations of theories with 16 supercharges, within this framework. These arise from non-compact 8d geometries given by fibrations of $\R^{4}/\Gamma_{ADE}$ over Biberbach 4-manifolds. The existence of consistent $Spin(7)$-structures on the 8d spaces requires the rotational holonomy of the Biberbach spaces to act on the $Sp(1)$-structure of the fibers. Furthermore, we analyze Higgsing the 7d $\N=1$ $ADE$ gauge theories induced by the action of a permutation group on the centres of the corresponding $\R^{4}/\Gamma_{ADE}$ spaces. We show that this operation admits a natural interpretation in terms of nilpotent, upper-triangular, Higgsing, although it breaks supersymmetry. Supersymmetry is restored by fibering the singular geometry over a compact internal space, whose structure group is chosen to coincide with the permutation group to implement the nilpotent Higgsing. We refer to such backgrounds as T-geometries, where ``T'' denotes the triangular nature of the nilpotent Higgsing. Within this framework, we investigate the nilpotent Higgsing of the 3d $\N=2^{\ast}$ and 4d $\N=1^{\ast}$ theories, where the rotational holonomy groups of the Bieberbach spaces realize the permutation groups. We demonstrate that the Higgs branch moduli are encoded by specific elements of the Slodowy slices associated with nilpotent elements. Moreover, we demonstrate that additional elements of the same slice give rise to non-chiral charged matter under the unbroken gauge algebra. We establish that both the Higgs branch moduli and the charged matter are massless and admit a natural interpretation as localized matter.}

\begin{document}

\maketitle

%%%%%%%%%%%%%%%%%%%%%%%%%%%%%%%%%%%%%%%%%%%%%%%%%%%%%%%%%%%%%%%%%%%%%%%%%%%%%%%%%%%%%%%%%

\section{Introduction and summary}

A landscape of supersymmetric quantum field theories (SQFTs) can be realized within the string theory, either via geometric engineering, brane configurations, or AdS/CFT correspondence, which provide powerful tools and methods to study them beyond the capability of perturbative methods in field theories. Geometric engineering|which has been around since the mid 90s \cite{Witten:1995ex,Witten:1996qb,Katz:1996fh,Katz:1996th,Klemm:1996bj,Bershadsky:1996nh,Ooguri:1997ih,Sen:1997js,Sen:1997kz,Acharya:1998pm}| encodes the construction and properties of lower-dimensional interacting field theories in the geometry of the extra spatial dimensions. Remarkably, compactifications of string/M-/F-theory on manifolds of special holonomy give us access to SQFTs with a wide variety of novel structures. For example, such framework enable us to discover and classify superconformal field theories as in, e.g., \cite{Seiberg:1996bd,Shapere:1999xr,Heckman:2013pva,Xie:2015rpa,Jefferson:2018irk,Caorsi:2018zsq}, study their flavor symmetries as in, e.g., \cite{Morrison:1996xf,Apruzzi:2019opn,Acharya:2021jsp,Najjar:2022eci,Najjar:2023hee}, realize different phases of gauge theories as in, e.g., \cite{Intriligator:1997pq,Acharya:2001hq,Closset:2020scj,Closset:2021lwy,Acharya:2020vmg,Acharya:2023xlx,Najjar:2025rgt}, and study global generalized symmetries of SQFTs as in, e.g., \cite{Apruzzi:2021nmk,Najjar:2024vmm,Najjar:2025htp, Khlaif:2025jnx}.

In M-theory geometric engineering, one constructs 7d $ADE$ gauge theories with 16 supercharges by placing M-theory on non-compact Calabi-Yau 2-folds with $ADE$ type singularities, which can be locally written as $\R^{4}/\Gamma_{ADE}$, see, e.g., \cite{Sen:1997js,Sen:1997kz}. Lower-dimensional theories are obtained by fibering these singularities over a compact internal manifold $Y$. To preserve supersymmetry, the internal space $Y$ is required to admit a spin structure, and ideally the total space should admit a metric with special holonomy. In practice, however, constructing such metrics explicitly is often intractable. Instead, one may rely on the following equivalent and more tractable criteria:
\begin{itemize}
  \item[(i)] The holonomy of the total space is determined by that of the constituent geometries. In the present case, this typically involves a (semi)direct product of $SU(2)$|the holonomy of $\R^{4}/\Gamma_{ADE}$| with the holonomy of $Y$, denoted by $H$.

\item[(ii)] The existence of a compatible $\mathscr{G}$-structure on the total space, such that the holonomy group(s) of the previous point can be embedded into $\mathscr{G}$. 

\item[(iii)] A field-theoretic interpretation in terms of twisted dimensional reduction of the 7d $ADE$ gauge theory on $Y$, see, e.g., \cite{Witten:1988ze,Yamron:1988qc,Vafa:1994tf,Acharya:1998pm,Beasley:2008dc,Najjar:2022eci,Najjar:2023hee}. In this framework, the geometry and topology of $Y$ determines the spectrum through the decomposition of fields into differential form bundles on $Y$, with zero modes classified by the corresponding cohomology groups. In particular, supersymmetry is encoded in some Betti numbers of $Y$.
\end{itemize}

We take the latter route in this paper where our internal spaces are Biberbach 4-manifolds|as classified in \cite{lambert2013}. This enable us to engineer 3d $\N=2^{\ast}$ and $\N=4^{\ast}$ $ADE$ gauge theories, which can be interpreted as mass-deformations of the 3d $\N=8$ theories associated with the first and trivial Biberbach space given simply by a 4-torus $T^{4}$. Such mass-deformation theories were explored earlier in \cite{Donagi:1995cf,Dorey:1999sj,Myers:1999ps,Polchinski:2000uf,Naculich:2001us}.

Gauge theories are partially characterized by their moduli space of vacua, which typically decomposes into Coulomb, Higgs, and mixed
branches, each admitting a distinct low-energy description. In the context of geometric engineering, the Coulomb branch is associated with crepant resolutions of the underlying singular geometry. From the field-theoretic perspective, this corresponds to giving diagonal vacuum expectation values (vevs) to scalar fields in vector multiplets. In contrast, the Higgs branch is interpreted in terms of deformations of the singularities, as discussed in \cite{Closset:2020scj,Closset:2021lwy}. An alternative geometric realization was proposed in \cite{Acharya:2023xlx}, which may be paraphrased as follows: The Higgs branch arises from the action of the holonomy group $H$ of the internal space on the singular fibers. More precisely, this action projects out a subset of the independent 2-spheres appearing in the crepant resolution of the singularities, thereby reduces the gauge algebra. However, this latter approach does not incorporate the mass-deformation perspective, nor does it clarify the origin of the Higgs fields and the role of nilpotent Higgsing. These aspects are central to the framework developed in this work.

Gauge theories with adjoint-valued scalar fields admit a rich class of Higgsing phenomena in which the vacuum expectation values need not be diagonal. In particular, one may consider strictly upper-triangular (Jordan block) vevs, corresponding to nilpotent elements $e$ of the complexified gauge algebra $\mk{g}$. A particularly interesting situation arises when one has three adjoint-valued fields obeying the commutation relations of $\mk{sl}(2,\C)$, as discussed, e.g., in \cite{Gaiotto:2008sa}. In this case, one is led to consider embeddings, or homomorphisms,  $\rho:\mk{sl}(2,\C)\hookrightarrow\mk{g}$. The choice of a nilpotent element $e$ is unique only up to the adjoint action of the associated adjoint group $G_{\mathrm{ad}}=G/\mathcal{Z}$, where $\mathcal{Z}$ denotes the center of $G$. This action defines the nilpotent orbit $\mathcal{O}_{e}$ associated with $e$. The tangent space to the nilpotent orbit $\mathcal{O}_{e}$ at a point $e$ is given by $\mathrm{im}(\mathrm{ad}_{e})$, while the transverse directions are captured by the Slodowy slice $\mathcal{S}_{e}$. The classification of such embeddings $\rho$ is provided by the Jacobson-Morozov theorem. In the case $\mk{g}=\mk{su}(N)$, the inequivalent embeddings are in one-to-one correspondence with partitions of $N$, which label the possible nilpotent orbits.\\
Further aspects of nilpotent Higgsing will be developed later in this work. For a comprehensive treatment of nilpotent orbits and Slodowy slices, see \cite{Collingwood1993,slodowy1980four,slodowy2006simple}.

Since we propose that the Higgs branch of the geometries we consider in this paper can be describe through nilpotent Higgsing, we refer to such quotient geometry by ``\textit{T-geometry}``, with T stand for the triangle of the nilpotent Higgs vev. This terminology is analogous to the T-brane one established in \cite{Cecotti:2010bp}, which builds on \cite{Donagi:2003hh}. For further work on T-branes one may consider, e.g., \cite{Donagi:2011jy,Anderson:2013rka,Collinucci:2014qfa,Collinucci:2014taa,Cicoli:2015ylx,Heckman:2016ssk,Collinucci:2016hpz,Bena:2016oqr,Marchesano:2016cqg,Anderson:2017rpr,Collinucci:2017bwv,Cicoli:2017shd,Cvetic:2018xaq,Heckman:2018pqx,Apruzzi:2018xkw,Carta:2018qke,Marchesano:2019azf,Barbosa:2019bgh,Bourget:2023wlb}.

%%%%%%%%%%%%%%%%%%%%%%%%%%%%%%%%%%%%%%%%%%%%%%%%%%%%%%%%%%%%%%%%%%%%%%%%%%%%%%%%%%%%%%%%%%%%%%%%%

\subsection{Summary of results}

We now summarize the main results and wisdom of our work. Note that the summary is organized in a different order from the main body of the paper, as this better serves its intended purpose.

\paragraph{Nilpotent Higgs in 7d and the need for compactification.}

In M-theory geometric engineering, we can construct 7d $\N=1$ $ADE$ SYM theories at the singular locus of $\Gamma_{ADE}\subset SU(2)$ of $\C^{2}\cong\R^{4}$, i.e., $\R^{4}/\Gamma_{ADE}$. These theories admit 3 real adjoint scalars associated with the hyperK\"ahler structure on $\R^{4}/\Gamma_{ADE}$. It is then natural to expect a Higgsing phenomena to occur due to these scalar fields. One interesting possibility is to define a holomorphic scalar field $\Phi$ and allow non-diagonal vacuum expectation value (vev) given by nilpotent elements the (complexfied) $ADE$ Lie algebra $\mk{g}$ as in \cite{Donagi:2003hh} and later in, e.g., \cite{Cecotti:2010bp,Donagi:2011jy,Collinucci:2014qfa,Collinucci:2014taa}. In the following, all Lie algebras are understood to be complexified unless stated otherwise.  

In section \ref{sec:CB-HB-7d}, we observe that the nilpotent vev can be given by restricting to the upper-triangle of a permutation matrix $P$ acting on the centres associated with $\R^{4}/\Gamma_{ADE}$. Permutations of these centres were examined in \cite{wright2011quotientsgravitationalinstantons}, where it is found that at most they from a cyclic group, which we denote by $H$. Unfortunately, one can show that quotienting $\R^{4}/\Gamma_{ADE}$ by the group $H$ generically lift the hyperk\"ahler structure of the space. Consequently, breaks the supersymmetry of the 7d theories. However, we argue that we can use the outlined Higgs phenomena for lower-dimensional field theories by compactifing M-theory further on an internal $Y_{n}$ space. In other words, we shall consider 
\begin{equation}
  X_{4+n}(\Gamma_{ADE},\,H,\,Y_{n})\,=\, \frac{(\R^{4}/\Gamma_{ADE}\,\times\, Y_{n})}{H}\,,
\end{equation}
where $H$ acts non-trivially on $Y_{n}$. Two essential requirements are expected to be satisfied:
\begin{itemize}
  \item $X_{4+n}$ admits a $\mathscr{G}$-structure compatible with the action of $H$.
  
\item The group $H$ must act freely on the space $Y_{n}$.
\end{itemize}

\paragraph{Effective theories with massive fields.}

An interesting class of examples of the $Y_{n}$ spaces are the Bieberbach $n$-manifolds $B_{n}$, which are quotients of an $n$-torus, $T^{n}/H$. Here, $H$ can be, in principle, any finite subgroup of $SO(n)$.

In section \ref{sec:3d-theories-GE}, we analyze an 8d geometry given as
\begin{equation}
  \R^{4}/\Gamma_{ADE}\, \hookrightarrow\, X_{8}\,\to\, B_{4}\,.
\end{equation}
Here, the finite group $H$ acts non-trivially on $\R^{4}/\Gamma_{ADE}$, on its centres and hyperK\"ahler structure, as mentioned above. $B_{4}$ are Biberbach 4-manifolds which are classified in \cite{lambert2013}. We review the $B_{4}$ spaces in appendix \ref{app:Bieberbach} and list them in Table \ref{Table:all-B4-spaces}. We determine the parallel $Spin(7)$-structre on such spaces, which is given in terms of the 2-forms of the hyperK\"ahler structure of the fibers and self-dual 2-forms of the base. We verify that the $Spin(7)$-structre is invariant under the action of the $H$ group. The metric on the total space can be written, at least for a first order, as a sum of that on the fibers and the torus. We further discuss the holonomy groups of the $X_{8}$ space.

We then turn our attention to study the 3d theories, denoted by $\mathcal{T}$, associated the $B_{4}$ spaces. We argue that the associated 3d theories can be interpret as deformations of 3d $\N=8$ theories such that we arrive at 3d $\N=2^{\ast}$ and $\N=4^{\ast}$. In particular, we have 
\begin{equation}
  \begin{split}
      \mathcal{T}(B_{(4;2)}^{(2-8)})\,\,&=\, \, \text{3d $\N=4^{\ast}$ $ADE$ gauge theory}\,,
      \\
      \mathcal{T}(B_{(4;1)}^{(9-27)})\,\,&=\, \, \text{3d $\N=2^{\ast}$ $ADE$ gauge theory}\,.
  \end{split}
\end{equation}
Here, we should exclude the non-spin $B_{4}$ spaces. 

In arriving at the 3d theories, we employ the standard twisted dimensional reduction procedure, which relates the fields of the theories to the topology and cohomology of the internal manifold $B_{4}$. One further observes that fine-tuning and supersymmetry, where the latter seems to imply the former, forces the radii of the $H$-twisted directions of $T^{4}$ to be equal. Consequently, the corresponding modes in the 3d theories acquire identical masses. These modes organize themselves into appropriate supermultiplets. It turns out that we have 3 massive adjoint-valued chiral multiplets in the case of 3d $\N=2^{\ast}$, whereas in the 3d $\N=4^{\ast}$ case we have a single massive adjoint-valued hypermultiplet.

In section \ref{sec:G-structure-co-seifert}, we examined the co-Seifert fibration structure of the $B_{4}$ spaces following the discussion in \cite{lambert2013}. This structure can be understood by analyzing the relation between the finite group $H$ acting on $B_{4}$ and the corresponding group acting on $B_{3}$. As a result, the $B_{4}$ spaces can be viewed as $B_{3}$ fibered over either a circle $\bbS^{1}$ or an interval $I$. We point out the cases where the 3d theories can be seen as direct $\bbS^{1}$ reduction of 4d theories associated with $B_{3}$ spaces, studied originally in \cite{Acharya:1998pm}. Following the above line of reasoning, one expects the corresponding 4d theories to arise as mass deformations of $\N=4$ supersymmetric theories associated with the first and trivial $B_{3}$ space, the 3-torus. More precisely, one obtains theories of the $\N=1^{\ast}$ and $\N=2^{\ast}$ type in 4 dimensions.

\paragraph{Higgs branch of 3d and 4d theories.}

In section \ref{sec:geometric-realization}, we provide the geometric understanding and realization of the 3d theories in terms of the behavior of the $p$-forms on $(\R^{4}/\Gamma_{ADE})/H$ and $T^{4}/H$ under the action of $H$. We refer to the $p$-forms invariant under $H$ as $H$-untwisted, whereas, to those which transform under $H$ as $H$-twisted.

The behavior of the harmonic 2-forms $\{\widetilde{\mathrm{h}}_{2}^{a}\}$ on $\widetilde{\R^{4}/\Gamma_{ADE}}$ under the action of the group $H$, with $a=1,\cdots,\mathrm{rank}(\mk{g}_{ADE})$, admits a natural interpretation in terms of the branches of the corresponding field theory \cite{Acharya:2023xlx}. Specifically, the Coulomb branch is realized when all such 2-forms are $H$-untwisted. Whereas having some, or all, of such 2-forms being $H$-twisted is interpreted as a partial or complete Higgsing of the gauge algebra. In the following we will not distinguish between Higgs and mixed branches and will refer to all such configurations simply as Higgs branche(s). The Higgs branch moduli are given in terms of the $H$-untwisted wedge product of $H$-twisted $\widetilde{\mathrm{h}}_{2}^{a}$ and other $H$-twisted 1-forms and 2-forms on $B_{4}$. Since these moduli correspond to harmonic $p$-forms in $X_{8}$, then they are expected to be massless. We refer to these moduli as the\textit{ geometric Higgs branch moduli}.

In section \ref{sec:HB}, we investigate the Higgs branch of particular 3d $\N=2^{\ast}$ and 4d $\N=1^{\ast}$ $\mk{su}(n)$ gauge theories within the general framework developed in \ref{sec:CB-HB-7d}. These theories arise from the Bieberbach $B_{4}^{(9)}$ and $B_{3}^{(6)}$, respectively, and come with 3 adjoint-valued massive chiral multiplets $\Phi_{i}$ with $i=1,2,3$, as explained above. In both cases, the finite group $H$ is $\Z_{2}\times\Z_{2}$.

We propose that these fields play a crucial role in the phenomenon of nilpotent Higgsing of the gauge algebra $\mk{g}$. In particular, their vacuum expectation values can be interpreted as specifying different embeddings of $\mk{sl}(2,\C)$ into $\mk{g}$. This occurs as the superpotential of these theories imposes $\mk{sl}(2,\C)$  commutation relation on these fields \cite{Donagi:1995cf,Dorey:1999sj,Myers:1999ps,Polchinski:2000uf,Naculich:2001us}. Equivalently, we identify the set $\{\Phi_{i}\}$ with the $\mk{sl}(2,\C)$ triple $\{e,h,f\}$ such that the triplet takes values in the adjoint representation of the gauge algebra $\mk{g}$.

In this work we specialize to the case $\mk{g}=\mk{su}(N)$. The starting point is the action of the finite group $H$ on the $N$ centres of the resolved space $\widetilde{\R^{4}/\Z_{N}}$. This action is not necessarily faithful: the corresponding $N\times N$ representation of $H$ may be reducible and may include trivial components acting on some of the centres. Given such representations of $H$, we construct candidate nilpotent vacuum expectation values, denoted by $e$, by restricting to the strictly upper--triangular part of these matrices. Schematically, this amounts to embedding the elementary nilpotent block
\begin{equation}
\begin{pmatrix}
  0 & 1 \\
  0 & 0
\end{pmatrix}
\ \hookrightarrow \ 
   \begin{pmatrix}
0 & 1 & 0 & \cdots & \ast \\
0 & 0 & \ast & \cdots &  \ast \\
\vdots &  \vdots & \vdots & \ddots & \vdots \\
0 & 0 & 0 & \cdots & 0
\end{pmatrix}\,,
\end{equation}
where the entries $\ast$ may take the values $0$ or $1$. Once the nilpotent element $e$ is specified, the corresponding $\mk{sl}(2,\C)$ triple is completed by constructing the associated semisimple element $h$ and the lowering operator $f$, which together satisfy the standard $\mk{sl}(2,\C)$ commutation relations.

For the $\mk{su}(n)$ case, the nilpotent Higgsing can be captured by integer partitions of $N$, see, e.g., \cite{Collingwood1993}. That we shall consider:
\begin{equation}
  N\,=\,\sum_{i=1}^{k}\,n_{i}\,\lambda_{i}\,,\quad \text{with}\quad \lambda_{1} > \lambda_{2} > \cdots > \lambda_{k}\,.
\end{equation}
Here, both of $n_{k}$ and $\lambda_{k}$ are positive integers. A given nilpotent element then breaks the $\mk{su}(N)$ gauge algebra down to 
\begin{equation}
 \mk{su}(N) \quad \xrightarrow[\text{Higgsing}]{\text{Nilpotent}} \quad \bigoplus_{i=1}^{k}\,\mk{su}(n_{k})\,\,\oplus\,\,\mk{u}(1)^{\oplus(k-1)}\,. 
\end{equation}

The geometric Higgs branch (HB) moduli can be captured by considering suitable physical fluctuations, denoted by $\delta\Phi_{\mathrm{phy}}$, subject to the linearized gauge transformation. We observe and conjecture that the geometric HB moduli  correspond to particular elements $\delta\Phi_{\mathrm{phy}}$ of the Slodowy slice $\mathcal{S}_{e}$ associated with the nilpotent background $e$ as defined in \cite{slodowy1980four,slodowy2006simple}.

For instance, in the $\mk{su}(2)$ case, the geometric Higgs branch modulus is a complex field associated with the wedge product of the $H$-twisted $\widetilde{\mathrm{h}}_{2}$ and $H$-twisted 1-form and 2-form. This modulus is identified with the physical fluctuations 
\begin{equation}
 \delta\Phi_{\mathrm{phy}}\ =\  \begin{pmatrix}
    0 & 0\\
    \phi &0
  \end{pmatrix}\,,\quad \text{with}\quad  \phi\,\in\,\C\,,
\end{equation}
which is the only non-trivial element in $\mathcal{S}_{e}$.

More generally, the lesson is that whenever an $\mk{su}(2)\subset \mk{su}(N)$ factor is Higgsed, we have a complex scalar field $\phi$ arising from the Slodowy slice. We find such parameters along the diagonal blocks of $N\times N$ representation of $\mathcal{S}_{e}$ where the embedded $\mk{su}(2)$'s are located. Geometrically, having $H$-twisted $\widetilde{\mathrm{h}}_{2}^{a}$ effectively amounts to twisting the dual independent 2-sphere in the resolved  $\widetilde{\R^{4}/\Z_{N}}$ space. Precisely, twisting the divisor given as $(\bbS^{2}\times T^{4})/H$. The observation is that, twisting such a divisor introduces two real parameters, which are combined into a single complex modulus. This modulus is identified with the physical fluctuation $\phi$ that belongs to $\mathcal{S}_{e}$.

\begin{figure}[H]
\centering{
\begin{tikzpicture}[
    % Define styles for nodes and arrows
    node distance=1.5cm,
    box/.style={rectangle, draw=black, thick, minimum width=2cm, minimum height=1cm, align=center},
    arrow/.style={-{Stealth[scale=1.2]}, thick}
]

% Place the nodes (boxes)
\node[box] (A) {Upper triangle of $N\times N$\\
                 representation(s) of $H$};
\node[box, right=of A] (B) {Nilpotent element(s)\\
of (complexified) $\mk{g}$};
\node[box, right=of B] (C) {Certain elements of\\ 
the Slodowy slice $\mathcal{S}_{e}$};

% Draw the arrows connecting the boxes
\draw[arrow] (A) -- (B);
\draw[arrow] (B) -- (C);
\end{tikzpicture}
}
\caption{The figure illustrates the correspondence between the action of the group $H$ on the centres of the resolution of $\R^{4}/\Z_{N}$ with nilpotent Higgsing and particular elements of the Slodowy slice $\mathcal{S}_{e}$. These elements correspond to the geometric Higgs branch moduli.}
\label{Fig:summary}
\end{figure}
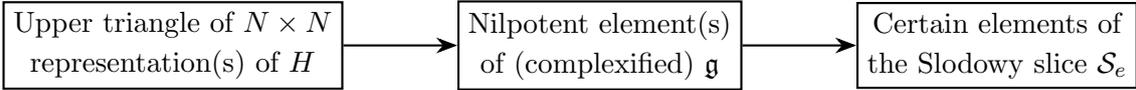

We summarize the above discussion in Figure \ref{Fig:summary}.

\paragraph{Non-chiral charged matter and trap point. }

In section \ref{sec:charged-matter}, we show that we can have non-chiral charged fields under the unbroken gauge algebra. Motivated by the correspondence between particular elements of $\mathcal{S}_{e}$ and the geometric Higgs branch moduli, we consider other elements of $\mathcal{S}_{e}$ which we find that they give rise to charged matter. We proceed according to the following simple algorithm\footnote{Note that, in some more complicated cases we may have to perform a change of basis such that the above algorithm may be applied.}:
\begin{itemize}
  \item For a given triplet $\{e,h,f\}$, we first consider the braking of the gauge algebra under the semisimple element $h$. It breaks $\mk{g}$ in a rank preserving manner.
  
  \item Since the physical fluctuations $\delta\Phi_{\mathrm{phy}}$ transform in the adjoint, we consider its decomposition according to the breaking given by $h$. This provide different bi-fundamental sectors between the non-abelian sub-algebra of $\mk{g}$ which are also charged under some $\mk{u}(1)$ factors. 

  \item At this point, we consider the nilpotent element and break the gauge algebra accordingly. This may occur before or after performing the $h$-breaking. 

  \item Finally, we consider the intersection between the Slodowy slice and the bi-fundamental sector of the decomposed $\delta\Phi_{\mathrm{phy}}$. Schematically, we can present it as:
  \begin{equation}
 \text{charged matter}\ = \    \mathcal{S}_{e} \  \cap \  \restr{\delta\Phi_{\mathrm{phy}}}{\text{bi-fund.}}
  \end{equation}
\end{itemize}
An equivalent procedure is given by first writing $\mathcal{S}_{e}$ as a sum of different terms according to the decomposition of the adjoint under the $h$-breaking. Then intersect those terms along the bi-fundamental sector with the decomposed $\delta\Phi_{\mathrm{phy}}$. We use the latter in the relevant section below for the examples.

Naively, one expects all modes associated with $\delta\Phi_{\mathrm{phy}}$ are massive as the expected scalar potential is given in terms of $\mathrm{ad}_{e}(\delta\Phi_{\mathrm{phy}})$. However, as it turns out that, we can interpret our work in the light of that given in \cite{Cecotti:2010bp}, and later in \cite{Barbosa:2019bgh}, insuring that both the Higgs branch moduli and the non-chiral charged matter are massless. Specifically, for the $B_{3}^{(6)}$ space, we make use of its co-Seifert structure as a 2-torus fibration over an interval $I$, with $\mathbb{D}_{2}$ being the structure group. On the 2-torus, the physical fluctuations $\delta\Phi_{\mathrm{phy}}$ can be represented as
\begin{equation}
  z^{m}\,\delta\Phi_{\mathrm{phy}}\,=\,\mathrm{ad}_{e}(\eta)\,,
\end{equation}
which is zero at the pole $z=0$. Here, $m$ is a positive integer, $z$ is a coordinate on a local patch $U\cong\C\subset T^{2}$, and $\eta$ is an adjoint-valued element of $\mk{g}$. Whereas, on the interval we have to consider the linearized gauge transformations involving $t$, the coordinate on the interval $I$, times the semisimple element $h$, to remove unphysical degrees of freedom. Note that only the physical degrees of freedom in $\delta\Phi_{\mathrm{phy}}$ survive at $z=0$. 

It turns out that the physical fluctuations in $\mathcal{S}_{e}$ are given by the localized, or trapped, charged matter at the trap point $z=0=t$ in the above setup. Note that the trap point is invariant under the action of $\mathbb{D}_{2}$. We refer to this setup as the ``\textit{trap matter framework}``. Such matter fields are massless as their scalar potential is given in terms of $\mathrm{ad}_{e}(\eta)$ which is identically zero at the trap point.

The above discussion can be carried to the $B_{4}^{(9)}$ space and the associated 3d $\N=2^{\ast}$ theories as we comment in the relevant section. 

%%%%%%%%%%%%%%%%%%%%%%%%%%%%%%%%%%%%%%%%%%%%%%%%%%%%%%%%%%%%%%%%%%%%%%%%%%%%%%%%%%%%%%%%%%%%%%%%

\section{3d \texorpdfstring{$\N=2^{*}$}{N=2} and \texorpdfstring{$\N=4^{*}$}{N=4} \texorpdfstring{$ADE$}{ADE} gauge theories}\label{sec:3d-theories-GE}

In this section, we begin by reviewing the construction of 7d $\N=1$ $ADE$ gauge theories in M-theory. We then perform a generic twisted reduction of these theories on oriented 4-manifolds $M_{4}$, expressing the resulting supermultiplets and the number of preserved supercharges in terms of the Betti numbers, and Euler characteristic, of $M_{4}$.

Specializing to the case where $M_{4}$ is a Bieberbach 4-manifold $B_{4}$, we further analyse the existence of parallel $Spin(7)$-structures on the associated 8-dimensional total spaces $X_{8}$, constructed as fibrations of  $\R^{4}/\Gamma_{ADE}$ over $B_{4}$. There are, up to affine equivalence, only 24 orientable and spin Bieberbach manifolds. We also interpret the corresponding 3d theories as having $\N=2^{\ast}$, or $\N=4^{\ast}$.

We conclude this section by exploring the co-Seifert structure of the 4d Bieberbach spaces and their relation to 3d Bieberbach spaces. In some case, the 3d theories can be seen as direct $\bbS^{1}$ reduction of 4d $\N=1^{\ast}$ and $\N=2^{\ast}$ theories. 

\paragraph{From M-theory to 7d $\N=1$ SYM theories.} 

Here, we review the construction of 7d $\N=1$ SYM theories with $\mk{g}_{ADE}$ gauge algebra following, e.g., \cite{Sen:1997kz,Sen:1997js,Acharya_2004}, and similar discussion in \cite{Najjar:2022eci,Khlaif:2025jnx}.

Placing M-theory on the non-compact $\R^{4}/\Gamma_{ADE}$ of $SU(2)$-holonomy, with $\Gamma_{ADE}$ being the finite $ADE$ subgroups of $SU(2)$,  breaks half of the M-theory supercharges. Further, the compactfication breaks the 11d spacetime symmetry $Spin(1,10)$ to the subgroup:
\begin{equation}\label{7dglobalsymm}
    Spin(3)\,\times\,Spin(1,6)\,.
\end{equation}
Here, $Spin(3)$ is identified with the R-symmetry group of the 7d theory and $Spin(1,6)$ is it's Lorentz symmetry. 

The 7d $\N=1$ theory degrees of freedom are given only by the vector multiplet, denoted by $\text{VM}_{\text{7d}}^{\N=1}$, which transform under the global symmetries in (\ref{7dglobalsymm}) as:  
\begin{equation}\label{eq:7d-VM}
  \text{VM}^{\N=1}_{\text{7d}} \quad :   \qquad   \mathsf{A}: (\textbf{1},\textbf{7}), \quad \mathsf{\Phi}: (\textbf{3}, \textbf{1}),
        \quad  \mathsf{\Psi}: (\textbf{2}, \textbf{8}), \quad \mathsf{Q}: (\textbf{2}, \textbf{8})\,.
\end{equation}
Here, $\mathsf{A}$ is the gauge field, $\mathsf{\Phi}$ denotes the scalar fields, $\mathsf{\Psi}$ is the gaugino, and $\mathsf{Q}$ represents the 16 supercharges.

The gauge theory algebra near the singularity $\{0\}\in\R^{4}/\Gamma_{ADE}$ is given by the corresponding $ADE$ Lie algebra. To arrive at this result, we follow the standard procedure:
\begin{itemize}
    \item Resolving the singularity, which introduces $\mathrm{rank}(\Gamma_{ADE})$ independent vanishing 2-cycles along with their Poincaré dual $L^{2}$-normalisable 2-form $\{\widetilde{\mathrm{h}}_{2}^{a}\}$, with $a=1$ , $\cdots$ , $\mathrm{rank}(\Gamma_{ADE})$.
    \item Expanding the M-theory $C_{3}$-field, along $\{\widetilde{\mathrm{h}}_{2}^{a}\}$, to get $\mathrm{rank}(\Gamma_{ADE})$ massless abelian gauge fields, i.e., photons.  
    \item Implementing the 2d McKay correspondence \cite{mckay}, which insures that the second homology of $\widetilde{\R^{4}/\Gamma_{ADE}}$ coincides with the root lattice of the $ADE$ algebra. Moreover, the intersection numbers between the 2-cycles coincide with the corresponding elements of the $ADE$ Cartan matrix.
    \item Introducing M2-branes that wrapping the vanishing 2-cycles, which correspond to massive charged particles, with their charges determined by the intersection matrix of the vanishing 2-cycles.
\end{itemize}
In the singular limit, a non-abelian $ADE$ gauge theory arises with adjoint-valued $\text{VM}_{\text{7d}}^{\N=1}$.

\subsection{Twisted reduction of 7d SYM theories on 4-manifolds}\label{sec:twisted-reduction}

In this subsection, we perform a twisted reduction of the 7d $\N=1$ $ADE$ gauge theory on a generic 4-dimensional manifolds $M_{4}$. From geometric perspective, the twisted reduction can be seen as the M-theory reduction on $\R^{4}/\Gamma_{ADE}$ bundle over $M_{4}$. In general, the bundle can be non-trivial. Examples of a similar reductions can be found in \cite{Najjar:2023hee}. Twisted reductions on manifolds of other dimensions can be found in \cite{Witten:1988ze,Yamron:1988qc,Vafa:1994tf,Acharya:1998pm,Beasley:2008dc}.

The direct reduction of the 7d theory on $M_{4}$ reduces the global symmetry as :
\begin{equation}
    Spin(3)\,\times\,Spin(1,6)\ \longrightarrow \ Spin(3)\,\times\,Spin(3)_{L}\,\times\,Spin(3)_{R}\,\times\,Spin(1,2) \,.
\end{equation}
Here, $Spin(3)_{L}\times Spin(3)_{R}$ is the rotational symmetry, or structure group, on $M_{4}$. In turn, the adjoint-valued $\text{VM}_{\text{7d}}^{\N=1}$ would be reduced as: 
\begin{equation}\label{eq:reduction-on-M4}
    \begin{split}
         & Spin(3)\,\times\,Spin(1,6)\ \longrightarrow \ Spin(3)\,\times\,Spin(3)_{L}\,\times\,Spin(3)_{R}\,\times\,Spin(1,2) 
         \\
         & \qquad \qquad (\textbf{1},\textbf{7})\qquad\quad  \ \longrightarrow \ \qquad  \qquad  (\textbf{1},\textbf{1},\textbf{1},\textbf{3}) + (\textbf{1},\textbf{2}_{L},\textbf{2}_{R},\textbf{1})\,, 
         \\
         & \qquad \qquad (\textbf{3},\textbf{1})\qquad\quad  \ \longrightarrow \ \qquad  \qquad  (\textbf{3},\textbf{1},\textbf{1},\textbf{1}) \,,
         \\
         & \qquad \qquad (\textbf{2},\textbf{8})\qquad\quad  \ \longrightarrow \ \qquad  \qquad  (\textbf{2},\textbf{2}_{L},\textbf{1},\textbf{2}) + (\textbf{2},\textbf{1},\textbf{2}_{R},\textbf{2}) \,.
    \end{split}
\end{equation}

To get a supersymmetric theory, the manifolds $M_{4}$ should admit parallel spinors, i.e., covariantly constant spinor. In mathematical terms, the manifold $M_{4}$ should admit a special holonomy group. Equivalently, the reduction of the 7d supercharges $Q$, given in \eqref{eq:7d-VM}, should have some components in the trivial representation of the `internal' global symmetry $Spin(3)\,\times\,Spin(3)_{L}\,\times\,Spin(3)_{R}$. However, as one observes from \eqref{eq:reduction-on-M4}, there are no parallel spinors; hence, the 3d theory is non-supersymmetric. 

A twisted compactification can be implemented by introducing a background gauge field for the R-symmetry. One modifies the covariant derivative on $M_{4}$ by including the R-symmetry connection, thereby, in principle, permitting the existence of parallel spinors. Equivalently, one can only consider the diagonal subgroup of the R-symmetry with part of the structure group on $M_{4}$. Without loss of generality, we consider the twist between the $Spin(3)$ R-symmetry and $Spin(3)_{L}$ of $M_{4}$ :
\begin{equation}\label{eq:twist:spin3-spin3L}
 Spin(3)\,\times\,Spin(3)_{L}\quad \xrightarrow{\quad \text{twisting}\quad} \quad   \widetilde{Spin(3)}_{L}\,.
\end{equation}

Under the twisted global symmetry, the field content on the right hand side of \eqref{eq:reduction-on-M4} now transform as
\begin{equation}\label{eq:twis-red-3d-fields-reps}
    \begin{split}
        & \widetilde{Spin(3)}_{L}\,\times\,Spin(3)_{R}\,\times\,Spin(1,2)
        \\
        & \quad \qquad (\widetilde{\textbf{1}}_{L},\textbf{1}_{R},\textbf{3}) + (\widetilde{\textbf{2}}_{L},\textbf{2}_{R},\textbf{1}) \,,
        \\
        & \quad \qquad (\widetilde{\textbf{3}}_{L},\textbf{1},\textbf{1})\,,
        \\
        &\quad \qquad (\widetilde{\textbf{3}}_{L},\textbf{1},\textbf{2}) + (\widetilde{\textbf{1}}_{L},\textbf{1},\textbf{2}) + (\widetilde{\textbf{2}}_{L},\textbf{2}_{R},\textbf{2})\,.
    \end{split}
\end{equation}
The spinor transforming in the representation $(\widetilde{\textbf{1}}_{L},\textbf{1},\textbf{2})$ is manifestly parallel on $M_{4}$. However, $M_{4}$ may admit additional parallel spinors beyond this representation. We now examine this possibility in detail.

Under the twisted internal symmetry group $\widetilde{Spin(3)}_{L}\times Spin(3)_{R}$, fermionic components transform as sections of the form bundle $\Omega^{p}(M_{4})$. Specifically, the massless degrees of freedom of the lower-dimensional theory are in one-to-one correspondence with harmonic $p$-form  $M_{4}$, denoted $\mathcal{H}^{p}(M_{4})$. By the Hodge theorem, $\mathcal{H}^{p}(M_{4})$ is isomorphic to the $p$-th de Rham cohomology group $H^{p}_{\rm{dR}}(M_{4})$ The dimension of both spaces is given by the Betti number $b_{p}(M_{4})$.

The identification between fermionic sector and $H^{p}_{\rm{dR}}(M_{4})$ works as the follow:
\begin{itemize}
  \item The spinors $\psi : (\widetilde{\textbf{1}}_{L},\textbf{1},\textbf{2})$ are sections of the $H^{0}_{\rm{dR}}(M_{4})$ bundle. The massless modes are given by $b_{0}(M_{4})$, which is one for a connected manifold $M_{4}$. Hence, we have only one massless spacetime spinor of this type.

  \item The spinors $\chi : (\widetilde{\textbf{2}}_{L},\textbf{2}_{R},\textbf{2})$ are sections of the $H^{1}_{\rm{dR}}(M_{4})$ bundle and the zero modes are determined by $b_{1}(M_{4})$.

  \item The space $\Omega^{2}(M_{4})$ splits into self-dual $\Omega^{2}_{+}(M_{4})$ and anti-self-dual $\Omega^{2}_{-}(M_{4})$ parts under the Hodge star operatoration. The same holds de Rham groups. The spinors $\xi : (\widetilde{\textbf{3}}_{L},\textbf{1},\textbf{2})$ are sections of the self-dual 2-form $H^{2}_{\rm{dR};\,(+)}(M_{4}):= H^{2}_{\rm{dR};\,L}(M_{4})$ bundle. The zero modes are given by $b_{2}^{+}(M_{4}):= b^{L}_{2}(M_{4})$. We note that there are no spinors in the $(\widetilde{\textbf{1}}_{L},\textbf{3}_{R},\textbf{2})$ representation, the would-be spinors are sections of the anti-self-dual 2-forms $H^{2}_{\rm{dR};\,(-)}(M_{4}):= H^{2}_{\rm{dR};\,R}(M_{4})$ bundle.    
\end{itemize}

The bosonic degrees of freedom are then given as
\begin{itemize}
    \item Originating from 7d gauge field : We have a massless 3d gauge field $A$ transform as $(\widetilde{\textbf{1}}_{L},\textbf{1}_{R},\textbf{3})$ and $b_{1}(M_{4})$ massless scalar fields $\phi : (\widetilde{\textbf{2}}_{L},\textbf{2}_{R},\textbf{1}) $.
    \item The 7d scalar fields $\sf{\Phi}$ decompose to $b_{2}^{L}(M_{4})$ scalar fields $\varphi : (\widetilde{\textbf{3}}_{L},\textbf{1},\textbf{1})$.  
\end{itemize}
Further discussion on these bosonic fields will be present in section \ref{sec:CB-HB-geometric}.

From the perceptive of 3d $\N=1$ supermultiplets, the above fields can be organized into a vector multiplet $(A,\psi)$, $b_{1}(M)$ scalar (or chiral) multiplets $(\phi,\chi)^{I}$, and $b_{2}^{L}(M_{4})$ scalar multiplets $(\varphi,\xi)^{J}$. Here, $I=1,\cdots b_{1}(M_{4})$ and $J=1,\cdots,b_{2}^{L}(M_{4})$. We shall note, though obvious, that all scalars are real and all fermions are Majorana. Since the original $\rm{VM}_{7d}^{\N=1}$ transform in the adjoint-representation of the $ADE$ gauge group, then all the 3d supermultiplets transform in the same representation under the 3d $ADE$ gauge group. However, since the number of the surviving 3d supercharges $Q$ matches that of the above $\psi$, $\chi^{I}$, and $\xi^{J}$ spinors, then the number of supersymmetries in 3d is given as
\begin{equation}\label{eq:susy-N-b1-b2}
    \N\,=\,\left(\, 1 \,+\,b_{1}(M_{4})\,+\, b_{2}^{L}(M_{4})\,\right) \,.
\end{equation}
Therefore, depending on the values of these Betti numbers, the 3d $\N=1$ supermultiplets will be merged and organized in supermultiplets of higher $\N$.

One can reexpress the above formula of $\N$ using only $b_{1}(M_{4})$ and the Euler characteristic number of $M_{4}$, denoted by $\chi(M_{4})$ as
\begin{equation}\label{eq:N=chi-b1}
    \N \, =\, \frac{1}{2}\left(\, \chi(M_{4})\,+\, 4 \, b_{1}(M_{4})   \,\right)\,.
\end{equation}
Here, one uses the fact that $b_{2}^{L}(M_{4})$ is given by
\begin{equation}
    b_{2}^{L}(M_{4})\,=\, \frac{1}{2}\,\left(\, \chi(M_{4})\,+\,2b_{1}(M_{4})\,-\,2  \,\right)\,.
\end{equation}

In Table \ref{table:examples-of-M4}, we present some examples of connected $M_{4}$ manifolds, the corresponding supersymmetries, and comments on the total 8-dimensional geometries. As mentioned earlier, the twisted reduction may give non-trivial $\R^{4}/\Gamma_{ADE}$ fibration over $M_{4}$. As one can notice from the table, we recover known 8-dimensional spaces with special holonomy. In particular, we have: 
\begin{itemize}
    \item The $\R^{4}$ bundle over $\bbS^{4}$ with $Spin(7)$ holonomy as first constructed in \cite{Bryant1989OnTC} (see also the discussion in \cite{Gibbons:1989er}). With the $ADE$ quotient on the fiber direction, one arrives at 3d $\N=1$ $ADE$ gauge theories. 
    \item The $\R^{4}$ bundle over $\bbS^{2}\times\bbS^{2}$ with $SU(4)$ holonomy, see, e.g., \cite{Oh:1998qi,Herzog:2000rz} and \cite[section 2]{Najjar:2025rgt}. In geometric engineering, one arrives at 3d $\N=2$ $ADE$ gauge theories as considered in \cite{Najjar:2023hee,Najjar:2025rgt}.

    \item For the third case with $M_{4}=T^{4}$, the bundle fibration is trivial and the total 8d space can be expressed as $\widetilde{\R^{4}/\Gamma_{ADE}}\times T^{4}$. Hence, it preserve 16 supercharges and the higher 7d $ADE$ gauge theory. 

    \item The $\R^{4}$ bundle over $\bbS^{2}\times\Sigma_{g=1}$, i.e., over $\bbS^{2}\times T^{2}$. In fact, this is equivelent to having the resolved conifold reduced on $T^{2}$, see, e.g., \cite[section 4.3]{Najjar:2023hee}. Hence, the total geometry has $SU(3)$ holonomy inherited from the resolved conifold.   
\end{itemize}

The primary focus of this work lies in the application of 4-dimensional Bieberbach spaces. We devote the following subsection to a detailed examination of their properties and relevance to our construction.

\begin{table}[H]
\centering
\begin{tabular}{|c|c|c|c|}
\hline
$M_{4}$ & $\chi(M_{4})$ & $b_{1}(M_{4})$  & $\N$     \\
\hline
\hline
$\bbS^{4}$ & $2$ & $0$  & $1$  \\
$\bbS^{2}\times\bbS^{2}$& $4$ & $0$ & $2$ \\
$T^{4}$ & $0$ & $4$  & $8$  \\
% $\chi(\bbS^{2}\times T^{2})=0$ & $2$ & $1$  & $2$ & --- \\
$\bbS^{2}\times \Sigma_{g}$ & $4-4g$ & $2g$  & $2+2g$  \\
% $\Sigma_{g_{1}}\times \Sigma_{g_{2}}$ & $4(1-g_{1})(1-g_{2})$ & $2g_{1}+2g_{2}$   & $2+2g_{1}g_{2}$ &  --- \\
$B_{(4;2)}$ & $0$ &  $2$  & $4$  \\
$B_{(4;1)}$ & $0$ & $1$  & $2$  \\
\hline
\end{tabular}
\caption{Here, we have several examples of $M_{4}$ manifolds and the corresponding amount of supercharges $\N$. We denote the 4-dimensional Bieberbach manifolds by their $b_{1}$ Betti number as $B_{(4;b_{1})}$. For the fourth case, we restrict $g\leq 3$, as otherwise, $\N$ would be larger than $8$.}
\label{table:examples-of-M4}
\end{table}

%%%%%%%%%%%%%%%%%%%%%%%%%%%%%%%%%%%%%%%%%%%%%%%%%%%%%%%%%%%%%%%%%%%%%%%%%%%%%%%%%%%%%%%%%%%%%

\subsection{\texorpdfstring{$\R^{4}/\Gamma_{ADE}$}{R4/Gamma-ADE} bundle over Bieberbach 4-manifolds}\label{sec:R4-bundle-over-B4}

Let us discuss the 3d gauge theories arises when considering 4-dimensional Bieberbach manifolds, which are closed and flat 4-manifolds. We review the construction and several details of generic Bieberbach manifolds in appendix \ref{app:Bieberbach}. A Bieberbach $n$-manifold is defined by the quotient of $\R^{n}$ by a non-trivial discrete subgroup of its isometry group, i.e., $\Gamma\subset \R^{n}\rtimes SO(n)$, 
\begin{equation}
    B_{n}\,=\, \R^{n}/\Gamma\,\cong\,T^{n}/H\,.
\end{equation}
$\Gamma$ is also known as the $n$-dimensional crystallographic group, or $n$-space group. The group $\Gamma$ is given, in general, by a non-splittable group extension of the form
\begin{equation}
    1\,\to\,\Z^{n}\,\to\,\Gamma\,\to\,H\,\to\,1\,.
\end{equation}
Here, $\Z^{n}$ is the group of translation, i.e., lattice shifts, and  $H$ is the holonomy of $B_{n}$.

In four dimensions, Bieberbach manifolds fall into two classes: orientable and non-orientable. This work focuses exclusively on the orientable case. The classification of orientable flat 4-manifolds yields $27$ distinct spaces \cite{lambert2013}, tabulated in Table \ref{Table:all-B4-spaces}. Among these, only $24$ admit a spin structure. Our analysis will be restricted to some of these $24$ spin-compatible Bieberbach manifolds.

In this subsection, we examine the $Spin(7)$-structure on the bundle $\mathbb{R}^4/\Gamma_{ADE}$ over $B_4$ and its covering space. Further, we discuss the effective field theory interpretation and content.

%%%%%%%%%%%%%%%%%%%%%%%%%%%%%%%%%%%%%%%%%%%%%%%%%%%%%%%%%%%%%%%%%%%%%%%%%%

\subsubsection*{Parallel \texorpdfstring{$Spin(7)$}{Spin(7)}-structures}\label{sec:spin-7-strucutre}

Let us first consider the parallel $Spin(7)$-structure on the covering space of the $\mathbb{R}^4/\Gamma_{ADE}$ bundle over $T^{4}/H$. Topologically, we write the covering space as:
\begin{equation}
 \widetilde{X}_{8}\,:=\,\left(\R^{4}/\Gamma_{ADE}\right)\,\times \, T^{4}\,.   
\end{equation}
There exist a natural closed 4-form, denoted by $\Phi_{4}$, on $\widetilde{X}_{8}$, which can be expressed as:
\begin{equation}\label{eq:Phi-4-form}
    \Phi_{4}\,= \, \sum_{i,j\,=\,1}^{3} \,\delta^{ij}\,\omega_{i}\,\wedge\, \alpha_{j}\,-\, \mathrm{vol}(T^{4})\,-\, \vol(\R^{4}/\Gamma_{ADE})\,.
\end{equation}
Here, $\{\omega_{i}\}$ are the $Sp(1)$ hyper-K\"ahler 2-forms on $\R^{4}/\Gamma_{ADE}$. The notation $\mathrm{vol}(\bullet)$ denotes the top-degree volume-form on the corresponding space. For the torus $T^{4}$, this is simply given by $\dd y_{1}\wedge\cdots\wedge\dd y_{4}$, while for $\R^{4}/\Gamma_{ADE}$ it is expressed in terms of the associated form-fields.

For the case $\Gamma_{A}=\Z_{N}$, we have an explicit expression for the $Sp(1)$ structure. This is due to the fact that the space $\R^{4}/\Z_{N}$ admit a multi-centered as given in, which is a metric on the $U(1)$ bundle over $\R^{3}$ base \cite{GIBBONS1978430}. The reader can refer to appendix \ref{sec:harmonic-2-forms} for a review. The the $Sp(1)$ structure is given in terms of self-dual 2-forms, as presented in \eqref{eq:SD-ASD-2-forms}. For this case the volume-form is given as $e^{1}\wedge\cdots \wedge e^{4}$ where the form-fields are presented in \eqref{eq:form-fields}.

Furthermore, $\{\alpha_{i}\}$ are the set of the self-dual 2-forms on $T^{4}$. For $(y_{1},y_{2},y_{3},y_{4})$ being coordinates on $T^{4}$, the triplet of 2-form $\alpha_{i}$ can be represented as:
\begin{equation}
    \begin{split}
         &\alpha_{1}\,=\, \dd y_{1}\,\wedge\, \dd y_{4}\,+\,\dd y_{2}\,\wedge\,\dd y_{3}\,,
        \\
        &\alpha_{2}\,=\, \dd y_{1}\,\wedge\, \dd y_{3}\,-\,\dd y_{2}\,\wedge\,\dd y_{4} \,, 
         \\
       &\alpha_{3}\,=\, \dd y_{1}\,\wedge\, \dd y_{2}\,+\,\dd y_{3}\,\wedge\,\dd y_{4} \,.
    \end{split}
\end{equation}

% \footnote{We can also think of a compact version given as $K_{3}\times T^{4}$. Here, $K_{3}$ being the known $K_{3}$-surface; the compact 4-manifolds with $Sp(1)$-structure.}

Observe, we have used the self-dual 2-forms $\{\alpha_{i}\}$ on $T^{4}$; this reflect the twisted reduction performed in \eqref{eq:twist:spin3-spin3L}. The hyper-K\"ahler triplet transforms in the $\textbf{3}$-representation of $Spin(3)$, while $\{\alpha_{i}\}$ transform under $Spin(3)_{L}$ in the $\textbf{3}_{L}$ representation. This makes the combination $\sum_{i}\omega_{i}\wedge\alpha_{i}$ invariant under the twisted $\widetilde{Spin(3)}_{L}$ group.

One can realize that the 4-form $\Phi_{4}$ is self-dual, i.e., a Cayley 4-form, by observing the following properties.
\begin{itemize}
  \item The wedge product of the triplet of the 2-forms on each of the constitute spaces of $\widetilde{X}_{8}$ defines $\vol(\bullet)$, i.e., 
\begin{equation}
    \omega_{i}\,\wedge\,\omega_{j}\, =\, 2\delta^{ij} \,\mathrm{vol}(\R^{4}/\Gamma_{ADE})\,,\qquad \alpha_{i}\,\wedge\,\alpha_{j}\,=\, 2\delta^{ij}\,\mathrm{vol}(T^{4})\,.
\end{equation}

\item Since each of the triplet of 2-forms are self-dual, then we expect that their wedge product to be self-dual as well, i.e.,
\begin{equation}
  \ast_{8}(\omega_{i}\,\wedge\,\alpha_{j})\,=\,\omega_{i}\,\wedge\,\alpha_{j}\,.
\end{equation}
Furthermore, it is natural to demand that the  Hodge stare operation exchange the volume-form of the constitute space, i.e., 
\begin{equation}
 \ast_{8}(\mathrm{vol}(\R^{4}/\Gamma_{ADE}))\,=\, \mathrm{vol}(T^{4})\,,\qquad \ast_{8}(\mathrm{vol}(T^{4}))\,=\,\mathrm{vol}(\R^{4}/\Gamma_{ADE})\,.
\end{equation}
\end{itemize}
These conditions can be verified explicitly for the case of $\Gamma_{A}=\Z_{N}$. However, the general properties hold for the general case. Therefore, we conclude that the 8d space $\R^{4}/\Gamma_{ADE}\times T^{4}$ admits a parallel $Spin(7)$-structure. 

We now address whether the $Spin(7)$-structure naturally descends to the quotient space for non-trivial holonomy group $H$.

\paragraph{$Spin(7)$-structure on quotient spaces and holonomy extensions.}

When incorporating the quotient group $H$, the total 8-dimensional space admits two topologically distinct realizations:
\begin{equation}
 X_{8}\,=\, \frac{\left(\R^{4}/\Gamma_{ADE}\right)\,\times\,T^{4}}{H}  \,\,, \qquad\text{or}\qquad \  \widehat{X}_{8}\,=\,\frac{\R^{4}}{\Gamma_{ADE}}\,\times\,\frac{T^{4}}{H}\,.
\end{equation}
In both cases, the total space is Ricci-flat as it consist of two Ricci-flat sub-spaces. In the first case, the holonomy group $H$ acts on both factors of the product space, while in the second, it acts only on the torus $T^4$.

Let $\{x_{i}\}$ denote local coordinates on $\R^{4}/\Gamma_{\mathrm{ADE}}$ and $\{y_{j}\}$ local coordinates on $T^4$ as before. The quotient by $H$ is generally implemented as:
\begin{equation}
h \ :\  (x_{i}, y_{j})\  \mapsto\  (\rho(h) \cdot x_{i}\,,\, \sigma(h) \cdot y_{j}) \sim (x_{i}, y_{j})\,,\qquad \forall h\,\in\, H\,.
\end{equation}
With $\rho$ and $\sigma$ are representations of $H$ acting on the respective coordinate sets. For $X_8$, both $\rho$ and $\sigma$ are non-trivial, whereas for $\widehat{X}_8$, $\rho$ is trivial.

The general principle is that:
\begin{equation}\label{eq:Spin(7)-structure-Phi4}
  \parbox{11cm}{The $Spin(7)$-structure defined by the 4-form $\Phi_{4}$ in \eqref{eq:Phi-4-form} descends to the quotient space, provided $\Phi_{4}$ is invariant under the induced action of $H$.}.
\end{equation}
 In this light, we observe the following:
\begin{itemize}
\item For $\widehat{X}_{8}$, only the self-dual 2-forms $\{\alpha_{i}\}$ transform non-trivially under $H$, so $\Phi_{4}$ is not invariant and does not descend to the quotient. Consequently, we will exclude $\widehat{X}_{8}$ from further discussion.

\item For $X_{8}$, both the hyper-K\"ahler 2-forms $\{\omega_{i}\}$ and $\{\alpha_{i}\}$ are affected by the induced action of $H$. The $Spin(7)$-structure defined by $\Phi_4$ remains invariant under $H$ provided:\\
$(i)$ The volume 4-forms on both $\R^{4}/\Gamma_{ADE}$ and $T^{4}$ are $H$-invariant.\\
$(ii)$ The combination $\sum_{i}\omega_{i}\wedge\alpha_{i}$ is preserved.
\end{itemize}
This second condition implies that $H$ should be a subgroup of the twisted $\widetilde{SO}(3)_{L}$, with the double cover is $\widetilde{Spin(3)}_{L} \cong SU(2)$. Hence, the possible choices for $H$ include the cyclic groups $\Z_{N}$, the dihedral groups $\bbD_{N}$ of order $2N$, and the polyhedral groups: tetrahedral $\mathbb{T}$ (order 12), octahedral $\mathbb{O}$ (order 24), or icosahedral $\mathbb{I}$ (order 60), as well as their corresponding double covers.

This argument is general and applies to any 4-manifold, not just to $T^{4}$ or $B_{4}$. For the specific case of the Bieberbach spaces $B_{4}$, the admissible subgroups $H$ are restricted to
\begin{equation}\label{eq:list-holonomy-group-H}
    H\, \in \, \{\,\mathds{1}\,, \Z_{2},\,\Z_{3},\,\Z_{4},\,\Z_{6},\, \Z_{2}\times\Z_{2},\,\bbD_{3},\, \bbD_{4},\, \bbD_{6},\, \mathbb{T}\,  \}\,,
\end{equation}
as summarized in Table \ref{Table:all-B4-spaces} and discussed in \cite{lambert2013}. In Table \ref{Table:all-B4-spaces}, the column labeled “BBNWZ” gives the classification symbol for the holonomy matrices $H$, following the notation established in \cite{brown1978}. The explicit matrix representations can be found in that reference. We note that some of the matrices listed are not orthogonal. Nevertheless, it is always possible to find an appropriate change of basis that yields an orthogonal representation for each holonomy group $H$.

In this case, we take $\rho(h)$ and $\sigma(h)$ to be in orthogonal representations of $H$ such that: 
\begin{equation}\label{eq:invariance-H-omega-alpha}
    \sum_{i}\,\omega_{i}\,\wedge\,\alpha_{i}\,\to\, \sum_{j}\,\omega_{i}\,\rho(h)^{\ast}_{ji}\,\wedge\,\sigma(h)^{\ast}_{ik}\,\alpha_{k}\,=\,\sum_{i}\,\omega_{i}\,\wedge\,\alpha_{i}\,,
\end{equation}
which implies that $\sigma^{\ast}(h)=(\rho^{\ast}(h))^{T}$. Therefore, for the space(s) $X_{8}$ the parallel $Spin(7)$-structure descend naturally from the covering space $\widetilde{X}_{8}$. For the case of $\Gamma_{A}=\Z_{N}$, one can verify this conclusion by explicitly work out the induced transformations on $\{\omega_{i}\}$ and $\{\alpha_{i}\}$ for all the cases in Table \ref{Table:all-B4-spaces}.

\paragraph{Notation.}

We could denote the 8-dimensional space by its finite holonomy group as $X_{8}=X_{8}(H)$. However, as evident from Table \ref{Table:all-B4-spaces}, multiple distinct Bieberbach 4-manifolds may share the same finite holonomy group $H$.  To uniquely specify the geometry, it is therefore more appropriate to label $X_{8}$ by both the finite $ADE$ subgroup $\Gamma_{ADE}$ groupand the Bieberbach 4-manifold $B_{(4;b_{1})}^{(k)}$:
\begin{equation}
    X_{8}\,=\, X_{8}(\Gamma_{ADE},B_{(4;b_{i})}^{(k)})
\end{equation}
We have updated the notation for these 4-dimensional Bieberbach manifolds to $B_{(4;b_{i})}$, where the second entry $b_{1} =1$ or $2$, denoting the number of free 1-cycles in each case, respectively.

In what follows, we may suppress the explicit $\Gamma_{ADE}$ label for notational simplicity, as the $ADE$ type will be clear from context.

\paragraph{Singularities and free action.}
Since Bieberbach manifolds are smooth, the action of $H$ on $T^{4}$ is free of singularities. Consequently, the group $H$ acts freely on the total space $X_{8}(B_{(4;b_{i})}^{(k)})$, and the only singularities present are the codimension-4 $ADE$ singularities $\{0\}\times B_{4}$ originating from the $\R^{4}/\Gamma_{ADE}$ sector.

\paragraph{Holonomy groups on $X_{8}(B_{(4;b_{i})}^{(k)})$.}

We turn our attention to holonomy groups of the quotient spaces $X_{8}$. Naively, one might expect the holonomy to be simply $SU(2) \cong Sp(1)$, originating from the $\mathbb{R}^{4}/\Gamma_{ADE}$ component\footnote{In general, one transitions from an $Sp(1)$-structure to an $SU(2)$-structure by  selecting a specific complex structure. For example, one may choose $\omega_{3}$ to be identified with the K\"ahler 2-form, while $\Omega^{(2,0)} = \omega_{1} + i\,\omega_{2}$ defines the holomorphic top form on $\mathbb{R}^{4}/\Gamma_{ADE} \cong \mathbb{C}^{2}/\Gamma_{ADE}$.}. That is, one might expect the holonomy to coincide with that of the covering space $\widetilde{X}_{8}$.

However, as argued earlier, the action of $H$ induces non-trivial transformations on the hyper-K\"ahler 2-forms $\{\omega_{i}\}$. Consequently, the $Sp(1)$-structure |and hence the holonomy| of the covering space is not preserved under $H$, and the naive expectation does not hold. Further discussion on this observation is given in section \ref{sec:CB-HB-7d}.

In general, we expect that the the holonomy of $X_{8}$ to contain the holonomy group of the covering space $\widetilde{X}_{8}$, i.e.,  
\begin{equation}\label{eq:HoltildX-inside-X}
    \mathrm{Hol}(\widetilde{X}_{8})\,\subseteq\, \mathrm{Hol}(X_{8})\,.
\end{equation}
The actual relation between $\mathrm{Hol}(X_{8})$, $\mathrm{Hol}(\widetilde{X}_{8})$, and the group $H$ is given by, see, e.g., \cite{kobayashi1963I,kobayashi1969II,Besse1987,Berger2003,clarke2012,RudolphSchmidt2017},
\begin{equation}\label{eq:Hol-extension-H}
    \begin{split}
        \mathrm{Hol}(X_{8}(B_{(4;b_{i})}^{(k)}))\,&=\, \mathrm{Hol}(\widetilde{X}_{8})\,\rtimes\, H\,
        \\
        \,&=\, SU(2)\,\rtimes \, H\,,\qquad \forall k\,.
    \end{split}
\end{equation}
Meaning that $\mathrm{Hol}(X_{8})$ is an extension of $\mathrm{Hol}(\widetilde{X}_{8})=SU(2)$ by the finite $H$ subgroup. The possible finite holonomy groups $H$ are listed in Table \ref{Table:all-B4-spaces} and \eqref{eq:list-holonomy-group-H}. For completness, we present the general argument of the preceding equation in Appendix \ref{app:holonomy}.

In all cases, the total holonomy group has at least rank equals two. Indeed, these are subgroups of $Spin(7)$. There are, at least, $4$ possible chains of $\mathrm{Hol}(X_{8})$ embeddings:
\begin{equation}\label{eq:embedd-group-4-cases}
    \begin{split}
        &\mathrm{Case -}\, 1\qquad  SU(3)\,\subset\, SU(4)\,\subset \, Spin(7)\,,
        \\
        &\mathrm{Case -}\, 2\qquad Sp(1)\,\times\, Sp(1)\,\subset\, Sp(2)\,\subset\, SU(4)\,\subset \, Spin(7)\,,
        \\
        &\mathrm{Case -}\, 3\qquad   SU(3)\,\subset\, G_{2}\,\subset \, Spin(7)\, ,
        \\
        &\mathrm{Case -}\, 4\qquad   SU(2)\,\times\,SU(2) \,\subset\, G_{2}\,\subset \, Spin(7)\, .
    \end{split}
\end{equation}
For a given $X_{8}$, the possibly holonomy group $\mathrm{Hol}(X_{8})$ is a subgroup of one of the above subgroups, depending on the amout of 3d supercharges survive the compactification.

Before concluding this paragraph, it should be noted that the spaces 
\begin{equation}\label{eq:general-our-X8-topology}
X_{8}(B_{(4;b_{i})}^{(k)}) \,=\,    \frac{\left(\R^{4}/\Gamma_{ADE}\right)\,\times\,T^{4}}{H}
\end{equation}
admit $Spin(7)$-structures, but not necessary metrics with $Spin(7)$-holonomy. For all possible choices of $H$, the 3d theories always have more than two real supercharges. Therefore, our 8d spaces do not admit metrics with $Spin(7)$-holonomy.

\paragraph{The effective 3d theories and their field content.}

We now determine the degrees of freedom and the number of invariant supercharges for the two classes of Bieberbach spaces $B_{(4;2)}$ and $B_{(4;1)}$ mentioned in Table \ref{table:examples-of-M4} and listed in Table \ref{Table:all-B4-spaces}.

Given that the Euler characteristic satisfies $\chi(B_{4})=0$, the number of preserved supercharges, via \eqref{eq:N=chi-b1}, depends solely on $b_{1}(B_{4})$:
\begin{equation}\label{eq:Supercharges-N-B4-b1-N=2-4}
B_{(4;2)}\ :\quad \N\,=\, 4\,,\qquad  B_{(4;1)}\ :\quad \N\,=\, 2\,.
\end{equation}
The general criteria for having massless degrees of freedom can be summarized by the following: The spaces of $H^{1}_{\mathrm{dR}}(T^{4}/H)$ and  $H^{2}_{\mathrm{dR};\,L}(T^{4}/H)$ are precisely the subspaces of $H$-invariant (untwisted) forms on $T^{4}$ and give rise to massless fields, as discussed previously. In contrast, modes associated with $H$-variant (twisted) $p$-forms are projected out at the level of zero modes and therefore become massive. Further discussion on $H$-twisted versus $H$-untwisted $p$-forms and their physical interpretation is given in section \ref{sec:geometric-realization}.

We now consider the three possible cases:
\begin{itemize}

\item For $B_{(4;4)}^{(1)}=T^{4}$, we have 3d $\N=8$ $ADE$ gauge theories consist only of a vector multiplet(s). Using the terminology of the 3d $\N=1$ supermultiplets, the $\N=8$ vector multiplet is given as
\begin{equation}\label{eq:VM-3d-N=8}
  \mathrm{VM}_{3d}^{\N=8} \ : \ (A,\psi)\,\oplus\, \bigoplus_{i=1}^{4} \,\,(\phi^{(i)},\chi^{(i)})\,\oplus\, \bigoplus_{a=1}^{3} \,\, (\varphi^{(a)},\xi^{(a)})\,.
\end{equation}

\item For $B_{(4;2)}^{(2-8)}$, the compactification yields 3d $\N=4$ $ADE$ gauge theories. In terms of $\N=1$ supermultiplets, the degrees of freedom are organized as:
\begin{equation}\label{eq:VM-3d-N=4}
  \mathrm{VM}_{3d}^{\N=4} \ : \ (A,\psi)\,\oplus\,(\phi^{(1)},\chi^{(1)})\,\oplus\,(\phi^{(2)},\chi^{(2)})\,\oplus\,(\varphi,\xi)\,.
\end{equation}
The fields $\phi^{(1)}$ and $\phi^{(2)}$ naturally pair up to give a complex scalar field within the $\mathrm{VM}_{3d}^{\N=4}$. 

Comparing to $\mathrm{VM}_{3d}^{\N=8}$ above, four of the 3d $\N=1$ scalar multiplets become massive. These modes can have identical mass due to the following reasoning:
\begin{itemize}
  \item Fine-tuning the radii along which the $H$ group acts. Requiring that the radii along the $H$-twisted directions are equal, insures that the Kaluza-Klein masses of the corresponding fields degenerate.

For example, suppose that we have $H=\Z_{2}$ which acts by exchanging two coordinates. In this case, one notices that having identical radii is a required to make sense of the $H$ transformation.   
    
  \item The preserved supersymmetry requires that these massive degrees of freedom to be organized in $\N=4$ supermultiplet. Since we have four real scalar, then it is natural to organize these fields in one massive 3d $\N=4$ adjoint hypermultiplet, which is given by:
\begin{equation}
\mathrm{HM}_{3d}^{\N=4}\,=\,  (\phi^{(3)},\chi^{(3)})\,\oplus\, (\phi^{(4)},\chi^{(4)}) \,\oplus \, (\varphi^{(2)},\xi^{(2)})\,\oplus\, (\varphi^{(3)},\xi^{(3)})\,.
\end{equation} 
Hence, supersymmetry, in some sense, forces the previous fine-tuning point. 
\end{itemize}
Therefore, one expects that the correct interpretation of the effective 3d theory is given as:
\begin{equation}\label{eq:effective-3d-N=4*}
  \mathcal{T}(B_{(4;2)}^{(2-8)})\,\, =\, \, \text{3d $\N=4^{\ast}$ $ADE$ gauge theory.}
\end{equation}
    
The holonomy group of the total 8d space must be a subgroup of $Spin(7)$ that preserves $1/4$ of the supercharges, leading to the possibilities:
\begin{equation}\label{eq:Hol-for-N=4}
    \mathrm{Hol}(X_{8})\,\subset \,SU(2)\,\times\,SU(2)\,,\quad \mathrm{or}\quad SU(3)\,.
\end{equation}
Since the general structure of $\mathrm{Hol}(X_{8})$ is as given in \eqref{eq:Hol-extension-H}, then the first possibility is excluded\footnote{Another way to exclude this possibility, is to recall that the hyper-K\"ahler $Sp(1)$-structure $\{\omega_{i}\}$ is not invariant under the action of $H$ and so can not extend to the total $X_{8}$ space.}.

\item For $B_{(4;1)}^{(9-27)}$, excluding the cases with $k=16,17,24$ as they are non-spin manifolds, the compactification yields 3d $\N=2$ $ADE$ gauge theories, with degrees of freedom:
\begin{equation}\label{eq:VM-3d-N=2}
        \mathrm{VM}_{3d}^{\N=2} \ : \ (A,\psi)\,\oplus\,(\phi,\chi)\,.
\end{equation}
Comparing to \eqref{eq:VM-3d-N=8}, there are 6 adjoint-valued scalar multiplets become massive, 3 of the type $(\phi,\chi)$ and the other of the $(\varphi,\xi)$ type. As in the previous case, these fields can be taken to have identical masses, a feature dictated by supersymmetry together with fine-tuning. In particular, supersymmetry organizes the six scalar multiplets into three massive 3d $\N=2$ adjoint-valued complex chiral multiplets. Since, a priori, there is a freedom in how these scalars are paired, consistency with supersymmetry requires that all such pairings to be equivalent, which also seems to force the fine-tuning of the radii. Hence, we have a degeneracy in the spectrum, i.e., the six scalar multiplets acquire identical masses. Therefore, the effective field theories can then be described as:
\begin{equation}\label{eq:effective-3d-N=2*}
  \mathcal{T}(B_{(4;1)}^{(9-27)})\,\,=\, \, \text{3d $\N=2^{\ast}$ $ADE$ gauge theory.}
\end{equation}

One can consider the following possible pairing:
\begin{equation}\label{eq:massive-chiral}
  \Phi_{1}\,=\,\varphi^{(1)}\,+\,i\,\phi^{(1)}\,,\quad \Phi_{2}\,=\,\varphi^{(2)}\,+\,i\,\phi^{(2)}\,,\quad \Phi_{3}\,=\,\varphi^{(3)}\,+\,i\,\phi^{(3)}\,,
\end{equation}
with $\Phi_{i}$ being the massive chiral multiplets. Recall that $\varphi^{(i)}$ are the scalar fields obtained from the reduction of the triplet 7d Higgs field $\mathsf{\Phi}$. These chiral multiplets will play a crucial role in section \ref{sec:HB}. 
    
The embedding of holonomy groups that preserve $1/8$ of the M-theory supercharges is:
\begin{equation}\label{eq:Hol-for-N=2}
    \mathrm{Hol}(X_{8}) \ \subset \ SU(4)\,.
\end{equation}
It can be further embedded in $G_{2}$, where the $G_{2}$-structure is defined on a 7d subspace of $X_{8}$. 
\end{itemize}
We also note that the field theories of the latter two points are mass-deformations of the theory associated with the first point, i.e., with the trivial Biberbach space. 

In the next subsection, we consider the co-Seifert structure of the Bieberbach spaces and extend it to the 8d $X_{8}$ spaces. We also discuss the relevant $SU(4)$-structure on the $X_{8}$ spaces along with the correct $G_{2}$-structure on 7d subspaces.

\subsection{Co-Seifert fibrations and other \texorpdfstring{$\mathscr{G}$}{G}-structures on \texorpdfstring{$X_{8}$}{X8}}\label{sec:G-structure-co-seifert}

Bieberbach manifolds possess rich geometric structures, including natural fibration patterns described by the co-Seifert constructions. This framework has been extensively studied in \cite{Ratcliffe_2010,ratcliffe2012,lambert2013} and is reviewed in Appendix \ref{sec-co-Seifert-fibration} for completeness.

The co-Seifert fibration provides a systematic method to decompose Bieberbach manifolds. For a given 4-dimensional space group $\Gamma$, there exists a 3-dimensional subgroup $\Gamma'$, such that the corresponding Bieberbach 4-manifold admits a topological decomposition given as:
\begin{equation}
    B_{4}\,=\,\frac{B_{3}\,\times\, \bbS^{1}}{G}\,.
\end{equation}
Here, $B_{3}$ is one of the 6 cases of Bieberbach 3-manifolds as listed in Table \ref{Table:all-B3-spaces} and plays the role of a typical fiber. $\bbS^{1}$ is the base, and $G$ is the structure group.

Following the discussion around \eqref{eq:Bn=TnmHtilde-TmG} and \eqref{eq:H-subgroup-Htilda-G}, this decomposition can be expressed more explicitly as:
\begin{equation}\label{B4=T3Htilde-S1G}
    B_{4}\,=\,\frac{\left(T^{3}/\widetilde{H}\right)\,\times\, \bbS^{1}}{G}\,.
\end{equation}
Here, $\widetilde{H}$ is the holonomy group of the Bieberbach 3-manifold $B_{3}=T^{3}/\widetilde{H}$, with
\begin{equation}
    \widetilde{H}\,\in\, \{\,\mathds{1},\,\Z_{2},\,\Z_{3},\,\Z_{4},\,\Z_{6},\,\Z_{2}\,\times\,\Z_{2} \,\}\,,
\end{equation} 
as seen from Table \ref{Table:all-B3-spaces}. One can read a representation for the $\widetilde{H}$ group from \cite{brown1978} using the the column labeled “BBNWZ” in the same table. We also note that one can always find an orthogonal representation of the $\widetilde{H}$ matrices. The groups $G$ are given in tables \ref{Table:fiberB31}-\ref{Table:fiberB36} along with $B_{4}$ and the corresponding typical fibers. Consequently, the holonomy group $H$ of the 4-manifold $B_{4}=T^{4}/H$ satisfies:
\begin{equation}
    H\,=\, \widetilde{H}\,\rtimes\,G\,.
\end{equation}
See the discussion in Appendix \ref{app:holonomy} for general discussion and derivation. Since we work exclusively with orthogonal representations of both $H$ and $\widetilde{H}$, the structure group $G$ must likewise admit an orthogonal representation.

\paragraph{Co-Seifert decomposition of $X_{8}$ and a $G_{2}$-structure on $Y_{7}$.} 

The co-Seifert fibration structure described above can be carried to the 8-dimensional space:
\begin{equation}
X_{8}(B_{(4;b_{1})}^{(k)}) \,=\,    \frac{\left(\R^{4}/\Gamma_{ADE}\right)\,\times\,T^{4}}{H}\,.
\end{equation}
$B_{4}$ has typical fiber $B_{3}=T^{3}/\widetilde{H}$, up to a $G$ action. In particular, we may reinterpret $X_{8}$ as a fibration of a 7-dimensional space $Y_{7}(\widetilde{H})$ over the base $\bbS^{1}$, modulo the $G$-action, where:
\begin{equation}
   Y_{7}(\widetilde{H})\,=\,  \frac{\left(\R^{4}/\Gamma_{ADE}\right)\,\times\,T^{3}}{\widetilde{H}}\,.
\end{equation}
These spaces and their holonomy groups have been established in \cite{Acharya:1998pm}, which are given as $SU(2)\rtimes \widetilde{H}$. It was also shown that the $Y_{7}(\widetilde{H})$ spaces admit parallel $G_{2}$-structure given as:
\begin{equation}\label{eq:G2-structure-Y7}
    \varphi_{3}\,=\,\sum_{i=1}^{3}\,\omega_{i}\,\wedge\,\dd y_{i}\,+\, \mathrm{vol}(T^{3})\,,
\end{equation}
which is invariant under the $\widetilde{H}$ action, in analogues way to the invariance of the parallel $Spin(7)$-structure discussed around \eqref{eq:invariance-H-omega-alpha}, provided that we take orthogonal representations for $\widetilde{H}$.

The holonomy groups of the $Y_{7}$ spaces are embedded as \cite{Acharya:1998pm}, 
\begin{equation}\label{eq:holonomy-B3}
  \mathrm{Hol}(Y_{7})\,=\,
    \begin{cases}
        SU(2)\,\rtimes\,\Z_{n}\, \subset\, SU(3) \qquad  &\text{for \quad $B_{3}^{(2-5)}$}\,,\\
         SU(2)\,\rtimes\,(\Z_{2}\,\times\,\Z_{2})\,\subset\, G_{2} \qquad   &\text{for \quad $B_{3}^{(6)}$}\,.
    \end{cases}
\end{equation}
In the former case, the total space preserve $1/4$ of the M-theory supercharges, while in the later case, it preserve $1/8$. The space $B_{3}^{(1)}$ is nothing but $T^{3}$ and the total holonomy is $SU(2)$ preserving $1/2$ of the M-theory supercharges.

The holonomy group of the quotient space $X_{8}$ can be then reexpressed by:
\begin{equation}\label{eq:docomposed-holonomy}
 \begin{split}
     \mathrm{Hol}(X_{8})\,&\,=\, \mathrm{Hol}(Y_{7})\,\rtimes\,G\,
     \\
     \,&\,=\,  SU(2)\,\rtimes\,\widetilde{H}\,\rtimes\,G\,,
     \\
     \,&\,=\, SU(2)\,\rtimes\, H\,.
 \end{split}
\end{equation}
As evident from the tables \ref{Table:fiberB31}-\ref{Table:fiberB36} and Table \ref{Table:all-B3-spaces}, both groups  $\widetilde{H}$ and $G$ are cyclic groups.

As noted earlier, the $G_{2}$-structure $\varphi_{3}$ is invariant under the action of $\widetilde{H}$. Moreover, one can verify its invariance under the structure group $G$ with orthogonal representation. Consequently, $\varphi_{3}$ remains invariant under the full holonomy group $H$ and therefore\footnote{One can also argue that it can be extends naturally to the total 8-dimensional space $X_{8}$. However, such a claim is not needed in our discussion here.} it can be defined on a generic fiber $Y_{7}$ over $\bbS^{1}$. 

As we will observe, in many cases, the structure group $G$ is trivial and the total $X_{8}$ space can be think of as a direct $\bbS^{1}$ reduction, i.e., 
\begin{equation}
  X_{8}\,=\, Y_{7}\,\times\, \bbS^{1}\,,\qquad \text{for}\quad G \,=\,  \mathrm{id}\,.
\end{equation}
In these cases, the $G_{2}$-structure accommodates all possible holonomy groups of the form $SU(2)\rtimes H\,=\,SU(2)\rtimes \widetilde{H}$. Thus, we note that the corresponding 3d theories seen as an $\bbS^{1}$ reduction of 4d $\N=2^{\ast}$ and $\N=1^{\ast}$ theories which are associated with $Y_{7}$ spaces. The analysis of \cite{Acharya:1998pm} does not interpret the resulting 4d theories as mass deformations of 4d $\N=4$ theories. However, our arguments presented in the previous subsection apply equally well in this context, and naturally lead to such an interpretation.

\paragraph{An $SU(4)$-structure on $X_{8}$.}

One possible $SU(4)$-structure on the total space can be defined by first considering the following complex differential 1-forms on $X_{8}(\Z_{N},B_{4})$:
\begin{equation}
  \begin{split}
        &\dd Z_{1}\,=\,e^{1}\,+\,i\,\dd y_{1}\,,\qquad \dd Z_{2}\,=\,e^{2}\,+\,i\,\dd y_{2}\,,
        \\
        &\dd Z_{3}\,=\,e^{3}\,+\,i\,\dd y_{3}\,,\qquad \dd Z_{4}\,=\,e^{4}\,+\,i\,\dd y_{4}\,,
  \end{split}
\end{equation}
Here, $(e^{1},\cdots,e^{4})$ are the set of 1-forms, or form-fields, associated with the metric on $\R^{4}/\Z_{N}$ and presented in \eqref{eq:form-fields}.

The K\"ahler 2-form and the holomorphic top-form can then have the canonical expressions as: 
\begin{equation}\label{eq:su4-structure}
    \omega^{(1,1)}\,=\, \frac{i}{2}\,\sum_{i=1}^{4}\, \dd Z_{i}\,\wedge\,\dd \overline{Z}_{i}\,,\qquad    \Omega^{(4,0)}\,=\, \dd Z_{1}\,\wedge\,\dd Z_{2}\,\wedge\,\dd Z_{3}\,\wedge\, \dd Z_{4}\,.
\end{equation}

For this $SU(4)$-structure to be consistent, it must be compatible with the $Spin(7)$-structure defined in \eqref{eq:Phi-4-form}. One can explicitly work out that the $SU(4)$-structure obeys the compatibility condition, 
\begin{equation}
  \Phi_{4}\,=\,  \, \frac{1}{2}\,\omega^{(1,1)}\,\wedge\,\omega^{(1,1)}\,+\, \mathrm{Re}(\Omega^{(4,0)})\,.
\end{equation}
Additionally, the $SU(4)$-structure is invariant under the rotational holonomy group $H$. This invariance can be further verified using the groups $\widetilde{H}$ and $G$. This is another consistency check for the proposed $SU(4)$-structure. Hence, it is a well-defined structure on the $X_{8}(\Z_{N},B_{4})$ spaces\footnote{Furthermore, one expect this conclusion to hold for the more general cases of $X_{8}(\Gamma_{DE},B_{4})$. However, for these cases, the explicit metric and the form-fields are not known.} and the holonomy groups in \eqref{eq:Hol-for-N=2} can be embedded naturally in $SU(4)$.

%%%%%%%%%%%%%%%%%%%%%%%%%%%%%%%%%%%%%%%%%%%%%%%%%%%%%%%%%%%%%%%%%%%

\subsection*{Co-Seifert fibrations of the 24 \texorpdfstring{$X_{8}$}{X8} spaces}

In the following, we examine in some details the co-Seifert fibration of the $X_{8}$ spaces. We provide selected examples for illustrations throughout this discussion and restrict attention to the action of the structure group $G$ on the fiber directions. The action on the base $\bbS^{1}$ plays no role in our analysis and can be found in \cite{lambert2013}.

%%%%%%%%%%%%%%%%%%%%%%%%%%%%%%%%%%%%%%%%%%%%%%%%%%%%%%%%%%%%%%%%%%%%%%%%%%

\subsubsection*{Cases with typical fiber \texorpdfstring{$B_{3}^{(1)}$}{B31}}

These represent the simplest scenarios, where the co-Seifert fibers are copies of $\R^{4}/\Gamma_{ADE}\times T^{3}$ with $SU(2)$ holonomy of the $\R^{4}/\Gamma_{ADE}$ subspace. The 8-dimensional space takes the form:
\begin{equation}
   \left( \frac{\R^{4}/\Gamma_{ADE}\times T^{3}}{\mathds{1}}\,\times \, \bbS^{1}\right)/\Z_{n}\,,  \quad \text{with } \,\,\, n = 2, 3, 4, 6.
\end{equation}
Table \ref{Table:fiberB31} enumerates the 8 distinct realizations of this geometric structure.

The holonomy group of the total space is $SU(2)\,\rtimes\,\Z_{n}$, which embeds naturally into $SU(3)$, i.e.,  
\begin{equation}
    SU(2)\,\rtimes\,\Z_{n}\,\subset\,SU(3)\,.
\end{equation}
Consequently, we expect the holonomy to preserve $1/4$ of the M-theory supercharges. This result agrees with the twisted reduction analysis presented in \eqref{eq:N=chi-b1} and Table \ref{table:examples-of-M4}.

\begin{table}[H]
\centering{
\begin{tabular}{|c|c|c|}
\hline	
 4-manifold  &  Structure group $G$ &  $\mathrm{Hol}(X_{8}(B_{(4;b_{1})}^{(k)}))$ \\
 \hline	
 \hline	
$B_{(4;2)}^{(1)}$   & $\mathds{1}$ & $SU(2)$ 
	  	  \\ \hline
$B_{(4;2)}^{(2)}$  &  $\Z_{2}$ & $SU(2)\rtimes \Z_{2}\subset SU(3)$
		    \\ \hline
$B_{(4;2)}^{(3)}$  &  $\Z_{2}$ & $SU(2)\rtimes \Z_{2}\subset SU(3)$
 		   \\ \hline
$B_{(4;2)}^{(4)}$  &  $\Z_{3}$ & $SU(2)\rtimes \Z_{3}\subset SU(3)$
 		    \\ \hline
 $B_{(4;2)}^{(5)}$  &  $\Z_{3}$ & $SU(2)\rtimes \Z_{3}\subset SU(3)$
 		    \\ \hline
$B_{(4;2)}^{(6)}$  &  $\Z_{4}$ & $SU(2)\rtimes \Z_{4}\subset SU(3)$
 	   \\ \hline
$B_{(4;2)}^{(7)}$  &  $\Z_{4}$ & $SU(2)\rtimes \Z_{4}\subset SU(3)$
 	   \\ \hline
$B_{(4;2)}^{(8)}$  &  $\Z_{6}$ & $SU(2)\rtimes \Z_{6}\subset SU(3)$
 	    \\ \hline
\end{tabular}
}
\caption{The holonomy group of $B_{3}^{(1)}$ is $\mathds{1}$. The matrix representations of the $G$ groups can be found in table 10 of \cite{lambert2013}.}
\label{Table:fiberB31}
\end{table}

%%%%%%%%%%%%%%%%%%%%%%%%%%%%%%%%%%%%%%%%%%%%%%%%%%%%%%%%%%%%%%%%%%%%%%%%%%%%%%%%

\subsubsection*{Cases with typical fiber \texorpdfstring{$B_{3}^{(2)}$}{B32}}  

We now examine the scenario where the typical fiber is $B_{3}^{(2)}$, which has holonomy $\widetilde{H}=\Z_{2}$. The 7-dimensional fibers are given by:
\begin{equation}
    \frac{\R^{4}/\Gamma_{ADE}\,\times\,T^{3}}{\Z_{2}}\,,
\end{equation}
with $SU(2)\rtimes \Z_{2}$ holonomy embedded in $SU(3)$.

As indicated in Table \ref{Table:fiberB32}, the first five cases spaces exhibit $SU(2)\rtimes\Z_{n}$ holonomy with $n=1,4,6$. These cases appeared in the previous scenario with $B_{3}^{(1)}$-fiber. However, here, they admit a distinct geometric realization, i.e., co-Seifert fibrations. Note that $B_{(4;2)}^{(2)}$ and $B_{(4;2)}^{(3)}$ have trivial structure groups; hence, the associated 3d $\N=4^{\ast}$ field theories can be seen as $\bbS^{1}$ reduction of 4d $\N=2^{\ast}$ theories. As a non-tricial example, we will analyze the $X_{8}(B_{(4;2)}^{(6)})$ space.

The remaining $X_{8}$ spaces in Table \ref{Table:fiberB32} possess holonomy $SU(2)\rtimes (\Z_{2}\times\Z_{2})$. As discussed earlier, this group embeds into $SU(4)$, in agreement with the twisted-reduction prediction from equation \eqref{eq:N=chi-b1}. This follows from the observation that the fibers $Y_{7}$ in these cases has holonomy $SU(2)\rtimes \Z_{2}\subset SU(3)$, leading to the chain:
\begin{equation}
    SU(2)\rtimes (\Z_{2}\times\Z_{2})\,\subset\, SU(3)\,\rtimes\,\Z_{2}\,\subset SU(4)\,.
\end{equation}
To illustrate this, we will analyse $X_{8}(B_{(4;1)}^{(9)})$ as a representative examples in detail.

\paragraph{\underline{Example}: the space $X_{8}(B_{(4;2)}^{(6)})$.}

The geometry under consideration admits the co-Seifert fibration structure:
\begin{equation}
    X_{8}(B_{(4;2)}^{(6)})\,=\,\left( \frac{\R^{4}/\Gamma_{ADE}\, \times \, T^{3}}{\Z_{2}}\,\times  \bbS^{1}\right)/\Z_{4}\,.
\end{equation}
The structure group $\Z_{4}$ is generated by the matrix:
\begin{equation}
A\,=\,        \begin{pmatrix}
        1&0&0\\
        0&0&-1\\
        0&1&0
    \end{pmatrix}\,.
\end{equation}
which corresponds to the $\Z_{4}$ holonomy generator of $B_{(4;2)}^{(6)}$, denoted by $B$ below. Specifically, it is identified with the lower $(3\times 3)$ block diagonal of  
\begin{equation}
B\,=\,    \begin{pmatrix}
        1&0&0&0\\
        0&1&0&0\\
        0&0&0&-1\\
        0&0&1&0
    \end{pmatrix}\,.
\end{equation}
Further, the $\Z_{2}$ holonomy of the $Y_{7}=\frac{\R^{4}/\Gamma_{ADE}\, \times \, T^{3}}{\Z_{2}}$ subspace(s) is generated by $C=\mathrm{diag}(1,-1,-1)$. We observe that $C=A^{2}$, indicating that the $\Z_{2}$ holonomy forms a subgroup of the $\Z_{4}$ structure group. Consequently, the holonomy of the total space is $ SU(2) \rtimes \Z_{4} \subset SU(3)$. Combined with the fact that the effective 3d theory possesses 8 supercharges, we conclude that the holonomy of the total space is indeed a subgroup of $SU(3)$.

\paragraph{\underline{Example}: the space $X_{8}(B_{(4;1)}^{(9)})$.}

The co-Seifert representation of the total space is given as:
\begin{equation}
    X_{8}(B_{(4;1)}^{(9)})\,=\,\left( \frac{\R^{4}/\Gamma_{ADE}\, \times \, T^{3}}{\Z_{2}}\,\times  \bbS^{1}\right)/\Z_{2}
\end{equation}
In contrast to the previous example, the two $\Z_{2}$ factors here act independently. One $\Z_{2}$ factor acts solely on the fibers and generated by $A=\mathrm{diag}(-1,-1,1)$. The other factor acts on the total space as a structure group and generated by $A=\mathrm{diag}(-1,1,-1)$. This implies that the holonomy of the fibers is $SU(2) \rtimes \Z_{2}\subset SU(3)$ and the holonomy of the total space is $SU(2) \rtimes \Z_{2} \rtimes \Z_{2}$.

\begin{table}[H] 
\centering{
\begin{tabular}{|c|c|c|}
\hline	
 4-manifold  &  Structure group $G$ &  $\mathrm{Hol}(X_{8}(B_{(4;b_{1})}^{(k)}))$ \\
 \hline	
 \hline	
$B_{(4;2)}^{(2)}$   & $\mathds{1}$ & $SU(2)\,\rtimes\,\Z_{2}\,\subset\, SU(3)$ 
	  	  \\ \hline
$B_{(4;2)}^{(3)}$  &  $\mathds{1}$ & $SU(2)\rtimes\Z_{2}\subset SU(3)$
		    \\ \hline
$B_{(4;2)}^{(6)}$  &  $\Z_{4}$ & $SU(2)\rtimes\Z_{4}\subset SU(3)$
 		   \\ \hline
$B_{(4;2)}^{(7)}$  &  $\Z_{4}$ & $SU(2)\rtimes\Z_{4} \subset SU(3)$
 		    \\ \hline
 $B_{(4;2)}^{(8)}$  &  $\Z_{6}$ & $SU(2)\rtimes\Z_{6}\subset  SU(3)$ \\ \hline
$B_{(4;1)}^{(9)}$  &  $\Z_{2}$ & $SU(2)\rtimes\Z_{2} \rtimes \Z_{2}$ \\ \hline
$B_{(4;1)}^{(10)}$  &  $\Z_{2}$ & $SU(2)\rtimes\Z_{2} \rtimes \Z_{2}$ \\ \hline
$B_{(4;1)}^{(11)}$  &  $\Z_{2}$ & $SU(2)\rtimes\Z_{2} \rtimes \Z_{2}$ \\ \hline
$B_{(4;1)}^{(12)}$  &  $\Z_{2}$ & $ SU(2)\rtimes\Z_{2} \rtimes \Z_{2}$ \\ \hline
$B_{(4;1)}^{(13)}$  &  $\Z_{2}$ & $ SU(2)\rtimes\Z_{2} \rtimes \Z_{2}$ \\ \hline
\end{tabular}
}
\caption{The holonomy group of $B_{3}^{(2)}$ is $\Z_{2}$. The matrix representations of the $G$ groups can be found in table 15 of \cite{lambert2013}.}
\label{Table:fiberB32}
\end{table}

%%%%%%%%%%%%%%%%%%%%%%%%%%%%%%%%%%%%%%%%%%%%%%%%%%%%%%%%%%%%%%%%%%%%%%%%%%

\subsubsection*{Cases with typical fiber $B_{3}^{(3)}$.}

From table \ref{Table:fiberB33}, we observe that there are two generic classes. The first, for the spaces $X_{8}(B_{(4;2)}^{(4)})$ and $X_{8}(B_{(4;2)}^{(5)})$, which were previously analysed in a different co-Seifert realization, exhibit holonomy groups contained in $SU(3)$. These geometries preserve $1/4$ of the M-theory supercharges, in agreement with the twisted reduction analysis presented in the previous subsection. 

In the second case, we have the spaces $X_{8}(B_{(4;1)}^{(k)})$ for $k=18,19$, and $20$ have the following co-Seifert fibration:
\begin{equation}
 \frac{\R^{4}/\Gamma_{ADE}\, \times \, T^{4}}{\bbD_{3}}\,  \,=\,\,\left( \frac{\R^{4}/\Gamma_{ADE}\, \times \, T^{3}}{\Z_{3}}\,\times  \bbS^{1}\right)/\Z_{2}\,.
\end{equation}
The holonomy of the 7d fibers is given as $SU(2) \rtimes \Z_{3} \subset  SU(3)$. The holonomy of the total space is a $\Z_{2}$ extension of that, which can be embedded either in $G_{2}$ or $SU(4)$. The corresponding effective theory admits $4$ real supercharges.

For the case with typical fibers $B_{3}^{(3)}$, we note that $B_{(4;2)}^{(4)}$ and $B_{(4;2)}^{(5)}$ have trivial structure groups. Hence, the corresponding 3d $\N=4^{\ast}$ field theories can be seen as $\bbS^{1}$ reduction of 4d $\N=2^{\ast}$ theories corresponding to the space $Y_{7}(B_{3}^{(3)})$.

\begin{table}[H]
\centering{
\begin{tabular}{|c|c|c|}
\hline	
 4-manifold  &  Structure group $G$ &  $\mathrm{Hol}(X_{8}(B_{(4;b_{1})}^{(k)}))$ \\
 \hline	
 \hline	
$B_{(4;2)}^{(4)}$   & $\mathds{1}$ & $SU(2)\rtimes \Z_{3}\subset SU(3)$ 
	  	  \\ \hline
$B_{(4;2)}^{(5)}$  &  $\mathds{1}$ & $SU(2)\rtimes \Z_{3} \subset SU(3)$
		    \\ \hline
$B_{(4;2)}^{(8)}$  &  $\Z_{2}$ & $SU(2)\rtimes \Z_{6} \subset SU(3)$
 		   \\ \hline
$B_{(4;1)}^{(18)}$  &  $\Z_{2}$ & $ SU(2)\rtimes \Z_{3} \rtimes \Z_{2}$
 		    \\ \hline
 $B_{(4;1)}^{(19)}$  &  $\Z_{2}$ & $ SU(2)\rtimes \Z_{3} \rtimes \Z_{2} $ \\ \hline
  $B_{(4;1)}^{(20)}$  &  $\Z_{2}$ & $ SU(2)\rtimes \Z_{3} \rtimes \Z_{2} $ \\ \hline
\end{tabular}
}
\caption{The holonomy group of $B_{3}^{(3)}$ is $\Z_{3}$. The matrix representations of the $G$ groups can be found in table 19 of \cite{lambert2013}.}
\label{Table:fiberB33}
\end{table}

%%%%%%%%%%%%%%%%%%%%%%%%%%%%%%%%%%%%%%%%%%%%%%%%%%%%%%%%%%%%%%%%%%%%%%%%%%

\subsubsection*{Cases with typical fiber $B_{3}^{(4)}$.}

Similar to the previous scenarios, we have two generic cases. The first give holonomies contained in $SU(3)$ and preserve $1/4$ supercharges, which are the first two cases of Table \ref{Table:fiberB34}. In fact, the associated 3d theories can be seen as $\bbS^{1}$ reduction of 4d $\N=2^{\ast}$ theories constructed via $Y_{7}(B_{3}^{(4)})$.

The second case gives holonomies embedded in $G_{2}$ or $SU(4)$. In particular, the spaces $X_{8}(B_{(4;1)}^{(21)})$ and $X_{8}(B_{(4;1)}^{(22)})$ have the following with co-Seifert fibration:
\begin{equation}
 \frac{\R^{4}/\Gamma_{ADE}\, \times \, T^{4}}{\bbD_{4}}\,  \,=\,\,\left( \frac{\R^{4}/\Gamma_{ADE}\, \times \, T^{3}}{\Z_{4}}\,\times  \bbS^{1}\right)/\Z_{2}\,.
\end{equation}
The holonomy of the 7-dimensional fiber is $   SU(2)\rtimes\Z_{4}\subset SU(3)$. Therefore, the holonomy of the total 8-dimensional space is understood as
\begin{equation}
    SU(2)\,\rtimes\,\bbD_{4}\,=\, SU(2)\,\rtimes\,\Z_{4}\,\rtimes\,\Z_{2}\,\subset\, SU(3)\,\rtimes\, \Z_{2}\,.
\end{equation}

\begin{table}[H]
\centering{
\begin{tabular}{|c|c|c|}
\hline	
 4-manifold  &  Structure group $G$ &  $\mathrm{Hol}X_{8}(B_{(4;b_{1})}^{(k)})$ \\
 \hline	
 \hline	
$B_{(4;2)}^{(6)}$   & $\mathds{1}$ & $SU(2)\rtimes \Z_{4} \subset SU(3)$ 
	  	  \\ \hline
$B_{(4;2)}^{(7)}$  &  $\mathds{1}$ & $SU(2)\rtimes \Z_{4} \subset SU(3)$
		    \\ \hline
$B_{(4;1)}^{(21)}$  &  $\Z_{2}$ & $SU(2)\rtimes \Z_{4}\rtimes\Z_{2}$
 		   \\ \hline
$B_{(4;1)}^{(22)}$  &  $\Z_{2}$ & $SU(2)\rtimes \Z_{4}\rtimes\Z_{2}$
 		    \\ \hline
\end{tabular}
}
\caption{The holonomy group of $B_{3}^{(4)}$ is $\Z_{4}$. The matrix representations of the $G$ groups can be found in table 21 of \cite{lambert2013}.}
\label{Table:fiberB34}
\end{table}

%%%%%%%%%%%%%%%%%%%%%%%%%%%%%%%%%%%%%%%%%%%%%%%%%%%%%%%%%%%%%%%%%%%%%%%%%%

\subsubsection*{Cases with typical fiber $B_{3}^{(5)}$.}

In this class, there are only two possibilities. The first arises from the Bieberbach four-manifold $B_{4}^{(8)}$, which admits several different typical co-Seifert fibers, as listed in tables \ref{Table:fiberB31}-\ref{Table:fiberB33}. In all these instances, our results are mutually consistent: the resulting eight-dimensional space has holonomy contained in $SU(3)$ and therefore preserves only eight real supercharges.

The genuinely new case is the eight-dimensional space $X_{8}(B_{(4;1)}^{(25)})$ with the $B_{(4;1)}^{(25)}$ Bieberbach space. Its rotational holonomy group is generated by:
\begin{equation}
   A\,=\, \begin{pmatrix}
        1&0&0&0\\
        0&1&0&0\\
        0&0&0&-1\\
        0&0&1&1
    \end{pmatrix}\,\,,
    \qquad \qquad 
   B\,=\, \begin{pmatrix}
        1&0&0&0\\
        0&-1&0&0\\
        0&0&0&1\\
        0&0&1&0
    \end{pmatrix}\,.
\end{equation}
We note that the generator $A$ is not orthogonal. However, after an appropriate change of basis, one obtains an orthogonal representative $\widetilde{A}$ of the form:
\begin{equation}
 \widetilde{A}\,=\,
 \begin{pmatrix}
        1&0&0&0\\
        0&1&0&0\\
        0&0&\frac{1}{2}&\frac{\sqrt{3}}{2}\\
        0&0&-\frac{\sqrt{3}}{2}&\frac{1}{2}
    \end{pmatrix}\,.
\end{equation}

Using this orthogonal form, one verifies that the $SU(4)$-structure, as introduced in \eqref{eq:su4-structure}, is preserved under the finite group generated by $\widetilde{A}$ and $B$. Consequently, the holonomy group of the full eight-dimensional space is:
\begin{equation}
    SU(2)\,\rtimes\,\Z_{6}\,\rtimes\, \Z_{2}\,\subset SU(3)\,\rtimes\, \Z_{2}\,\subset\, SU(4)\,.
\end{equation}

Moreover, the 7d fiber $Y_{7}(B_{3}^{(5)})$ possesses a $G_{2}$-structure that is invariant under the $\Z_{2}$ action on the total space. Consequently, all $Y_{7}$ fibers of the 8d space $X_{8}(B_{(4;1)}^{(25)})$ have compatible $G_{2}$-structures. Hence, the holonomy group $SU(2)\rtimes \bbD_{6}$ embeds naturally in the group $G_{2}$.

\begin{table}[H] 
\centering{
\begin{tabular}{|c|c|c|}
\hline	
 4-manifold  &  Structure group $G$ &  $\mathrm{Hol}(X_{8}(B_{(4;b_{1})}^{(k)}))$ \\
 \hline	
 \hline	
$B_{(4;2)}^{(8)}$   & $\mathds{1}$ & $SU(2)\rtimes\Z_{6} \subset SU(3)$ 
	  	  \\ \hline
$B_{(4;1)}^{(25)}$  &  $\Z_{2}$ & $SU(2)\rtimes\Z_{6}\rtimes\Z_{2}$
		    \\ \hline
\end{tabular}
}
\caption{The holonomy group of $B_{3}^{(5)}$ is $\Z_{6}$. The matrix representations of the $G$ groups can be found in table 23 of \cite{lambert2013}.}
\label{Table:fiberB35}
\end{table}

%%%%%%%%%%%%%%%%%%%%%%%%%%%%%%%%%%%%%%%%%%%%%%%%%%%%%%%%%%%%%%%%%%%%%%%%%%

\subsubsection*{Cases with typical fiber $B_{3}^{(6)}$.}

Two main themes arise in this class of examples.

\paragraph{(i) Trivial structure group.} For configurations with trivial structure group, as summarized in Table \ref{Table:fiberB36}, the holonomy group of the eight-dimensional space $X_{8}$ coincides precisely with the holonomy of the typical seven-dimensional fiber
\begin{equation}
    Y_{7}\;=\;\frac{\left(\,\R^{4}/\Gamma_{\mathrm{ADE}}\,\right)\,\times\, T^{3}}{\Z_{2}\times \Z_{2}}\,.
\end{equation}
M-theory compactification on $Y_{7}$ yields 4d $ADE$ $\N=1^{\ast}$ gauge theories preserving four real supercharges\footnote{These spaces where considered originally in \cite{Acharya:1998pm} with 4d $\N=1$ interpretation rather than $\N=1^{\ast}$; this distinction plays a crucial role in section \ref{sec:HB}.} More explicitly, the holonomy group takes the form
\begin{equation}
    \mathrm{Hol}(Y_{7}) \;=\; SU(2)\rtimes(\Z_{2}\times \Z_{2}) \;\subset\; G_{2}\,.
\end{equation}
In this case, the 8d space is given by the direct product between the above $Y_{7}$ and $\bbS^{1}$. Hence, the $G_{2}$-structure on $X_{8}$ is inherited directly from the $G_{2}$-structure of the fiber $Y_{7}$.

The resulting 3d $\N=2^{\ast}$ theory may equivalently be viewed as an $\bbS^{1}$ reduction of the effective 4d $\N=1^{\ast}$ theory, so that the preserved supersymmetry and field content follow directly from the 4d description.

\paragraph{(ii) Non-trivial structure group.}
The second theme arises when the structure group of the fibration is non-trivial, namely for $G=\Z_{2}$ or $G=\Z_{3}$. In these cases, one can show that the $SU(4)$ structure defined by the Kähler form and holomorphic four-form in
\eqref{eq:su4-structure}, as well as the $G_{2}$ structure on the 7-dimensional fibers given in \eqref{eq:G2-structure-Y7}, are invariant under the action of $G$. Consequently, the holonomy group of $X_{8}$ admits an embedding into either structure groups.

\begin{longtable}{|c|c|c|}
\caption{The holonomy group of $B_{3}^{(6)}$ is $\mathbb{Z}_{2}\times\mathbb{Z}_{2}$. The spaces $B_{4}^{(16)}$, $B_{4}^{(17)}$, and $B_{4}^{(24)}$ do not admit a spin-structure \cite{lambert2013}; however, we are adding them for completeness. The matrix representations of the $G$ groups can be found in table 24 of \cite{lambert2013}.}
\label{Table:fiberB36} \\
\hline
4-manifold  &  Structure group $G$ &  $\mathrm{Hol}(X_{8}(B_{(4;b_{1})}^{(k)}))$ \\
\hline
\hline
\endfirsthead
\multicolumn{3}{c}{{\tablename\ \thetable{} -- continued from previous page}} \\
\hline
4-manifold  &  Structure group $G$ &  $\mathrm{Hol}(X_{8}(B_{(4;b_{1})}^{(k)}))$ \\
\hline
\hline
\endhead
\hline
%\multicolumn{3}{r}{{Continued on next page}} \\
\endfoot
\hline
\endlastfoot
$B_{(4;1)}^{(14)}$   & $\mathds{1}$ & $SU(2)\rtimes(\mathbb{Z}_{2}\times\mathbb{Z}_{2})$ \\
$B_{(4;1)}^{(15)}$  &  $\mathds{1}$ & $SU(2)\rtimes(\mathbb{Z}_{2}\times\mathbb{Z}_{2})$ \\
$B_{(4;1)}^{(16)}$   & $\mathds{1}$ & $SU(2)\rtimes(\mathbb{Z}_{2}\times\mathbb{Z}_{2})$ \\
$B_{(4;1)}^{(17)}$  &  $\mathds{1}$ & $ SU(2)\rtimes(\mathbb{Z}_{2}\times\mathbb{Z}_{2})$ \\
$B_{(4;1)}^{(23)}$   & $\mathbb{Z}_{2}$ & $SU(2)\rtimes(\mathbb{Z}_{2}\rtimes\mathbb{Z}_{2})$ \\
$B_{(4;1)}^{(24)}$  &  $\mathbb{Z}_{2}$ &  $SU(2)\rtimes(\mathbb{Z}_{2}\rtimes\mathbb{Z}_{2})$ \\
$B_{(4;1)}^{(26)}$   & $\mathbb{Z}_{3}$ & $SU(2)\rtimes(\mathbb{Z}_{3}\rtimes \mathbb{Z}_{2}\rtimes\mathbb{Z}_{2})$ \\
$B_{(4;1)}^{(27)}$  &  $\mathbb{Z}_{3}$ &  $SU(2)\rtimes(\mathbb{Z}_{3}\rtimes \mathbb{Z}_{2}\rtimes\mathbb{Z}_{2})$ \\
\end{longtable}

%\begin{table}[H]
%\centering{
%\begin{tabular}{|c|c|c|}
%\hline	
% 4-manifold  &  Structure group $G$ &  $\mathrm{Hol}(X_{8}(B_{(4;b_{1})}^{(k)}))$ \\
% \hline	
% \hline	
%$B_{(4;1)}^{(14)}$   & $\mathds{1}$ & $SU(2)\rtimes(\Z_{2}\times\Z_{2})$ 
%	  	  \\ \hline
%$B_{(4;1)}^{(15)}$  &  $\mathds{1}$ & $SU(2)\rtimes(\Z_{2}\times\Z_{2})$
%		    \\ \hline
%$B_{(4;1)}^{(16)}$   & $\mathds{1}$ & $SU(2)\rtimes(\Z_{2}\times\Z_{2})$ 
%	  	  \\ \hline
%$B_{(4;1)}^{(17)}$  &  $\mathds{1}$ & $ SU(2)\rtimes(\Z_{2}\times\Z_{2})$
%		    \\ \hline
%$B_{(4;1)}^{(23)}$   & $\Z_{2}$ & $SU(2)\rtimes(\Z_{2}\rtimes\Z_{2})$ 
%	  	  \\ \hline
%$B_{(4;1)}^{(24)}$  &  $\Z_{2}$ &  $SU(2)\rtimes(\Z_{2}\rtimes\Z_{2})$ 
%		    \\ \hline
%$B_{(4;1)}^{(26)}$   & $\Z_{3}$ & $SU(2)\rtimes(\Z_{3}\rtimes \Z_{2}\rtimes\Z_{2})$   
%	  	  \\ \hline
%$B_{(4;1)}^{(27)}$  &  $\Z_{3}$ &  $SU(2)\rtimes(\Z_{3}\rtimes \Z_{2}\rtimes\Z_{2})$ 
%		    \\ \hline
%\end{tabular}
%}
%\caption{The holonomy group of $B_{3}^{(6)}$ is $\Z_{2}\times\Z_{2}$. The spaces $B_{4}^{(16)}$, $B_{4}^{(17)}$ ,and $B_{4}^{(24)}$ do not admit a spin-structure \cite{lambert2013}; however, we are adding them for completness. The matrix representations of the $G$ groups can be found in table 24 of \cite{lambert2013}. }
%\label{Table:fiberB36}
%\end{table}

%%%%%%%%%%%%%%%%%%%%%%%%%%%%%%%%%%%%%%%%%%%%%%%%%%%%%%%%%%%%%%%%%%%%%%%%%%%%%%%%%%%%%%%%

\section{T-geometry and Higgs branches}\label{sec:CB-HB-geometric}

In this section, we investigate the possibility of realizing a Higgs branch for geometrically engineered 7d $\N =1$ SYM theories. We show that, cyclic permutations of the centers of the $\R^{4}/\Gamma_{ADE}$ space, generated by a group $H$, can in principle induce a Higgs branch through the introduction of a 7d nilpotent Higgs field. However, such a background does not preserve supersymmetry in the 7d theory by itself. Supersymmetry can be restored by compactifying the 7d theory on an internal space $Y_{n}$. For this construction to be consistent, the action of $H$ must extend to the internal space $Y_{n}$ in such a way that the total space
\begin{equation}
  X_{4+n} \;=\;
  \frac{\left(\,\R^{4}/\Gamma_{ADE}\,\times\, Y_{n}\,\right)}{H}
\end{equation}
preserves a non-trivial number of supercharges. We specify our discussion to key Biberbach 4-manifolds and the corresponding 3d $\N=2^{\ast}$. The discussion can also be applied to 4d $\N=1^{\ast}$ as done below. 

In the resulting lower-dimensional theories, the Higgsing is interpreted in terms of a nilpotent Higgs field. Within this framework, we determine the Higgs branch moduli and identify additional non--chiral charged matter transforming under the unbroken gauge algebra. As we will show, these fields are massless provided that the system is analyzed in the trapped matter framework developed in \cite{Cecotti:2009zf,Cecotti:2010bp} and later in \cite{Barbosa:2019bgh}.

%%%%%%%%%%%%%%%%%%%%%%%%%%%%%%%%%%%%%%%%%%%%%%%%%%%%%%%%%%%%%%%%%%%%%%%%%%%%

\subsection{Nilpotent Higgs and the 7d SYM theories}\label{sec:CB-HB-7d}

As outlined at the beginning of section \ref{sec:3d-theories-GE}, the codimension-four singularity $X_{4}=\R^{4}/\Gamma_{ADE}$ admits a crepant resolution
\begin{equation}\label{eq:resolution-X4}
    \widetilde{X}_{4}\,\xrightarrow[]{\,\,\pi_{C}\,\,}\, X_{4}\,,
\end{equation}
where $\widetilde{X}_{4}$ denotes the resolved geometry. This resolution introduces a bouquet of $r=\mathrm{rank}(G_{ADE})$ distinct 2-spheres $ \lor_{r}\bbS^{2}:=\bbS^{2}_{1}\lor\bbS^{2}_{2}\lor\cdots\lor\bbS^{2}_{r}$, whose intersection matrix is given by the negative of the Cartan matrix of the corresponding semi-simple $ADE$ Lie algebra. In the singular limit, we recover the 7d $\N=1$ Yang–Mills theory with $ADE$ gauge symmetry, as previously discussed.

From the perspective of the effective 7d $\N=1$ theory, the resolution corresponds to assigning a diagonal vacuum expectation value (vev) to the adjoint scalar field in the 7d vector multiplet. In particular, the 7d theory admit three adjoint scalars $ \mathsf{\Phi} = (\phi_{1},\phi_{2},\phi_{3})$ as can be read from \eqref{eq:7d-VM}. The fields $(\phi_{1},\phi_{2},\phi_{3})$ are in one-to-one correspondence with the hyperK\"ahler 2-forms $(\omega_{1},\omega_{2},\omega_{3})$. For example, in the $\mk{su}(n)$ Lie algebra case we can assign a vev as
\begin{equation}\label{eq:vev-phi1-diagonal}
\begin{split}
   \langle\phi_{1}\rangle\,&:=\, \langle\phi\rangle
   \\
\,&\,=\, \mathrm{diag}(\lambda_{1}\,,\lambda_{2}\,,\cdots\,,\lambda_{n}) \,,\ \ \text{with} \ \ \sum_{i=1}^{n}\,\lambda_{i}\,=\,0\,.
\end{split}
\end{equation}
Here, $\lambda_{i}$ are fixed eigenvalues and $\langle\phi\rangle$ is refer to as a semi-simple element of the Lie algebra. The most generic case with $\lambda_{i}\neq\lambda_{j}$ for all $i$ and $j$ the gauge symmetry is broken to the maximal torus $\mk{u}(1)^{\oplus(n-1)}$.

For any $ADE$ Lie algebra, a generic non-zero vev breaks the gauge symmetry to its maximal torus, thereby interpreting the resolution in \eqref{eq:resolution-X4} as the Coulomb branch (CB) of the 7d YM theories. This justifies the notation $\pi_{C}$ in \eqref{eq:resolution-X4}, where the projection maps the resolved geometry (i.e., the Coulomb branch) back to the singular geometry (i.e., the SYM description), by setting $\lambda_{i}=0\,, \forall i$. In other words, the vev is geometrically identified with the sizes of the  $\lor_{r}\bbS^{2}$ spheres.

Now, we would like to consider non-diagonaly Higgsing for the 7d theories. 

\paragraph{Nilpotent Higgsing.} In the 7d theories, a holomorphic adjoint scalar field $\Phi$ can be defined as
\begin{equation}\label{eq:holomorphic-Phi}
  \Phi\,=\, \phi_{2}\,+i\,\phi_{3}\,.
\end{equation}
This choice is equivalent to have an $SU(2)$-structure constructed from the hyperK\"ahler 2-forms. Non-diagonal vev's, i.e., nilpotent, can be assigned to the field $\Phi$ in contrast to $\phi_{2}$ and $\phi_{3}$ which are Hermitian. These vevs were analyzed in \cite{Donagi:2003hh} and later in a number of works including \cite{Cecotti:2010bp,Donagi:2011jy,Anderson:2013rka,Collinucci:2014qfa,Collinucci:2014taa,Barbosa:2019bgh}. In particular, one can assign a nilpotent, i.e., a triangular or a Jordan block vev, for $\Phi$ which takes values in the complixified gauge algebra $\mk{g}_{\C}$. In this work we use the abuse notation and drop the $\C$ subscript in $\mk{g}_{\C}$. Such a vev would Higgs the gauge algebra and reduce its rank. For example, consider an $\mk{su}(2)$ 7d SYM with a Higgs field acquiring a nilpotent vev as
\begin{equation}\label{eq:nilpotetn-vev-su(2)}
  \langle \Phi \rangle\,=\, \begin{pmatrix}
    0&1\\
    0&0
  \end{pmatrix}\,, \ \ \ \text{with}\ \ \ \langle \Phi \rangle^{2}\,=\,0\,.
\end{equation}
In this background, the gauge algebra is broken completely. We also note that we can turn on non-trivial vev for the $\phi_{1}$ field as in \eqref{eq:vev-phi1-diagonal}; in both cases, the gauge algebra is totally broken.

As noted, for instance, in \cite{Collinucci:2014qfa}, the $ADE$ singularities are insensitive to nilpotent Higgsing. For example, the $A_{1}$-singularity, described algebraically by
\begin{equation}\label{eq:resolving-A1-spectral-curve}
  uv \, = \, \det(\,z\,\mathds{1}_{2\times 2}\,-\,\langle \phi \rangle \, - \,\langle \Phi \rangle\,)\,, \ \ \text{with} \ \ u,v,z\,\in\,\C\,,
\end{equation}
is indeed unchanged under such Higgsing. Nevertheless, we argue that while the defining equation remains unchanged, a natural geometric interpretation of the nilpotent Higgs field can still be given.

\begin{figure}[H]
\centering{
\includegraphics[scale=0.4]{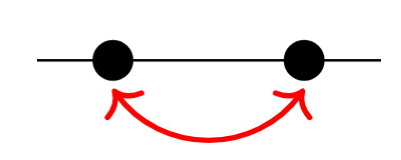}
}
\caption{The nodes represents the two centres of the $\widetilde{\R^{4}/\Z_{2}}$ geometry. The line between the two centres represent a 2-sphere of the creapant resolution. The red arrow represent the $\Z_{2}^{H}$ action on the two centres, which can also be understood as $\vec{x}_{1}\sim\vec{x}_{2}$   .}
\label{Fig:Z2-on-2-centres}
\end{figure}

The codimension-four singularity $X_{4}=\R^{4}/\Gamma_{ADE}$ can be described as $U(1)$ bundle over a base $\R^{3}$, see  \cite{GIBBONS1978430} and appendix \ref{sec:harmonic-2-forms}. The points in $\R^{3}$ where the $U(1)$ bundle degenerate are called centres, i.e., monopoles, and between any two centres we have a degenerate 2-sphere. In other words, the bouquet $\lor_{r}\bbS^{2}$ of 2-spheres that resolves the $ADE$ singularities can be equivalently described by the corresponding centres. For the case of $A_{1}$-singularity, there are exactly two centers. We propose that, Higgsing the $\mk{su}(2)$ gauge theory can be performed through the nilpotent vev in \eqref{eq:nilpotetn-vev-su(2)}, which is equivalent to the following geometric procedure depicted in Figure \ref{Fig:Z2-on-2-centres}:
\begin{itemize}
  \item Begin with two distinct labeled centers, say $\vec{x}_{1}$ and $\vec{x}_{2}$,
  \item allow them to undergo a permutation,
  \item and then take a quotient, which identifies centres with different labels, i.e., $\vec{x}_{1}\sim\vec{x}_{2}$. 
\end{itemize}
The action on the two centres can be represented by the matrix
\begin{equation}\label{eq:action-on-centers-su2-P2}
P \,=\,  \begin{pmatrix}
        0&1\\
        1&0
    \end{pmatrix}\,.
\end{equation}

At this stage, we observe that the upper triangle of the matrix $P$ correspond exactly to the nilpotent Higgs in \eqref{eq:nilpotetn-vev-su(2)}. In other words, the nilpotent Higgs is tidily related to the quotient on the centres. More detailed discussion and examples will be given in section \ref{sec:HB}.

The prescribed algorithm above can be seen as a $\Z_{2}^{H}$ quotient on the $A_{1}$-singularity. The superscript $H$ stands for Higgs. First, we note that the resolution of the $A_{1}$-singularity is given by turning on a vev of $\phi$ as
\begin{equation}\label{eq:vev-phi1-su(2)-case}
  \langle\phi\rangle \,=\, \mathrm{diag}(\,\lambda_{1}\,,-\,\lambda_{1}\,) \,,
\end{equation}
which give
\begin{equation}
  uv\,=\, (\,z\,+\,\lambda_{1}\,)\,(\, z\,-\,\lambda_{1} \,)\,.
\end{equation}
The eigenvalues $\pm\lambda_{1}$ correspond to the locations of the two centers. The permutation is performed by sending $\pm\lambda_{1}\to\mp\lambda_{1}$, which leave the equation invariant. This reminds us with the fact that adding nilpotent vev leave the $A_{1}$-singularity unchanged as in \eqref{eq:resolving-A1-spectral-curve}. The centres permutation induces an action on the vanishing 2-sphere of the resolved $A_{1}$-singulariy, which can be interpreted as flipping its orientation.

Upon taking the quotient, the $W^{\pm}$ bosons|which arise from M2-branes wrapping the corresponding 2-sphere with opposite orientations|are Higgsed. Intuitively, the quotient identifies $W^{+}$ and $W^{-}$; consistency of the quotient then requires that these modes be removed from the spectrum. Further, and most important, the quotient Higgs the photon associated with the 2-sphere itself. To fulfill our claim, we would like to consider the $L^{2}$ normalizable harmonic 2-forms denoted by  $\{\widetilde{\mathrm{h}}_{2}^{a}\}$ associated with the bouquet of 2-spheres $ \lor_{r}\bbS^{2}$. These 2-forms are reviewed in appendix \ref{sec:harmonic-2-forms}.

\paragraph{Harmonic 2-forms and the Higgs branch.}

The bosonic fields of the 7d SYM theories can be obtained through the aforthmentioned harmonic 2-forms in the following way. First, the Cartan 1-form gauge fields are given by the reduction of the M-theory $C_{3}$-form along $\{\widetilde{\mathrm{h}}_{2}^{(a)}\}$ as:
\begin{equation}
  C_{3}\, =\, \sum_{a=1}^{r}\, A_{1}^{a}\,\wedge\,\widetilde{\mathrm{h}}_{2}^{(a)} \,.
\end{equation}
While the Cartan-valued triplet Higgs field $\mathsf{\Phi}=(\phi_{1},\phi_{2},\phi_{3})$ can be expressed in terms of the $Sp(1)$-structure on $\R^{4}/\Gamma_{ADE}$ as:
\begin{equation}
 \sum_{i=1}^{3}\,\sum_{a=1}^{r}\,\,\phi_{i}^{a}\,\,\widetilde{\mathrm{h}}_{2}^{(a)}\,\wedge\, \omega_{i}\,.
\end{equation}

Let us now consider the nilpotent Higgsing and their action on the 2-forms. We will be focusing on the $\mk{su}(2)$ gauge theory given above. First, we note from \eqref{eq:tilde-h2}, the 2-forms $\{\widetilde{\mathrm{h}}_{2}^{a}\}$ are written in terms of other 2-forms $\{h_{2}^{a}\}$ that are directly related to the centres. For the example of $\mk{su}(2)$ gauge theory, the expression of the 2-form and its transformation under $\Z_{2}^{H}$ is given as: 
\begin{equation}\label{eq:Z2H-on-h2-7d-theory}
  \widetilde{\mathrm{h}}_{2}\,=\,(\, h_{2}^{(1)}\,-\,h_{2}^{(2)} \,) \  \xrightarrow[]{\,\,\Z_{2}^{H}\,\,} \  (\,h_{2}^{(2)}\,-\,h_{2}^{(1)}\,)\,=\, -\,   \widetilde{\mathrm{h}}_{2}\,,
\end{equation}
which is not invariant, i.e., $\Z_{2}^{H}$-twisted. The general requirement is that:
\begin{equation}\label{eq:7d-only-Z2H-invariant-forms}
  \parbox{10cm}{Only $\Z_{2}^{H}$-invariant $p$-forms are allowed on $(\R^{4}/\Z_{2})/\Z_{2}^{H}\,$.}
\end{equation}
This exclude $\widetilde{\mathrm{h}}_{2}$ and amount to Higgsing the gauge field corresponding to the Cartan of $\mk{su}(2)$, i.e., the photon. However, we note that by allowing the $\Z_{2}^{H}$ to have an induced action on the $Sp(1)$-structure, then certain field of $\{\phi_{i}\}$ survive the quotient and parametrize the Higgs branch of the theory.

One way to achieve the induced action is to embed the $\Z_{2}^{H}$ in the $Sp(1)$ and consider its action on the real coordinates $(x_{1},\cdots,x_{4})$ of $\widetilde{\R^{4}/\Z_{2}}$. This space has the description of $U(1)$ bundle over $\R^{3}$, we can take $(x_{1},x_{2},x_{3})$ along the base $\R^{3}$ and $x_{4}\sim x_{4}+1$ along the $U(1)$ fiber directions. Let us focus on a trivial action of $\Z_{2}^{H}$ the fiber directions, and locate the centers along the $x_{1}$-direction, then the induced action on $(x_{1},\cdots,x_{4})$ would be given as:
\begin{equation}\label{eq:Z2H-action-on-x1234}
  \Z_{2}^{H}\ : \ (x_{1},\,x_{2},\,x_{3},\,x_{4})\, \, \to\,\, (-\,x_{1},\,-\,x_{2},\,x_{3},\,x_{4})\,.
\end{equation}
In this case, we can read the induced action on $\{\omega_{i}\}$, which coincide with the self-dual 2-forms in \eqref{eq:SD-ASD-2-forms}, explicitly as
\begin{equation}\label{eq:omegai-trans-Z2H}
  (\omega_{1},\omega_{2},\omega_{3})\,\,\to\,\, (-\omega_{1},-\,\omega_{2},\,\omega_{3})\,.
\end{equation}
Hence, the space $(\R^{4}/\Z_{2})/\Z_{2}^{H}$ admit two real scalar modules
\begin{equation}\label{eq:2-real-scalar-modules-Geo-HB}
  \phi_{1}\,\,\widetilde{\mathrm{h}}_{2}\,\wedge\,\omega_{1}\,\,+\,\,\phi_{2}\,\,\widetilde{\mathrm{h}}_{2}\,\wedge\,\omega_{2}\,,
\end{equation}
that parameterize the Higgs branch of the theory.

However, the $Sp(1)$-structure defined in  \eqref{eq:omegai-trans-Z2H} does not survive the quotient. Consequently, the quotient space is no longer hyperK\"ahler; it reduces to a K\"ahler structure with $\omega_{3}$ being the K\"ahler 2-form. Despite this reduction, the orbifolded space remains Ricci-flat, as demonstrated in \cite{wright2011quotientsgravitationalinstantons}. Therefore, the holonomy of the quotient space is $U(2)$ and the 7d theory can not be consistently defined. Nevertheless, we proceed by examining the Higgs branch of the 7d theory, because our ultimate goal is to understand how insights from the 7d Higgs mechanism can inform the construction of consistent supersymmetric lower‑dimensional theories. Further discussion will be presented around \eqref{eq:theorem-A-of-wright-2011}.

\paragraph{Higgs branch modules and nilpotent fluctuations.}

We will now argue that the $(\phi_{1},\phi_{2})$ modules can be obtained from fluctuations of the nilpotent Higgs given in \eqref{eq:nilpotetn-vev-su(2)}. First, note that our choice of the $\Z_{2}^{H}$ action above retain the $(\phi_{1},\phi_{2})$ modules appear in the holomorphic Higgs field $\Phi$ in \eqref{eq:holomorphic-Phi}. Therefore, this justify looking at fluctuations of the nilpotent Higgs vev $\langle\Phi\rangle$. Let us write $\Phi$ as 
\begin{equation}
  \Phi\,=\, \langle\Phi\rangle\, +\, \delta\Phi\,,
\end{equation}
with $\delta\Phi$ is given by
\begin{equation}\label{eq:delta-Phi-7d}
\delta\Phi\,=\,  \begin{pmatrix}
    \phi_{0}\,\,&\,\,\phi_{+}\\
    \phi_{-}\,\,&\,\,-\phi_{0}
  \end{pmatrix}
  \,\,\in\mk{su}(2)\,.
\end{equation}
Here, $(\phi_{0},\phi_{\pm})$ are complex valued scalar fields and the subscript represent the charges of under the Cartan $\mk{u}(1)\subset\mk{su}(2)$. The fluctuations are subject to linearized gauge transformations:
\begin{equation}\label{eq:linearized-gauge-transformations}
  \delta\Phi\,\,\sim\,\, \delta\Phi \,+\, [\,\langle\Phi\rangle,\chi\,]\,,
\end{equation}
with $\chi\in\mk{su}(2)$ give by
\begin{equation}
   \chi\,=\, \begin{pmatrix}
    \chi_{0}\,\,&\,\,\chi_{+}\\
    \chi_{-}\,\,&\,\,-\chi_{0}
  \end{pmatrix}\,.
\end{equation}
The linearized gauge transformations enable us to gauge away $\phi_{0}$ and $\phi_{+}$, and the physical fluctuations is then given as
\begin{equation}\label{eq:delta-Phi-phys-su(2)}
\delta\Phi_{\mathrm{phy}}  \,=\,\begin{pmatrix}
    0\,\,&\,\,0\\
    \phi_{-}\,\,&\,\,0
  \end{pmatrix}
  \,.
\end{equation}
The surviving $\phi_{-}$ can now be interpreted as the complex module of the Higgs branch. The degrees of freedom of the algebraic module matches that of the geometric Higgs branch module given in \eqref{eq:2-real-scalar-modules-Geo-HB}. We aim to generalize these observations in the subsequent discussion.

For this case, one can show that the above physical parameter $\phi_{-}$ matches exactly the parameter one finds when considering the Slodowy slice \cite{slodowy1980four,slodowy2006simple}. The reader may also consult \cite{henderson2015}. In particular, having the above nilpotent vev $e:=\langle\Phi\rangle$, then there exist a matrix $f:=\langle\Phi\rangle^{T}$ such that we have the $\mk{sl}(2,\C)$ triple $\{e,h,f\}$ satisfying
\begin{equation}\label{eq:ehf-triple-commutators}
 [h,e]\,=\,2e\,,\quad [h,f]\,=\,-2f\,\quad [e,f]\,=\,h\,,
\end{equation}
where $h :=\langle\phi\rangle =  \mathrm{diag}(1,-1)$ as in \eqref{eq:vev-phi1-su(2)-case}. The Slodowy slice is defined as
\begin{equation}\label{eq:Slodowy-slice}
  \mathcal{S}_{e}\,=\,e\,+\,\mathrm{ker}(ad_{f})\,.
\end{equation}
Imposing that the fluctuation $\delta\Phi$ in
\eqref{eq:delta-Phi-7d} lies in the Slodowy slice $\mathcal{S}_{e}$|in particular, enforcing the constraint $ [f,\delta\Phi]=0$|selects the physical fluctuations, which are precisely those given in \eqref{eq:delta-Phi-phys-su(2)}.

The above can be applied to a more general semi-simple Lie algebra $\mk{g}$, via the embedding of the triplet $\{e,h,f\}\hookrightarrow\mk{g}$. For further discussion on this embedding the reader may consider section 3.2 of \cite{Collingwood1993}. However, at this stage, we are content with these observations and we postpone further discussion to section \ref{sec:HB} and toward the end of section \ref{sec:charged-matter}.

\paragraph{Geometric generalization and lower dimensional field theories.}

The more general situation is described Theorem A of \cite{wright2011quotientsgravitationalinstantons}, which we rephrase in the following way:
\begin{equation}\label{eq:theorem-A-of-wright-2011}
\parbox{12cm}{Consider the hyperK\"ahler space $\R^{4}/\Gamma_{ADE}$ with a configuration of centers, denote by $F$, whose center of mass is fixed at the origin. The maximal continuous isometry of $F$ is at most a cyclic group $H=\Z_{k}\subset SO(3)$, which acts as a permutation of the centers without fixed points. The resulting quotient space remains Ricci‑flat and K\"ahler.}
\end{equation}
Here, $SO(3)$  act as rigid rotations on $\R^{3}$ where the centres are located. The fact that only a K\"ahler structure remains can be already seen from 
\eqref{eq:omegai-trans-Z2H}. The theorem gives two limitations for dealing with the quotient space 
\begin{equation}
  X_{4}(\Gamma_{ADE},\, H)\,=\,(\R^{4}/\Gamma_{ADE}\,)/H\,,
\end{equation}
which are:
\begin{itemize}
\item The total space $X_{4}(\Gamma_{ADE},\, H)$ is non-K\"ahler. Hence, the 7d theory is, in general, non-supersymmetric and it is harder to deal with. 

\item The fact that $H$ should act freely on $F$, i.e., without fixed points or no singlets, limits the resulting possibilities of gauge symmetries after Higgsing. Further, $H$ is cyclic. 
\end{itemize}

In M-theory compactification, we can prevail over the limitation of the theorem in \eqref{eq:theorem-A-of-wright-2011} by enlarging the compactification space. Topologically, such spaces may be taken to be of the form
\begin{equation}
  X_{4+n}(\Gamma_{ADE},\,H,\,Y_{n})\,=\, \frac{(\R^{4}/\Gamma_{ADE}\,\times\, Y_{n})}{H}\,,
\end{equation}
where $\Gamma_{\mathrm{ADE}}\subset SU(2)$ is a finite ADE subgroup, $Y_{n}$ is an $n$-dimensional manifold, and $H$ is a discrete group acting on the product space. Here, $H$ should have a 3-dimensional representation that acts on the centres.

The general constrains on $X_{4+n}$ to define a consistent $(7-n)$d supersymmetric theory can be listed as the following:
\begin{itemize}
  \item The total space $X_{4+n}$ must admits a $\mathscr{G}$-structure compatible with the twisted dimensional reduction of the 7d SYM theory associated with $\R^{4}/\Gamma_{ADE}$ on $Y_{n}$, including the action of the quotient group $H$. 

The $\mathscr{G}$-structure can be constructed from the natural $Sp(1)$-structure on $\R^{4}/\Gamma_{ADE}$ (prior to quotienting by $H$), together with additional geometric
data on $Y_{n}$, such as a reduced structure group or differential forms. Our main example is provided by the $Spin(7)$-structure discussion around \eqref{eq:Spin(7)-structure-Phi4}.

\item The group $H$ must act freely on the space $Y_{n}$. As a consequence, the only singularities present in the quotient space $X_{4+n}$ arise from the co-dimension four singularities of $\R^{4}/\Gamma_{\mathrm{ADE}}$ itself.

Since $H$ acts freely on the total space $X_{4+n}$, it may still leave invariant a subset of the centers in the set $F$ associated with the resolution of $\R^{4}/\Gamma_{\mathrm{ADE}}$, provided that the combined action on $\R^{4}/\Gamma_{\mathrm{ADE}}\times Y_{n}$ has no fixed points.

Furthermore, the group $H$ is no longer a cyclic and can be arbitrary finite group depending on the geometry of $Y_{n}$.
\end{itemize}

In the subsequent sections, we focus on the 8d spaces introduced in section \ref{sec:3d-theories-GE}. We would first discuss the Coulomb branch of the effective 3d theories, then their Higgs and mixed branches.

%%%%%%%%%%%%%%%%%%%%%%%%%%%%%%%%%%%%%%%%%%%%%%%%%%%%%%%%%%%%%%%%%%%%%%%%%%%%%%%%%%%%%%%%%%%%

\subsection{Geometric realizations of the 3d theories}\label{sec:geometric-realization}

Our $X_{8}$ geometry naturally inherits harmonic forms from its constitute parts, modulo the action of $H$. As mentioned earlier, we have $L^{2}$-normalizable harmonic 2-form $\{\widetilde{\mathrm{h}}_{2}^{a}\}$, for $a=1,\cdots, r=\mathrm{dim}(\mk{h}_{ADE})$, due to the $\widetilde{X}_{4}$ subspace. Moreover, we have four harmonic 1-forms $\dd y_{n}$ and three self-dual harmonic 2-forms $\alpha_{i}$ from the $T^{4}/H$ subspace, as discussed in section \ref{sec:R4-bundle-over-B4}.

Each set of the above $p$-forms decomposes into $H$-invariant (untwisted) and $H$-variant (twisted) sets under the action of the holonomy group $H$. A natural consistency condition is that:
\begin{equation}
  \parbox{9cm}{Only $H$-invariant (untwisted) $p$-forms can be properly defined over the full eight-dimensional geometry $X_{8}$}.
\end{equation}
This is a direct generalization of the criterion given in \eqref{eq:7d-only-Z2H-invariant-forms}. The untwisted $p$-forms on $X_{8}$ can be obtained through:
\begin{itemize}
  \item Untwisted $p$-forms on the constitute spaces $\widetilde{X}_{4}$ and $T^{4}/H$.
  \item As wedge product of two untwisted $p$-forms, or two twisted $p$-forms of the constitute spaces. 
\end{itemize}

These $H$-invariant (untwisted) harmonic forms on $X_{8}$ provide a direct geometric understanding of the origin of the bosonic fields appearing in
\eqref{eq:VM-3d-N=4}, \eqref{eq:VM-3d-N=2}, along with extra massless degrees of freedom. Before entering into the detailed analysis, it is useful to emphasize the following general criterion concerning the effective field theories:
\begin{equation}\label{criterion:massless-dof}
\parbox{10cm}{
  Only $H$-untwisted harmonic $p$-forms on $X_{8}$ give rise to massless degrees of freedom in the associated effective 3d theory.}
\end{equation}

Let us now examine the role of the untwisted $p$-forms, relations to the Coulomb and Higgs branches, and dual cycles in more details.

\paragraph{Coulomb branch and untwisted $p$-forms.}

The Cartan-valued gauge fields of the 3d theories $A^{a}:$   $(\widetilde{\textbf{1}}_{L},\textbf{1}_{R},\textbf{3})$ and the scalar fields 
$\phi^{a} : (\widetilde{\textbf{2}}_{L},\textbf{2}_{R},\textbf{1})$, are originated from the reduction of the M-theory $C_{3}$-field as:
\begin{equation}\label{eq:C3-expansion}
    C_{3}\,=\, A_{1}^{a}\,\wedge\, \widetilde{\mathrm{h}}_{2}^{a}\,+\, \phi_{\alpha}^{a}\,\,\,\widetilde{\mathrm{h}}_{2}^{a}\,\wedge\,\dd y_{\alpha} \,
\end{equation}
Here, $\{\widetilde{\mathrm{h}}_{2}^{a}\}$ are untwisted harmonic 2-forms on $X_{8}$ that correspond to the Cartan sub-algebra $\mk{h}_{ADE}\subset \mk{g}_{ADE}$ with $a=1,\cdots, \mathrm{dim}(\mk{h}_{ADE})$. $\{\widetilde{\mathrm{h}}_{2}^{a}\,\wedge\,\dd y_{\alpha}\}$ is the set of untwisted harmonic 3-forms on $X_{8}$ with $a$ runs as before and $\alpha$ correspond to the $1^{st}$ Betti number of $T^{4}/H$ as listed in Table \ref{Table:all-B4-spaces}. In other words, $\{\dd y_{\alpha}\}$ are the untwisted 1-forms on $B_{4}=T^{4}/H$.

The scalar field $\varphi^{a} : (\widetilde{\textbf{3}}_{L},\textbf{1},\textbf{1})$, introduced in equation \eqref{eq:twis-red-3d-fields-reps}, has a geometric interpretation in terms of, see, e.g., \cite{Najjar:2023hee}:
\begin{itemize}
\item The K\"ahler moduli associated with vanishing 2-cycles in the crepant resolution of the $\R^{4}/\Gamma_{ADE}$ singularity. Thus, in terms of the corresponding $\{\widetilde{\mathrm{h}}_{2}^{a}\}$.
\item The self-dual 2-forms $\alpha_{\beta}$ on $T^{4}/H$.
\end{itemize}
This can be written as
\begin{equation}\label{eq:phi-h2-alpha-from-untwisted}
  \varphi_{\beta}\,\, J_{\beta}\,\wedge\,\alpha_{\beta}\,=\, \varphi^{a}_{\beta}\,\,\widetilde{\mathrm{h}}_{2}^{a}\,\wedge\, \alpha_{\beta}\,,
\end{equation}
with $\{\widetilde{\mathrm{h}}_{2}^{a}\,\wedge\, \alpha_{\beta}\}$ is the set of harmonic 4-forms on $X_{8}$ with dimension $b_{2}^{L}(B_{4})\times \mathrm{dim}(\mk{h}_{ADE})$, i.e., $\beta=1,\cdots, b_{2}^{L}(B_{4})$. In other words, $\{\alpha_{\beta}\}$ are the untwisted self-dual 2-forms on $B_{4}$.

For the 3d $\N=2^{\ast}$ theories associated with $B_{(4;1)}^{(9-27)}$, we can now draw a clear conclusion. The $\mathrm{VM}_{3d}^{\N=2}$ originates entirely from the expansion of the $C_{3}$ field. This precisely reproduces the effective description given in \eqref{eq:VM-3d-N=2}. On the other hand, the $\mathrm{VM}_{3d}^{\N=4}$ arises from the expansion of the $C_{3}$ field along with a scalar degrees of freedom obtained from from the untwisted sector as in \eqref{eq:phi-h2-alpha-from-untwisted}. Hence, one recovers \eqref{eq:VM-3d-N=4}.

For both 3d $\N=2^{\ast}$ and $\N=4^{\ast}$, the expansion of the $C_{3}$ field above capture the Coulomb branch providing that the $H$ action on the harmonic 2-forms $\widetilde{\mathrm{h}}_{2}^{a}$ is trivial $\forall a $. In terms of the dual centres of the $\R^{4}/\Gamma_{ADE}$ spaces, the full Coulomb branch correspond to trivial action on these centres. Here, the full Coulomb branch means that the Lie algebra is given by $\mk{u}(1)^{\oplus r}$ with $r=\mathrm{dim}(\mk{h}_{ADE})$.

Let us specialize to $\mk{a}_{N-1}$ singularities and we realize $\R^{4}/\Z_{N}$ as a $U(1)$ bundle over $\R^{3}$. For example, in the case of $H=\Z_{2}^{H}$ with action given in \eqref{eq:Z2H-action-on-x1234}, the Coulomb branch would correspond to align all the centres along the third axis; with $x_{4}\sim x_{4}+1$ correspond to the $U(1)$ bundle direction. In general, $H\subset SO(3)$ that acting on the base $\R^{3}$ where the centres are located. The cases where $H$ acts on all directions of $\R^{3}$ implies that the effective 3d theories admits a partial Coulomb branch, i.e., a mixed branch.

With this discussion in mind, we point out that we have two distinct cases:
\begin{itemize}
  \item The 3d $\N=4^{\ast}$ always admit a full Coulomb branch. This is the case since $H$ leaves one direction in $\R^{3}$ invariant. This correspond to the spaces $B_{4}^{(k)}$ for $k=1,\cdots,8$ with $b_{1}(B_{4})=2$. 

  \item The 3d $\N=2^{\ast}$ do not admit a full Coulomb branch. The group $H$ acts non-trivially on all directions of the $\R^{3}$ base. This correspond to the rest of the $B_{4}$ spaces with $b_{1}(B_{4})=1$. 
\end{itemize}

\paragraph{Higgs branch and twisted $p$-forms.}

We now turn to the analysis of twisted harmonic forms of the constitute space $\widetilde{X}_{4}$ and $B_{4}$. In particular, we consider twisted harmonic 2-forms on $\widetilde{X}_{4}$ together with twisted 1-forms and 2-forms on the Bieberbach space $T^{4}/H$. One immediately observes that such twisted forms cannot be combined to produce field-theoretic gauge fields $A_{1}^{a}:$  $(\widetilde{\textbf{1}}_{L},\textbf{1}_{R},\textbf{3})$ in 3d. This is the case since the $C_{3}$ field can only be expanded along untwisted $p$-forms. 

On the other hand, the only possible degrees of freedom arising from these sectors are scalar fields. These scalars come from untwisted wedge product between twisted $\widetilde{\mathrm{h}}_{2}^{\vardiamond}$, twisted $\dd y_{\bullet}$, and twisted $\alpha_{2}^{\varheart}$ in the expansion of \eqref{eq:C3-expansion} and \eqref{eq:phi-h2-alpha-from-untwisted}. As these scalars are associated with harmonic untwisted $p$-form in $\widetilde{X}_{8}$, then these scalars are massless. From the examples in section \ref{sec:HB}, we learn that these correspond to massless Higgs branch moduli. Similar observations was found in \cite{Acharya:2023xlx}.

We now turn our attention to the different cycles of the geometry of $\widetilde{X}_{8}$.

\paragraph{The corresponding cycles.} The Poincaré duality assignment between the above untwisted $p$-forms and homological cycles of $\widetilde{X}_{8}$ is given as:
\begin{equation}\label{eq:p-forms-dual-cycles}
    \begin{split}
        \widetilde{\mathrm{h}}_{2}^{a} \ &\xleftrightarrow[]{\text{Poincaré \,dual}}\  \text{6-cycles in}\,\,\, \frac{\bbS^{2}_{a}\,\times\,T^{4}}{H}\,,
        \\
        \widetilde{\mathrm{h}}_{2}^{a}\,\wedge\,\dd y_{\alpha}\ &\xleftrightarrow[]{\text{Poincaré \,dual}}  \  \,\text{5-cycles in}\,\,\, \frac{\bbS^{2}_{a}\,\times\, T^{4}}{H}\,,
        \\
        \widetilde{\mathrm{h}}_{2}^{a}\,\wedge\,\alpha_{\beta}\ &\xleftrightarrow[]{\text{Poincaré \,dual}}\, \ \text{4-cycles in}\,\,\, \frac{\bbS^{2}_{a}\,\times\, T^{4}}{H}\,.
    \end{split}
\end{equation}

The preceding analysis demonstrates that 6-cycles can only correspond to untwisted $\widetilde{\mathrm{h}}_{2}^{a}$ 2-forms, as the space $B_{4}=T^{4}/H$ has a unique untwisted 4-cycle. Given the above discussion, these cycles correspond to the Coulomb branch of the theory. In addition, there are 4-cycles and 5-cycles correspond to the untwisted $\widetilde{\mathrm{h}}_{2}^{a}$ 2-forms which are associated with the scalar degrees of freedom in the vector multiplet of the 3d theories. For twisted $\widetilde{\mathrm{h}}_{2}^{a}$ 2-forms, the dual 4-cycles and 5-cycles correspond to the Higgs branch moduli discussed above.

We now turn our attention to examine such HB in details. 

%%%%%%%%%%%%%%%%%%%%%%%%%%%%%%%%%%%%%%%%%%%%%%%%%%%%%%%%%%%%%%%%%%%%%%%%%%%%%%%%%%%%%%%%%%%%

\subsection{Higgs branch of 3d \texorpdfstring{$\N=2^{\ast}$}{N=2*} and 4d \texorpdfstring{$\N=1^{\ast}$}{N=1*} theories}\label{sec:HB}

In this subsection, we study the Higgs branch of 3d $\N=2^{\ast}$ and 4d $\N=1^{\ast}$ theories within the framework developed in section \ref{sec:CB-HB-7d}. Specially, we focus on field theories associated with some of the Bieberbach spaces $B_{4}^{(\bullet)}$ with holonomy $\Z_{2}\times\Z_{2}$, as well as the $B_{3}^{(6)}$ space, which shares the same holonomy, as primary examples.

\subsection*{$\mk{su}(2)$ gauge theory: the $X_{8}(\Z_{2},B_{(4;1)}^{(9)})$ geometry}

Let us start with a simple example given by the geometry of $X_{8}(\Z_{2},B_{(4;1)}^{(9)})$. In this case, the total holonomy of the $X_{8}$ is given by $SU(2)\rtimes(\Z_{2}\times\Z_{2})$, as seen in Table \ref{Table:fiberB32}, and the 3d theory has 4 real supercharges. 

The rotational holonomy group of the Bieberbach $B_{(4;1)}^{(9)}$ space is given by the Klein group $H=\Z_{2}\times\Z_{2}$ and it is generated by 
\begin{equation}\label{eq:rotation-hol-B49}
  A\,=\,  \begin{pmatrix}
        -1&0&0&0\\
        0&-1&0&0\\
        0&0&1&0\\
        0&0&0&1
    \end{pmatrix}\,\,;
    \qquad \qquad 
   B\,=\, \begin{pmatrix}
        1&0&0&0\\
        0&-1&0&0\\
        0&0&-1&0\\
        0&0&0&1
    \end{pmatrix}\,.
\end{equation}
We note that the $3\times 3$ upper-block of $A$, $B$, and $AB$ acting non-trivially on all three coordinates of the base $\R^{3}$ of the $\R^{4}/\Z_{2}$ space. Therefore, any profile for the centres $\vec{x}_{a=1,2}$ would be affected by the $H$-action. There are three distinct non-trivial configurations of the centres along the three axes:
\begin{equation}\label{eq:x1-x2-for-B49}
    \vec{x}_{1}\,=\,-\,\vec{x}_{2}\,=\,
    \begin{pmatrix}
        |x_{1}|\\
        0\\
        0
    \end{pmatrix}\,, \ \ \text{or }\quad \begin{pmatrix}
        0\\
        |x_{1}|\\
        0
    \end{pmatrix}\,, \ \ \text{or }\quad \begin{pmatrix}
        0\\
        0\\
        |x_{1}|
    \end{pmatrix}\,.
\end{equation}

Before examining this example in detail, we note that an analogous analysis extends to the $X_{8}$ geometries constructed from the Bieberbach manifolds $B_{4}^{(10)}$, $B_{4}^{(11)}$, $B_{4}^{(12)}$, and $B_{4}^{(14)}$. Furthermore, as we learn from section \ref{sec:G-structure-co-seifert}, some of the 3d theories can be seen as an $\bbS^{1}$ reduction of 4d theories that are associated with Bieberbach 3-manifolds. In the current example, $B_{4}^{(14)}$ can be realized as $B_{3}^{(6)}\times \bbS^{1}$. After oxidation, the effective theory is 4d $\N=1^{\ast}$, i.e., with 3 massive chiral multiplet. Therefore, the following discussion applies to these 4d theories as well. 

\paragraph{The geometric Higgs branch.}

Following the analysis given around \eqref{eq:Z2H-on-h2-7d-theory}, we find that in any of the cases of \eqref{eq:x1-x2-for-B49}, the action on the harmonic 2-form is given as:
\begin{equation}\label{eq:h2-to-minus-h2}
    \widetilde{\mathrm{h}}_{2}\,\to\, -\, \widetilde{\mathrm{h}}_{2}\,,
\end{equation}
under the relevant generator. Thus, we conclude that $b_{2}(X_{8})=0$ and the theory has no Coulomb branch.  

However, the geometry permit other harmonic $p$-forms that are invariant under the $\Z_{2}\times\Z_{2}$ action. Explicitly, these are given by
\begin{equation}\label{eq:h2dy-h2alphs-B49}
    \widetilde{\mathrm{h}}_{2}\,\wedge\,\dd y_{2}\,,\qquad  \widetilde{\mathrm{h}}_{2}\,\wedge\,\alpha_{2}\,.
\end{equation} 
Here, we assumed the first profile in \eqref{eq:x1-x2-for-B49}; further, $\dd y_{2}$ and $\alpha_{2}$ transform exactly as $\widetilde{\mathrm{h}}_{2}$. The non-trivial Betti numbers are $b_{3}=b_{4}=1$. To each of the above $H$-inavariant $p$-form we associate a massless real scalar field, which are combined to give a $\C$-module for the Higgs branch of the 3d theory $\mathcal{T}(X_{8}(\Z_{2},B_{(4;1)}^{(9)}))$. This procedure should be repeated to the three profiles in \eqref{eq:x1-x2-for-B49} each gives us a $\C$-valued field, i.e., we have three distinct Higgs branches along with their own moduli.

\paragraph{Nilpotent Higgsing.}

Recall that the geometry of $X_{8}(\Z_{2},B_{(4;1)}^{(9)})$ can be seen as an $A_{1}$-fibration over the base $B_{(4;1)}^{(9)}$. In the current case, the $A_{1}$-fibration has a spectral equation involving the nilpotent Higgs vev|as given in \eqref{eq:nilpotetn-vev-su(2)}| which is already presented in \eqref{eq:resolving-A1-spectral-curve}. As before, the nilpotent Higgs is responsible for completely Higgsing the $\mk{su}(2)$ gauge symmetry. 

The parameters of the Higgs branch can be derived when considering a generic fluctuations around the nilpotent Higgs, subject to the linearized gauge transformations given in \eqref{eq:linearized-gauge-transformations}. The analysis around \eqref{eq:linearized-gauge-transformations} can be carried completely to the current case and one find that there is exactly one $\C$-field parametrizing the Higgs branch. 

However, it remains to clarify the 3d origin of the nilpotent Higgs field. As was noticed in section \ref{sec:R4-bundle-over-B4}, the appropriate interpretation of the effective 3d theory is given by 3d $\N=2^{\ast}$ gauge theory. We refer the reader to the discussion around \eqref{eq:effective-3d-N=2*}. The 3d theory contains three massive adjoint-valued chiral multiplets, which we denote by $\Phi_{i}$ with $i=1,2,3$. The superpotential of these theories, which can be seen as $\bbS^{1}$ reduction of the 4d $\N=1^{\ast}$ ones given in \cite{Donagi:1995cf,Dorey:1999sj,Myers:1999ps,Polchinski:2000uf}, insures that they satisfy the $\mk{sl}(2,\C)$ commutation relation. Since these fields are valued in the $\mk{su}(2)$ Lie algebra, we can identify them with the $\mk{sl}(2,\C)$ triplet $\{e,h,f\}$, as in \eqref{eq:ehf-triple-commutators}, after complexifying the gauge algebra. 

In this case, the vev of the triplet can be taken as
\begin{equation}
e\,=\,  \begin{pmatrix}
    0&1\\
    0&0
  \end{pmatrix}\,,\qquad h\,=\, \begin{pmatrix}
    1&0\\
    0&-1
  \end{pmatrix}\,,\qquad f\,=\, \begin{pmatrix}
    0&0\\
    1&0
  \end{pmatrix}\,,
\end{equation}
with $e$ correspond to taking the upper triangle of \eqref{eq:action-on-centers-su2-P2} as observed earlier.

With this in mind, we claim that the geometric Higgs branch moduli of \eqref{eq:h2dy-h2alphs-B49} correspond to the physical fluctuations around the nilpotent Higgs vev. These fluctuations can be determined via the Slodowy slice as
\begin{equation}\label{eq:3d-su2-HB-moduli}
\delta\Phi_{\mathrm{phy}}  \,=\,\begin{pmatrix}
    0\,\,&\,\,0\\
    \phi_{-}\,\,&\,\,0
  \end{pmatrix}
  \,.
\end{equation}
Since there are no gauge fields, the scalar $\phi_{-}$ is obviously neutral. 

For the more general Lie gauge algebra $\mk{g}$, instead of the above identification, we should now consider the reducible embeddings:
\begin{equation}
  \mk{sl}(2,\C)\, \hookrightarrow\, \mk{g}\,,
\end{equation}
where the gauge algebra is now complexified.

%%%%%%%%%%%%%%%%%%%%%%%%%%%%%%%%%%%%%%%%%%%%%%%%%%%%%%%%%%%%%%%%%%%%%%%%%%%%%%%%%%%%%

\subsection*{$\mk{su}(4)$ gauge theory: the $X_{8}(\Z_{4},B_{(4;1)}^{(9)})$ case}

\begin{figure}[H]
\centering{
\includegraphics[scale=0.3]{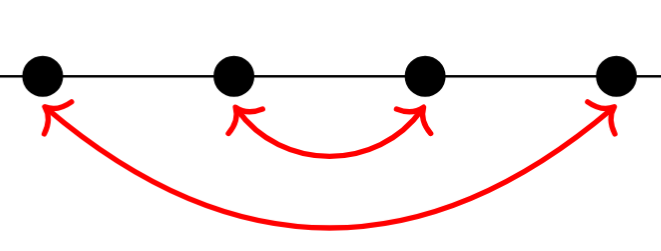}
}
\caption{Here, the nodes represent the four centres of the $\widetilde{\R^{4}/\Z_{4}}$ geometry. From left to right, we label the centers by $\vec{x}_{1}$, $\vec{x}_{2}$, $\vec{x}_{3}$, and $\vec{x}_{4}$. The lines between these nodes represent the 2-spheres corresponding to the Cartan of $\mk{su}(4)$. These centres are arranged in pairs under a $\Z_{2}$, say that generated by $A$ given in \eqref{eq:rotation-hol-B49}; that we have $\vec{x}_{1}\sim \vec{x}_{4} $ and $\vec{x}_{2}\sim \vec{x}_{3}$. Upon taking the $H$ quotient, only one 2-sphere survive. There are another two equivalent configurations given as: $\vec{x}_{1}\sim \vec{x}_{2} $ and $\vec{x}_{3}\sim \vec{x}_{4}$ and $\vec{x}_{1}\sim \vec{x}_{3} $ and $\vec{x}_{2}\sim \vec{x}_{4}$. Their corresponding $P_{[2,2]}$ can be easily constructed.}
\label{Fig:Z2Z2-on-4-centres}
\end{figure}

Let us now consider 3d $\N=2^{\ast}$ with $\mk{su}(4)$ Lie algebra, up to the $H$ action in \eqref{eq:rotation-hol-B49}. The theory possesses four centres, which can be arranged in terms of their behavior under the action of the $H=\Z_{2}\times\Z_{2}$ group, i.e., the centres are arranged as the partition of 4. This constraint explains the missing $4=3+1$ partition below. We have four cases of partitions given as:
\begin{itemize}
  \item All centers are invariant. In this case, all the centers are located at the origin of $\R^{3}$ as it is the only point remains invariant under the $H$ action. The partition of 4 is given as $[1,1,1,1]$ and the gauge Lie algebra remains $\mk{su}(4)$. 

  \item Two pairs. In this case, the centers can be arranged along, say, the $x_{1}$-axis of $\R^{3}$ and organized into two doublets under the $A$ and $AB$ operators defined in \eqref{eq:rotation-hol-B49}, as illustrated in Figure \ref{Fig:Z2Z2-on-4-centres}. The partition of $4$ is given as $[2,2]$. From the figure, we learn that the action on the centers is given by:
\begin{equation}\label{eq:action-on-centers-su4-P22}
P_{[2,2]}\,:=\,  \begin{pmatrix}
        0&0&0&1\\
        0&0&1&0\\
        0&1&0&0\\
      1&0&0&0
    \end{pmatrix}\,.
\end{equation}
There are another two equivalent actions as explained in the caption of Figure \ref{Fig:Z2Z2-on-4-centres}, which are given as:
\begin{equation}\label{eq:action-on-centers-su4-P22-other}
P_{[2,2]}^{'}\,:=\,  \begin{pmatrix}
        0&0&1&0\\
        0&0&0&1\\
        1&0&0&0\\
      0&1&0&0
    \end{pmatrix}\,,\qquad 
    P_{[2,2]}^{''}\,:=\,  \begin{pmatrix}
        0&1&0&0\\
        1&0&0&0\\
        0&0&0&1\\
      0&0&1&0
    \end{pmatrix}\,.
\end{equation}

The maximal torus of $\mk{su}(4)$ is $\mk{u}^{\oplus 3}$, geometrically represented by three harmonic 2-forms $\widetilde{\mathrm{h}}_{2}^{a}$ with $a=1,2,3$. Following the construction in Appendix \ref{sec:harmonic-2-forms}, the action of $P_{[2,2]}$ on the centres induces the following transformation on the harmonic 2-forms:
\begin{equation}
 \widetilde{\mathrm{h}}_{2}^{1}\,\to\,-\widetilde{\mathrm{h}}_{2}^{3}\,,\qquad \widetilde{\mathrm{h}}_{2}^{2}\,\to\,-\widetilde{\mathrm{h}}_{2}^{2} \,,\qquad \widetilde{\mathrm{h}}_{2}^{3}\,\to\, -\widetilde{\mathrm{h}}_{2}^{1} \,. 
\end{equation}
Consequently, only a single $\mk{u}(1)\subset \mk{u}(1)^{\oplus 3}$ survives the quotient, corresponding to the $(\Z_{2}\times\Z_{2})$-invariant harmonic 2-form $(\widetilde{\mathrm{h}}_{2}^{1}- \widetilde{\mathrm{h}}_{2}^{3})$. Since $B_{(4;1)}^{(9)}$ admits only one untwisted 1-form, namely $\dd y_{4}$, then the classical Coulomb branch is complex 1-dimensional. The vector multiplet originates from the decomposition:
\begin{equation}\label{eq:su4-u1-survive-B49}
 C_{3}\,=\, A_{1}\,\wedge\,  ( \widetilde{\mathrm{h}}_{2}^{1}\,-\, \widetilde{\mathrm{h}}_{2}^{3})\, + \, \phi\, ( \widetilde{\mathrm{h}}_{2}^{1}\,-\, \widetilde{\mathrm{h}}_{2}^{3})\,\wedge\,\dd y_{4}\,.
\end{equation}
Hence, there is a 6-cycle of the form presented in \eqref{eq:p-forms-dual-cycles}. The $\mk{u}(1)$ enhances to $\mk{su}(2)$ once the divisor collapse. In particular, M2-brane wrapping an $\bbS^{2}$ sub-cycle of the 6-cycle gives the $W^{\pm}$-bosons, which become massless upon collapsing the 2-cycle. 

On the total space $X_{8}$ there are additional invariant $(\Z_{2}\times\Z_{2})$ $p$-forms given as:
\begin{equation}\label{eq:b3-b4-harmonic-fields-B49-Z4}
    \widetilde{\mathrm{h}}_{2}^{2}\,\wedge\,\dd y_{2}\,,\qquad  (\widetilde{\mathrm{h}}_{2}^{1}\,+\,\widetilde{\mathrm{h}}_{2}^{3})   \,\wedge\,\dd y_{2}\,,\qquad  \widetilde{\mathrm{h}}_{2}^{2}\,\wedge\,\alpha_{2}\,,\qquad (\widetilde{\mathrm{h}}_{2}^{1}\,+\,\widetilde{\mathrm{h}}_{2}^{3})   \,\wedge\,\alpha_{2}\,.
\end{equation}
This introduces 4 real scalars in the effective 3d theory corresponding to $b^{3}=b^{4}=2$. The four real scalars combined to $\C^{2}$-module describing the breaking of $\mk{u}(1)^{\oplus 3}\to \mk{u}(1)$ and parameterizing the Higgs branch.

\item One pair and two invariants. In this case, we locate two centres at the origin and the other two centres form a pair under the action of $H$. The partition of 4 is given as $[2,1,1]$. The action on the centres can be represented as 
\begin{equation}\label{eq:action-on-centers-su4-P211}
P_{[2,1,1]}\,:=\,  \begin{pmatrix}
        1&0&0&0\\
        0&1&0&0\\
        0&0&0&1\\
      0&0&1&0
    \end{pmatrix}\,.
\end{equation}
Here, the invariant centres are acted upon by the identity. The action on the centres induces the following transformation on the harmonic 2-forms:
\begin{equation}
 \widetilde{\mathrm{h}}_{2}^{1}\,\to\,\widetilde{\mathrm{h}}_{2}^{1}\,,\qquad \widetilde{\mathrm{h}}_{2}^{2}\,\to\,\widetilde{\mathrm{h}}_{2}^{2} \,,\qquad \widetilde{\mathrm{h}}_{2}^{3}\,\to\, -\widetilde{\mathrm{h}}_{2}^{3} \,. 
\end{equation}
Therefore, preserving only $\mk{u}(1)^{\oplus 2}\subset \mk{u}(1)^{\oplus 3}$. There are two inequivalent collapsing 6-cycles that correspond to the above two untwisted harmonic 2-forms. In the collapsing limit, one finds $\mk{su}(3)$ Lie algebra, subject to non-trivial $\Z_{2}$ flat connection that can break it further to $\mk{su}(2)\oplus\mk{u}(1)$.

The Higgsing of $\mk{su}(4)$ down to $\mk{u}(1)^{\oplus 2}$ is parametrized by two real scalars that correspond to 
\begin{equation}\label{eq:geo-higgs-parameter-211}
\widetilde{\mathrm{h}}_{2}^{3}\,\wedge\,\dd y_{2}\,,\qquad   \widetilde{\mathrm{h}}_{2}^{3}\,\wedge\,\alpha_{2}\,,
\end{equation}
with $b^{3}=b^{4}=1$.

\item A quadruplet. In this case, the 4 centres transform to each other according to  
\begin{equation}\label{eq:action-on-centers-su4-P4}
P_{[4]}\,:=\,  \begin{pmatrix}
        0&1&0&0\\
        1&0&1&0\\
        0&1&0&1\\
      0&0&1&0
    \end{pmatrix}\,.
\end{equation}
The action on the harmonic 2-forms is given as
\begin{equation}
 \widetilde{\mathrm{h}}_{2}^{1}\,\to\,-\widetilde{\mathrm{h}}_{2}^{1}\,,\qquad \widetilde{\mathrm{h}}_{2}^{2}\,\to\,-\widetilde{\mathrm{h}}_{2}^{2} \,,\qquad \widetilde{\mathrm{h}}_{2}^{3}\,\to\, -\widetilde{\mathrm{h}}_{2}^{3} \,. 
\end{equation} 
Hence, break the $\mk{su}(4)$ completely. 

The Higgs branch in this case is parametrized by 6 real scalars correspond to $b^{3}=b^{4}=3$.

\end{itemize}

\paragraph{Interpretation in terms of Higgs fields.}

Understanding the effective theory as a 3d $\N=2^{\ast}$ enable us to give a field theoretic interpretation to the above geometric results. Here, we aim to realize the above four cases in terms of nilpotent vevs which correspond to the embedding of the generator $e$ of $\mk{sl}(2,\C)$ in $\mk{su}(4)$, along with semisimple vevs that correspond to the generator $h$ of $\mk{sl}(2,\C)$.

Interestingly, the nilpotent vacuum expectation values can be directly inferred from the action of $H$, i.e., from the geometry itself, through the following observation:
\begin{equation}\label{eq:nilpotetn-Higgs-upper-triangle-P}
  \parbox{10cm}{The nilpotent Higgs vacuum expectation value associated with each case is encoded in the strictly upper-triangular part of the corresponding matrix $P_{[\bullet]}$.}
\end{equation}
We expect this statement to hold for generic $\mk{su}(n)$ gauge Lie algebra. $[\bullet]$ refer to the partition of $n$. 

Now we analyze the four cases:
\begin{itemize}
  \item All centers are invariant. In this case, the Lie algebra is fully preserved as all the centres are located at the origin of the $\R^{3}$ space. This amount to vanishing vev for $\Phi_{i}$ for $i=1,2,3$. In other words, the matrix $P_{[1,1,1,1]}$ is the identity; hence, its upper-triangle is identically zero.

  \item  Two pairs. The nilpotent Higgs can be read from, say we take, $P_{[2,2]}^{''}$ of \eqref{eq:action-on-centers-su4-P22-other} as:
 \begin{equation}\label{eq:nilpotent-su4-P22}
\langle\Phi_{[2,2]}^{(n)}\rangle \,=\,  \begin{pmatrix}
        0&1&0&0\\
        0&0&0&0\\
        0&0&0&1\\
      0&0&0&0
    \end{pmatrix}\,.
\end{equation} 
Here, the superscript $(n)$ stand for the nilpotent part of the Higgs vev. The associated semisimple element is given by 
\begin{equation}
  \langle\Phi_{[2,2]}^{(s)}\rangle \,=\,  \begin{pmatrix}
        1&0&0&0\\
        0&1&0&0\\
        0&0&-1&0\\
      0&0&0&-1
    \end{pmatrix}\,.
\end{equation}
Here, the superscript $(s)$ stand for the semisimple Higgs vev. Together with $\langle\Phi_{[2,2]}^{(n)}\rangle^{T}$, the three matrices form a specific embedding of $\mk{sl}(2,\C)$ triplet $\{e,h,f\}$ in $\mk{su}(4)$. 

The nilpotent element comes in two $2\times 2$ block diagonal, with each block is a nilpotent of an $\mk{su}(2)$ algebra that break it compeletly. Therefore, the nilpotent breaks $\mk{su}(2)\oplus\mk{su}(2)$ inside $\mk{su}(4)$, and the unbroken Lie algebra is Higgsed down to
\begin{equation}
  \mk{g}_{e_{[2,2]}}\,=\,\mk{su}(2)\,.
\end{equation}

The physical fluctuations in this case are associated with each of the $\mk{su}(2)$ nilpotent as in \eqref{eq:delta-Phi-phys-su(2)}, and hence they are given as
\begin{equation}
 \delta\Phi_{\mathrm{phy}}^{[2,2]} \,=\, 
 \begin{pmatrix}
        0 & 0 & 0 & 0\\
        \varphi & 0 & 0 & 0\\
        0 & 0 & 0 & 0\\
        0 & 0 & \phi & 0
 \end{pmatrix}\,.
\end{equation}
We can think of this as Higgs moduli, which we observe that they match the geometric moduli in \eqref{eq:b3-b4-harmonic-fields-B49-Z4}. Further, we point out that $\delta\Phi_{\mathrm{phy}} $ belongs to the Slodowy slice as observed earlier.

 \item One pair and two invariants. In this case, the nilpotent Higgs field is given by
\begin{equation}\label{eq:nilpotent-su4-P211}
\langle\Phi_{[2,1,1]}^{(n)}\rangle \,=\,  \begin{pmatrix}
        0&0&0&0\\
        0&0&0&0\\
        0&0&0&1\\
      0&0&0&0
    \end{pmatrix}\,.
\end{equation}
One can show that the effect of this nilpotent is to break the $\mk{su}(4)$ Lie algebra down to $\mk{su}(2)\oplus \mk{u}(1)$ as we now proceed. 

The semisimple element that is compatible with $\langle\Phi_{[2,1,1]}^{(n)}\rangle $ is given by
\begin{equation}\label{eq:semisimple-su4-P211}
\langle\Phi_{[2,1,1]}^{(s)}\rangle \,=\,  \begin{pmatrix}
        0&0&0&0\\
        0&0&0&0\\
        0&0&1&0\\
      0&0&0&-1
    \end{pmatrix}\,\,. 
\end{equation}
Here, $\langle\Phi_{[2,1,1]}^{(n)}\rangle$, $\langle\Phi_{[2,1,1]}^{(s)}\rangle$, and $\langle\Phi_{[2,1,1]}^{(n)}\rangle^{T}$ defines the $\mk{sl}(2,\C)$ triple $\{e,h,f\}$ associated with the $[2,1,1]$ partition of $4$.

To understand the breaking pattern of the $\mk{su}(4)$ gauge theory, it is convenient to proceed in two steps. 
First, we place the theory in a background with a non-trivial $\Z_{2}$ flat connection, which induces the symmetry breaking
\begin{equation}
    \mk{su}(4)\;\longrightarrow\; \mk{su}(2)\,\oplus\,\mk{su}(2)\,\oplus\,\mk{u}(1)\,.
\end{equation}
Second, we observe that the $\mathfrak{sl}_{2}$ triple $\{e,h,f\}$ is embedded into the lower $2\times 2$ block of the $\mk{su}(4)$ matrices, which can be naturally identified with the second $\mk{su}(2)$ factor in the above decomposition. 

Turning on a nilpotent Higgs vev breaks this $\mk{su}(2)$ factor completely. As a result, the unbroken gauge algebra is
\begin{equation}
   \mk{g}_{e_{[2,1,1]}} \,=\, \mk{su}(2)\,\oplus\,\mk{u}(1)\,.
\end{equation}
This situation closely parallels the $\mk{su}(2)$ example discussed in Section~\ref{sec:CB-HB-7d}. Consequently, the physical parameter is given as
\begin{equation}
 \delta\Phi_{\mathrm{phy}}^{[2,1,1]} \,=\, 
 \begin{pmatrix}
        0 & 0 & 0 & 0\\
        0 & 0 & 0 & 0\\
        0 & 0 & 0 & 0\\
        0 & 0 & \phi & 0
 \end{pmatrix}\,,
\end{equation}
which is uncharged under the surviving gauge algebra $\mk{su}(2)\oplus\mk{u}(1)$. Further, the dof matches exactly the expected parameters from \eqref{eq:geo-higgs-parameter-211}. We observe that this parameter belongs to the Slodowy slice associated with the nilpotent of the current case. 

 \item A quadruplet. Once again, the nilpotent Higgs correspond to the upper-triangle of $P_{[4]}$ above and is given as:
\begin{equation}\label{eq:Phi-nilpotent-4}
\langle\Phi_{[4]}^{(n)}\rangle\,:=\,  \begin{pmatrix}
        0&1&0&0\\
        0&0&1&0\\
        0&0&0&1\\
      0&0&0&0
    \end{pmatrix}\,.
\end{equation}
The effect of this nilpotent vev is to break the gauge algebra completely. Hence, matching the above corresponding case.

The candidate physical fluctuations can be represented by a traceless $4\times 4$ matrix, i.e., an element of $\mk{su}(4)$, subject to the linearized gauge transformations given in \eqref{eq:linearized-gauge-transformations}. One finds that 9 complex degrees of freedom can be eliminated by gauge transformations, while an additional degree of freedom is removed by the tracelessness condition. Moreover, the remaining 6 complex degrees of freedom can be further reduced using residual gauge transformations, leaving only 3 independent physical modes. These can be brought to the following form:
\begin{equation}
 \delta\Phi_{\mathrm{phy}}^{[4]} \,=\, 
 \begin{pmatrix}
        0 & 0 & 0& 0 \\
        \phi & 0 & 0 & 0\\
        \varphi  & \phi & 0 & 0 \\
        \psi & \varphi  & \phi & 0
 \end{pmatrix}\,.
\end{equation}
which describes three complex scalar fields, in agreement with the expectation from the geometric analysis.

We also observe that, this matrix lies in the Slodowy slice associated with the corresponding nilpotent Higgs vev.
\end{itemize}

In the above examples, we observe that the Higgs branch moduli belongs to certain elements of the associated Slodowy slice. These observations lead to conjecture that: 
\begin{equation}\label{eq:conjecture-I}
\text{Conjecture--I} \ :\  \parbox{10cm}{Within the framework developed in this work, the geometric Higgs branch moduli of an $\mk{su}(N)$ gauge theory are captured by specific elements of the Slodowy slice associated with the corresponding nilpotent Higgsing.}
\end{equation}

Note that a refined version of this conjecture is given in \eqref{eq:conjecture-II} along with further discussion.

\subsection*{$\mk{su}(2N)$ gauge theory: the case of $X_{8}(\Z_{2N},B_{(4;1)}^{(9)})$}

We now consider a more general 3d $\N=2^{\ast}$, or 4d $\N=1^{\ast}$, with $\mk{su}(2N)$ Lie algebra. As discussed in the examples above, the different possible Higgsing are counted by
integer partitions, denoted by $\lambda_{i}\in\Z_{>0}$, of the $2N$ centres. In particular, we write 
\begin{equation}
  2N\,=\, \sum_{i=1}^{k}\,n_{i}\,\lambda_{i}\,, \quad \text{such that} \quad \lambda_{1} >  \lambda_{2} > \cdots > \lambda_{k}\,.
\end{equation}
with $n_{k}$ and $k \in \Z_{>0}$. We denote a possible partition by $\lambda$ and follow the convention:
\begin{equation}
  \lambda\,:=\, [\lambda_{1}^{n_{1}}\,,\lambda_{2}^{n_{2}}\,,\cdots\,,\lambda_{k}^{n_{k}}]\,.
\end{equation}

Since we are focusing on the scenario where all the centres are either at origin or organized in pairs or quadruplets under the action of $H=\Z_{2}\times\Z_{2}$, then the number $k$ can be at most three, i.e., $k\leq 3$.

We now discuss the three main possibilities:
\begin{itemize}
\item All centres are invariant. The partition in the case is $\lambda = [1^{2N}]$ and the nilpotent Higgs is identically zero. Hence, the gauge theory is unbroken.

\item $l$ pairs and $2N-2l$ invariant. The partition is given as $\lambda=[2^{l},1^{2N-2l}]$ with $l\leq N$. The associated nilpotent Higgs field is constructed by inserting the $2\times 2$ canonical nilpotent matrix
\begin{equation}\label{eq:e(2)}
  e_{(2)}
  \,=\,
  \begin{pmatrix}
    0 & 1 \\
    0 & 0
  \end{pmatrix}
\end{equation}
along $l$ diagonal blocks, together with $(2N-2l)$ trivial $1\times 1$ zero blocks, i.e., 
\begin{equation}\label{eq:nilpotent-l+2N-2l}
  \Phi_{[l,2N-2l]} \,=\, \begin{pmatrix}
    0 & 1 & 0 & \dots & 0 & 0 \\
    0 & 0 & 0 & 1 &\dots & 0 \\
    0 & 0 & 0 & 0 & \ddots   & \vdots \\
    \vdots & \vdots & \vdots & 0 & 0 &0  \\
    0 & 0 & 0 & \dots & 0 & 0 
\end{pmatrix}\,.
\end{equation}
Equivalently, we can write it as
\begin{equation}
   \Phi_{[l,2N-2l]} \,=\, \mathrm{diag}\left(\, \underbrace{e_{(2)}\,,\cdots\,,e_{(2)}}_{l\,\text{times}}\,, \underbrace{0,\,\dots,\,0}_{2N-2l\ \text{times}} \,\right)\,.
\end{equation}

One can verify that the above nilpotent is the upper-triangle of a $P_{[l,2N-2l]}$ matrix acting on the $(2N)$ centres. In other words, the $P_{[l,2N-2l]}$ matrix is obtained by mirroring the upper-triangle of \eqref{eq:nilpotent-l+2N-2l} along its diagonal. 

In this case, the unbroken Lie algebra is given as
\begin{equation}
  \mk{g}_{\mathrm{unbrok}}\,=\,\begin{cases}
     &\mk{su}(l)\,\oplus\, \mk{su}(2N-2l)\,\oplus\, \mk{u}(1)\,,\qquad \text{for}\ \ l\,<\,N\,,
     \\
     &\mk{su}(N)\,,\qquad\qquad \qquad  \qquad \qquad \quad \,\,\, \text{for}\ \ l\,=\,N\,.
  \end{cases}
\end{equation}
As an example, the reader may check the second and third cases of the above $\mk{su}(4)$ subsection. This result can be confirmed by checking the invariant subset of the harmonic 2-forms $\{\widetilde{\mathrm{h}}_{2}^{a}\}$ associated with the $\mk{su}(2N)$ gauge algebra according to the action of $H$, which is specified by $l$.

Following \eqref{eq:conjecture-I}, to find the physical fluctuations that correspond to the Higgs branch moduli, one considers the associated Slodowy slice. In particular, there is a complex field $\phi_{i}$ for each $e_{(2)}^{\,i}$ matrix in the block-diagonal, with $i=1,\cdots, l$\,, i.e., 
\begin{equation}
\mathrm{dim}_{\C}(\mathrm{HB})\,=\,  l\,.
\end{equation}
These correspond precisely to the geometric parameters which are given in the form of:
\begin{equation}
  \widetilde{\mathrm{h}}_{2}^{i}\,\wedge\,\dd y_{2}\,\qquad \, \widetilde{\mathrm{h}}_{2}^{i}\,\wedge\, \alpha_{2}\,,\quad i\,=\,1\,,\cdots\,,l\,.
\end{equation}
Some of the $\{\widetilde{\mathrm{h}}_{2}^{i}\}$ 2-forms are in fact given as linear combination of the $\{\widetilde{\mathrm{h}}_{2}^{1},\cdots,\widetilde{\mathrm{h}}_{2}^{2N-1}\}$ 2-forms associated with the $\mk{su}(2N)$ Lie algebra. For example, the reader may consider \eqref{eq:b3-b4-harmonic-fields-B49-Z4}.

The matching between the geometric Higgs branch moduli and the physical fluctuations parametrized by elements of the Slodowy slice can be verified in several explicit examples in the present setup, all of which provide evidence in support of our conjecture.

  \item Quadruplets. Recall that $H=\Z_{2}\times\Z_{2}$ can acts on 4 centres at once, which we refer to as a quadruplet of the group $H$. Including the quadruplet, the $(2N)$ centres can now have a partition in the form of: 
  \begin{equation}
    \lambda\,=\,[4^{n_{1}},\,2^{n_{2}}, \,1^{n_{3}}]\,,\qquad \text{with}\quad 4 n_{1}\,+\,2 n_{2}\,+\,n_{3}\,=\,2N\,.
  \end{equation}

In this case, the nilpotent Higgs would can also be presented in block-diagonal form as: (i) $n_{1}$ of $4\times 4$ matrices of the form given in \eqref{eq:Phi-nilpotent-4}, (ii) $n_{2}$ of $2\times 2$ matrices of the form given in \eqref{eq:e(2)}, (iii) $n_{1}$ of $1\times 1$ zero entries. Once again, such a nilpotent Higgs can be obtained from a corresponding $P_{[n_{1},n_{2},n_{3}]}$ as claimed in \eqref{eq:nilpotetn-Higgs-upper-triangle-P}.

The possible unbroken gauge algebra are then given as:
\begin{equation}
  \mk{g}_{\mathrm{unbrok}}\,=\,
  \begin{cases}
     &\mk{su}(n_{1})\,\oplus\,\mk{su}(n_{2})\,\oplus\,\mk{su}(n_{3})\, \oplus\, \mk{u}(1)^{\oplus 2} \qquad \text{non of $n_{i}$'s is zero}\,,
     \\
&\mk{su}(n_{1})\,\oplus\,\mk{su}(n_{j})\, \oplus\, \mk{u}(1) \qquad \text{either $n_{2}$ or $n_{3}$ is zero}\,,
     \\
&\mk{su}(n_{1})\, \qquad \text{both $n_{2}$ and $n_{3}$ are zero}\,.
\end{cases}
\end{equation}

We note that only the $\mk{su}(4)$ gauge algebra that is associated with the $B_{4}^{(9)}$ geometry would be Higgsed completely. Further, this braking pattern can be proven from the geometric point of view when considering the $H$-untwisted harmonic 2-forms $\{\widetilde{\mathrm{h}}_{2}^{\bullet}\}$.

Having the above $\mk{su}(4)$ example, we can deduce the number of physical fluctuations that belongs to the Slodowy slice and correspond the the HB moduli. Specifically, we have the following number of complex scalar fields:
\begin{equation}
\mathrm{dim}_{\C}(\mathrm{HB})\,=\,  3\,\times\,n_{1}\,+\, 1\,\times\,n_{2}\,.
\end{equation}
These are exactly the number of the geometric Higgs branch moduli that we can calculate as it has been done for the examples above. 
\end{itemize}

Before concluding this subsection, we should mention that an entirely analogous analysis applies to the case of $\Z_{2N+1}$ singularities.

%%%%%%%%%%%%%%%%%%%%%%%%%%%%%%%%%%%%%%%%%%%%%%%%%%%%%%%%%%%%%%%%%%%%%%%%%%%%%%%%%%%%%%%%%%%%%%%

\subsection*{An observation: the space $X_{8}(\Z_{N},B_{(4;1)}^{(26)})$}

In this case, the effective theory is a 3d $\N=2^{\ast}$ gauge theory, as indicated in Table~\ref{Table:fiberB36}. The rotational holonomy group of $B_{(4;1)}^{(26)}$ is the tetrahedral group $(\Z_{2}\times\Z_{2})\rtimes\Z_{3}$, which is generated by: 
\begin{equation}
   A\,=\, \begin{pmatrix}
        1&0&0&0\\
        0&-1&0&0\\
        0&0&-1&0\\
        0&0&0&1
    \end{pmatrix}\,\,,
     \qquad 
   B\,=\, \begin{pmatrix}
        1&0&0&0\\
        0&-1&0&0\\
        0&0&1&0\\
        0&0&0&-1
    \end{pmatrix}\,\,,
     \qquad 
    C\,=\, \begin{pmatrix}
        1&0&0&0\\
        0&0&0&1\\
        0&-1&0&0\\
        0&0&-1&0
    \end{pmatrix}\,.
\end{equation}
The lower $3\times 3$ block acts non-trivally on the base $\R^{3}$ where the centres of $\R^{4}/\Z_{N}$ are located, which we now consider. The elements  $A$ and $B$ generate the $\Z_{2}\times\Z_{2}$ subgroup and are analogous to those appearing in \eqref{eq:rotation-hol-B49}. Their action on the centres in $\R^{3}$ is therefore the same as \eqref{eq:x1-x2-for-B49}. Whereas, the new generator $C$ permutes the three distinct solutions of \eqref{eq:x1-x2-for-B49}. 

When we consider the action of the full group on the centres in $\R^{3}$, we note that we have two generic cases|apart from the trivial configurations:
\begin{itemize}
  \item First, we arrange the centres along one of the axes of $\R^{3}$. In this configuration, the theory is Higgsed as described above. We then take the quotient by the action of $C$, which identifies this non-trivial configuration with the other two axes, where no centres are present, i.e., with empty configurations. Hence, the gauge symmetry is completely broken.
  
  \item Naively, one might consider configurations with identical arrangements of centres along the three axes of $\R^{3}$. In such a case, it may appear that the quotient by $C$ produces a non-trivial effect. However, this configuration is inconsistent with the structure of the $\mk{su}(N)$ algebra. In particular, it does not correctly reproduce the $\mk{su}(N)$ Dynkin diagram, which is encoded by the intersection pattern of the $(N-1)$ independent 2-spheres between the centres. Hence, such a symmetric configuration
cannot be realized within the present framework.
\end{itemize}
We refrain from considering this geometry any further in this work. 

%%%%%%%%%%%%%%%%%%%%%%%%%%%%%%%%%%%%%%%%%%%%%%%%%%%%%%%%%%%%%%%%%%%%%%%%%%%%%%%%%%%%%%%%%

\subsection{Charged matter}\label{sec:charged-matter}

In the previous subsection, we saw that the geometric Higgs branch moduli, i.e., $H$-untwisted harmonic 3-forms and 4-forms, correspond to a particular fluctuations around the nilpotent vev that belong to the Slodowy slice. Now, the question is that: What do other fluctuations in the Slodowy slice correspond to? We claim that they can be interpret as charged matter under the gauge algebra $\mk{g}_{\mathrm{unbrok}}$.

In the following examples we will focus on particular elements of the Slodowy slice, those which provide fields in representations other that the adjoint of the surviving gauge algebra. However, generally speaking, one should consider all possible elements of $\mathcal{S}_{e}$ and examine their physical interpretation. Since, as we will argue toward the end of this subsection, all elements of $\mathcal{S}_{e}$ correspond to physical modes.  

We now proceed with the simplest non-trivial example.

\paragraph{$\mk{su}(3)$ gauge theory.} We consider both 3d $\N=2^{\ast}$ and 4d $\N=1^{\ast}$ $\mk{su}(3)$ gauge theories that are associated with the geometry of $X_{8}(\Z_{3},B_{4}^{(9)})$ and $Y_{7}(\Z_{3},B_{3}^{(6)})$, respectively. In both cases, we take the $H=\Z_{2}\times\Z_{2}$ action on the 3 centres of the corresponding GH space to leave one centre invariant and the other form a pair.

From the discussion of the previous subsection, we can immediately write down the nilpotent Higgs vev:  
\begin{equation}
\Phi_{[2,1]}^{(n)}\,=\,\left(\begin{array}{cc|c}
0 & 1 \,&\, 0 \\
0 & 0 \,&\, 0 \\
\hline
0 & 0 & 0
\end{array}\right)\,.
\end{equation}
This vev correspond to an $\mk{sl}(2,\C)$ triplet embedding in $\mk{su}(3)$ and breaks the gauge algebra down to $\mk{g}_{e_{[2,1]}}=\mk{u}(1)$. 

In this case, the Slodowy slice contains the following fluctuations:
\begin{equation}\label{eq:S-e-su(3)-example}
\mathcal{S}_{e}\  \supset\ \left( \begin{array}{cc|c}
0 & 0 \,&\, 0 \\
\psi & 0 \,&\, 0\\
\hline
0 & 0 & 0
\end{array}
\right)  \ + \ \left( \begin{array}{cc|c}
0 & 0 \,&\, 0 \\
0 & 0 \,&\, \phi_{-} \\
\hline
\varphi_{+} & 0 & 0
\end{array}
\right) \,.
\end{equation}
The first term gives a neutral complex scalar of a chiral multiplet, which gives the Higgs branch moduli of the theory and correspond to the expected geometric Higgs moduli. For the second term, we notice that these fields are charged under the surviving $\mk{u}(1)$ gauge algebra. In particular, they come in two chiral multiplets with opposite charges. In other words, we claim that the resulting low-energy description is
captured by
\begin{equation}
\parbox{10cm}{$\mk{u}(1)$ gauge theory with one neutral chiral multiplet and two chiral multiplets with opposite charges.}
\end{equation}

This structure is reminiscent of the F-theoretic derivation of charged matter as in \cite{Cecotti:2010bp}. Though, here, we have a geometric perspective of charged chiral multiplet rather than 7-branes. Further discussion on such relations will be given at the end of this subsection.

\paragraph{$\mk{su}(4)$ gauge theory.}

Let us now turn our attention to $\mk{su}(4)$ gauge theory with 4 real supercharges in 3d and 4d. We focus on the case of $4=[2,1,1]$, which has been analyzed in details above. In this case, the physical fluctuations contain the following: 
\begin{equation}
  \mathcal{S}_{e_{[2,1,1]}}\  \supset\ 
  \left(\begin{array}{cc|cc}
0 & 0 \,&\, 0 &0\\
\phi & 0 \,&\, 0 &0\\
\hline
0 & 0 & 0 &0 \\
0 & 0 & 0 &0 
\end{array}\right)\  +  \ \left(\begin{array}{cc|cc}
0 & 0 \,&\, 0 & 0  \\
0 & 0 \,&\, \psi_{1} &\psi_{2}\\
\hline
\varphi_{1} & 0 & 0 &0 \\
\varphi_{2} & 0 & 0 &0
\end{array}\right)\,.
\end{equation}
The first term correspond to the Higgs branch moduli as argued previously. The second term contains two doublets, namely $(\psi_{1},\psi_{2})$ and $(\varphi_{1},\varphi_{2})$, under the surviving $\mk{su}(2)$ factor of the Lie gauge algebra. This can be seen from considering the decomposition of an $\mk{su}(4)$ adjoint field under the unbroken $\mk{su}(2)$, which is given as    
\begin{equation}
\Phi_{(\mk{su}(4))}\,=\,  \left(\begin{array}{c|c}
 4\,\times\,\text{singlets} \,&\, 2\,\times\, \text{doublets} \\
 \hline
 2\,\times\,\text{doublets} \,&\, \Phi_{(\mk{su}(2))} 
\end{array}\right)\,.
\end{equation}
Here, the doublets in the upper-right block consist of rows, whereas, the doublets of the lower-left block are given by columns. The intersection of the Slodowy slice with the above representations, i.e.,
\begin{equation}
  \mathcal{S}_{e_{[2,1,1]}}\ \cap\   \Phi_{(\mk{su}(4))}\,,
\end{equation}
verify that we have two doublets under the $\mk{su}(2)\subset \mk{su}(2)\oplus\mk{u}(1)$. These doublets come with opposite charges under the $\mk{u}(1)$ factor. Seems to be an unexpected outcome!

\paragraph{$\mk{su}(5)$ gauge theory.}

Let us now consider the $\mk{su}(5)$ gauge theory, which can be realized in 3d $\N=2^{\ast}$ and 4d $\N=1^{\ast}$ as stated above. Further, we take the $H=\Z_{2}\times\Z_{2}$ action on the centres such that three centres are invariant and two form a pair. This choice amount to have a nilpotent Higgsing of $\mk{su}(2)\subset \mk{su}(5)$. In particular, we take the following semisimple element 
\begin{equation}
  h^{[1^{3},2]}\,=\,\mathrm{diag}(1,-1,0,0,0)\,,
\end{equation}
which break the Lie algebra as
\begin{equation}\label{eq:su5-su3-su2-u1}
  \mk{su}(5)\ \to \ \mk{su}(3)\,\oplus\,\mk{su}(2)\,\oplus\,\mk{u}(1)\,.
\end{equation}
The associated nilpotent Higgs vev is given by
\begin{equation}
  \Phi_{[1^{3},2]}^{(n)}\,:=\, e^{[1^{3},2]}\,=\,\left(\begin{array}{cc|ccc}
0 & 1 \,&\, 0 &0&0\\
0 & 0 \,&\, 0 &0&0\\
\hline
0 & 0 & 0 &0 & 0\\
0 & 0 & 0 &0 & 0\\
0 & 0 & 0 &0 & 0
\end{array}\right)\,.
\end{equation}
and it breaks the $\mk{su}(2)$ factor above.

In this case, the Slodowy slice contains the following fluctuation: 
\begin{equation}
  \mathcal{S}_{e}\  \supset\ 
  \left(\begin{array}{cc|ccc}
0 & 0 \,&\, 0 &0&0\\
\phi & 0 \,&\, 0 &0&0\\
\hline
0 & 0 & 0 &0 & 0\\
0 & 0 & 0 &0 & 0\\
0 & 0 & 0 &0 & 0
\end{array}\right)\  +  \ \left(\begin{array}{cc|ccc}
0 & 0 \,&\, 0 &0&0\\
0 & 0 \,&\, \psi_{1} &\psi_{2}&\psi_{3}\\
\hline
\varphi_{1} & 0 & 0 &0 & 0\\
\varphi_{2} & 0 & 0 &0 & 0\\
\varphi_{3} & 0 & 0 &0 & 0
\end{array}\right)\,.
\end{equation}

A generic $\mk{su}(5)$ adjoint matter, say a  Higgs field, would be decompose under \eqref{eq:su5-su3-su2-u1} as the following:
\begin{equation}
\Phi_{(\mk{su}(5))}\,=\,  \left(\begin{array}{c|c}
 \Phi_{(\mk{su}(2))} \,&\, \psi_{(\textbf{3},\textbf{2})} \\
 \hline
 \varphi_{(\overline{\textbf{3}},\textbf{2})} \,&\, \Phi_{(\mk{su}(3))} 
\end{array}\right)\,.
\end{equation}
Here, the bifundamental field  $\psi_{(\mathbf{3},\mathbf{2})}$ transforms in the $(\mathbf{3},\mathbf{2})$ representation of $\mk{su}(3)\oplus\mk{su}(2)$. It is a $2\times 3$ matrix, whose rows transform in the fundamental of  $\mk{su}(3)$ and whose columns transform in the fundamental of $\mk{su}(2)$. It carries $\mk{u}(1)$ charge $q$. Similarly,  $\varphi_{(\overline{\mathbf{3}},\mathbf{2})}$ transforms in the  $(\overline{\mathbf{3}},\mathbf{2})$ representation and can be represented as a $3\times 2$ matrix whose columns transform in the anti-fundamental of $\mk{su}(3)$ and whose rows transform in the fundamental of $\mk{su}(2)$. It carries $\mathfrak{u}(1)$ charge $(-q)$.

The intersection between the Slodowy slice and the above representations of the Higgs field, which considered as fluctuations, i.e.,
\begin{equation}
  \mathcal{S}_{e_{[3,2]}}\ \cap\   \Phi_{(\mk{su}(5))}\,,
\end{equation}
shows that the surviving dof are along a triplet $(\psi_{1},\psi_{2},\psi_{3})$ and anti-triplet $(\varphi_{1},\varphi_{2},\varphi_{3})$ of the unbroken $\mk{su}(3)$ gauge algebra. Thus, we have a non-chiral gauge theory. 

The above are surprising results, at least to the author.

%%%%%%%%%%%%%%%%%%%%%%%%%%%%%%%%%%%%%%%%%%%%%%

\subsection*{Are these massless fluctuations?} 

The correspondence we point out between the geometric HB and certain elements of $\mathcal{S}_{e}$, as seen in section \ref{sec:HB}, suggests that these modes are massless. One would like to be sure that those elements are indeed massless, as we claim that they are identified with untwisted harmonic 3-forms and 4-forms. The natural question is then, do we expect all elements of $\mathcal{S}_{e}$ to be massless? In
particular, one would like to understand whether the charged matter fields described above are also massless.

To answer this question, we link our work to the F-theoretic realization of similar non-chiral trapped matter in \cite{Cecotti:2010bp} and M-theoretic one in  \cite{Barbosa:2019bgh}. We use techniques developed in these references, as well as, in \cite{Donagi:2003hh,Cecotti:2009zf}, to argue that the Higgs branch moduli and the charged matter are indeed massless at special points in the geometry of $B_{3}^{(6)}$. Eventually, leading us to trapped, or localized, massless matter fields interpretation of our results above. Earlier work on localized matter can be found in \cite{Donagi:2008ca,Beasley:2008dc,Beasley:2008kw}

\paragraph{From co-Seifert fibration structure to localized matter.}

Let us examine the geometry of $B_{3}^{(6)}$ in more details. In geometric engineering, this geometry is associated with the 4d $\N=1^{\ast}$|and 3d $\N=2^{\ast}$ seen as $\bbS^{1}$ reduction of the 4d theories| which is one of the main examples in this section. Following the discussion in section 9 of \cite{lambert2013} and appendix \ref{sec-co-Seifert-fibration}, $B_{3}^{(6)}$ admits a co-Seifert fibration over an interval $I$. In particular, it can be realized as
\begin{equation}
  T^{2}\,\hookrightarrow\,B_{3}^{(6)}\, \to\, I\,,
\end{equation}
with $\mathbb{D}_{2}$ as the structure group. Here, we think of $T^{2}$ as a complex Riemann surface.

The above fibration structure suggests that we can think of the 7d triplet of Higgs fields, $\mathsf{\Phi}:=(\Phi_{1},\Phi_{2},\Phi_{3})$, see \eqref{eq:7d-VM}, in the following way. First, we can restrict two fields to live on the torus and define a holomorphic Higgs fields $\Phi(z)$, e.g., as
\begin{equation}
  \Phi(z)\,=\, \Phi_{2}\,+\,i\,\Phi_{3}\,,
\end{equation}
where $z$ is a coordinate on a local patch $U\cong\C \subset T^{2}$, up to the action of $\mathbb{D}_{2}$ group. Second, we can restrict the remaining Higgs field, $\Phi_{1}$, to the interval $I$. We now denote $\Phi_{1}$ by $\Phi(t)$ with $t$ being the coordinate on $I$. These choices are motivated in spirit by those taken in \cite{Barbosa:2019bgh}. The holomorphic field $\Phi(z)$ can be a nilpotent element of the complixified Lie algebra $\mk{g}$, while $\Phi(t)$ serve as the semisimple element.

From the discussion around \eqref{eq:effective-3d-N=2*}, the reduction of the 7d triplet of Higgs fields gives rise to a triplet of scalar fields in 3d, denoted by $\varphi^{(i)}$, with $i=1,2,3$. In the 3d $\N=2^{\ast}$ and 4d $\N=1^{\ast}$ theories, these fields are massive and are organized into chiral multiplets as in \eqref{eq:massive-chiral}, which we also denote by $\Phi_{i}$. Thus, the profile of the 7d Higgs fields  is inherited by the lower-dimensional theories. Supersymmetry further ensures that the accompanying scalar fields $\phi^{(i)}$ appearing in \eqref{eq:massive-chiral} exhibit the same behavior.

Similar to our discussion around \eqref{eq:delta-Phi-phys-su(2)}, one expects to get physical fluctuations around the vev,
\begin{equation}
  \Phi(z)\,=\,\langle\Phi\rangle\,+\,\delta\Phi(z)_{\mathrm{phy}}\,,
\end{equation}
In this context, the vev of $\Phi(z)$, which we will denote by $\Phi$, depend on $z$ and it is given by the embeddings of 
\begin{equation}
  \begin{pmatrix}
    0& 1
    \\
    z&0
  \end{pmatrix}\ \hookrightarrow \ \mk{g}\,,
\end{equation}
which is only nilpotent at $z=0$, i.e., $\Phi(0)=e$. As in \eqref{eq:linearized-gauge-transformations}. the physical modes $\delta\Phi(z)_{\mathrm{phy}}$ are defined modulo a linearized gauge transformation,
\begin{equation}\label{eq:Lin-Gaug-Trans}
  \delta\Phi(z)_{\mathrm{phy}}\,\sim\, \delta\Phi(z)_{\mathrm{phy}}\,+\, \mathrm{ad}_{\Phi}(\chi)\,.
\end{equation}

Following the discussion in \cite{Cecotti:2010bp}, we now focus on $\delta\Phi(z)_{\mathrm{phy}}$ which naively looks like a pure gauge, 
\begin{equation}\label{eq:phys-mod-like-pure-gauge}
  \delta\Phi(z)_{\mathrm{phy}}\,=\, \mathrm{ad}_{\Phi}(\xi)\,,
\end{equation}
with $\xi\in\mk{g}$. Indeed, this representation of $\delta\Phi(z)_{\mathrm{phy}}$ is a pure gauge, and so gauge equivalent to zero, having $\xi$ admits no zeros on the local patch $U$. However, once $\xi$ develops zeros on $U$, then it is no longer globally trivial and it can not be gauged away. We consider the case where $\xi$ develop poles at $z=0$ of order $m$; hence, we write $\delta\Phi(z)_{\mathrm{phy}}$ as
\begin{equation}
  \delta\Phi(z)_{\mathrm{phy}}\,=\, \mathrm{ad}_{\Phi}\left(\frac{\eta}{z^{m}}\right)\,,
\end{equation}
with $m$ is a positive integer. We also note that the $z=0$ point is singlet under the $\mathbb{D}_{2}$ group.

The question is how to use the above discussion practically? The following algorithm provides the answer.
\begin{itemize}
  \item Suppose we start with a gauge theory with a semisimple $\mk{g}$ Lie algebra. In this work we focus on $\mk{a}_{N-1}$ algebra. The theory is equipped with the above $\Phi(z)$ and $\Phi(t)$. $\Phi(t)$ takes the form of a semisimple element of $\mk{sl}(2,\C)$ algebra weighted by the coordinate on the interval. We decompose the adjoint valued field of $\mk{g}$ according to the semisimple element. This is similar to the procedure we performed in the above examples. We then choose a candidate for the charged matter. 

For example, we consider the $\mk{su}(3)$ theory equipped with:
\begin{equation}
  \Phi(t)\,=\,
  \begin{pmatrix}
    t &0&0\\
    0&t&0\\
    0&0&-2t
  \end{pmatrix} \,,\qquad \Phi(z)\,=\,  \left( \begin{array}{cc|c}
0 & 1 \,&\, 0 \\
z & 0 \,&\, 0\\
\hline
0 & 0 & 0
\end{array}
\right)\,.
\end{equation}
In this case, a candidate for the charged matter is given by
\begin{equation}
\delta\Phi(z)_{\mathrm{cand}}\,=\,   \left( \begin{array}{cc|c}
0 & 0 \,&\, \psi_{1} \\
0 & 0 \,&\, \psi_{2}\\
\hline
\psi_{3} & \psi_{4} & 0
\end{array}
\right)\,.
\end{equation}

\item As the second step, we impose the  linearized gauge transformation given in \eqref{eq:Lin-Gaug-Trans} such that our candidate is gauge equivalent to zero. In other words, we try to write it as in \eqref{eq:phys-mod-like-pure-gauge}. 

For the $\mk{su}(3)$ example, we impose
\begin{equation}
  0\,=\,\delta\Phi(z)_{\mathrm{cand}}\,+\, \mathrm{ad}_{\Phi(z)}(\chi)\,,
\end{equation}
with $\chi$ takes the same form as $\delta\Phi(z)_{\mathrm{cand}}$. Such a condition is referred to as the annihilator condition in \cite{Barbosa:2019bgh}. The above leads to 
\begin{equation}
\begin{split}
    \psi_{1}\,+\,\chi_{2}\,&=\,0\,,\qquad \psi_{2}\,+\,z\,\chi_{1}\,=\,0\,,
    \\
     \psi_{4}\,+\,\chi_{3}\,&=\,0\,,\qquad \psi_{3}\,+\,z\,\chi_{4}\,=\,0\,.
\end{split}
\end{equation}
We can radially see that $\psi_{1}$ and $\psi_{4}$ can be gauged away, i.e., it is not genuine physical degrees of freedom. However, this is not the case for $\psi_{2}$ and $\psi_{3}$ at $z=0$. This implies two things: (i) $\psi_{2}$ and $\psi_{3}$ are physical dof. (ii) $\psi_{2}$ and $\psi_{3}$ are trapped, or localized, at $z=0$. Further, we note that $\psi_{2}$ and $\psi_{3}$ are element of the Soldowy slice, see \eqref{eq:S-e-su(3)-example}, that is associated with the nilpotent element, i.e., the vev, at $z=0$.

\item We can verify, or determine, the physical degrees of freedom by observing that \cite{Cecotti:2010bp}
\begin{equation}\label{eq:annhilation-zm-delta-Phi-phys}
  z^{m}\,\delta\Phi(z)_{\mathrm{phy}}\,=\, \mathrm{ad}_{\Phi}(\eta(z))\,=\,0 \, \quad( \text{at $z=0$})\,.
\end{equation}
In other words, we can multiply $\delta\Phi(z)_{\mathrm{phy}}$ by the location of the poles such that $\mathrm{ad}_{\Phi}\left(\eta/z^{m}\right)$ has no poles. The physical dof are then given by 
\begin{equation}
\restr{\eta(z)}{z=0}\,.
\end{equation}

For the example at hand, $m=1$ and one can show that
\begin{equation}
  \eta(z)\,=\, 
\left( \begin{array}{cc|c}
0 & 0 \,&\, \psi_{2} \\
0 & 0 \,&\, z\,\psi_{1}\\
\hline
z\,\psi_{4} & \psi_{3} & 0
\end{array}
\right)\,,
\end{equation}
where $\psi_{2}$ and $\psi_{3}$ survive at $z=0$ as expected. 
\end{itemize}
A similar analysis can be done for the field $\Phi(t)$. That we shall consider the annihilation condition given as
\begin{equation}
  0\,=\,\delta\Phi_{\mathrm{cand}}\,+\,\mathrm{ad}_{\Phi(t)}(\chi)\,.
\end{equation}
which implies that the physical dof are located at $t=0$, see \cite{Barbosa:2019bgh}. In practice, the actual constrains on $\delta\Phi_{\mathrm{cand}}$ are coming from the annihilator condition with involving $\Phi(z)$. 

We should mention that, the above procedure, maybe supplemented with the traceless condition, enable us to determine the Higgs branch moduli as, e.g., in \eqref{eq:3d-su2-HB-moduli}. This can be performed to the example at hand where we arrive at the expected result. Hence, in general, giving the elements of $\mathcal{S}_{e}$ that correspond to the geometric Higgs branch moduli an interpretation as trapped matter.

On physical ground, one expects that the fluctuations in \eqref{eq:phys-mod-like-pure-gauge} to have a scalar potential, see, e.g., \cite{Cecotti:2010bp}, given as
\begin{equation}\label{eq:potential-eta}
  V(\eta)\,=\, \Tr\left(  \mathrm{ad}_{\Phi}(\eta)^{2}  \right)\,.
\end{equation}
In general, one expects them to get massive away from the trap point $z=0=t$. However, at $z=0=t$, $\eta$ commutes with $\Phi$ as can be seen from \eqref{eq:annhilation-zm-delta-Phi-phys}. Therefore, the localized matter, both the HB moduli and charged fields, are indeed massless at the trap point. The reader is invited to verify this for the $\mk{su}(3)$ example. 

It is somewhat puzzling that the geometric framework developed in the present and preceding subsections does not explicitly capture the presence of the trap point. Rather, the appearance of the trap point emerged indirectly when we analyzed whether the relevant fluctuations are massless or massive. Nevertheless, in the light of the above discussion, we observe that our analysis is most appropriately interpreted as being valid at the trap point\footnote{These observations work in both directions. In particular, the framework developed here can also be used to provide a geometric interpretation of the Higgsing analysis presented in \cite{Barbosa:2019bgh}.} $z=0=t$.

\paragraph{The $B_{4}^{(9)}$ geometry and the localized matter.}

For the $B_{4}^{(9)}$ space, the associated theories are genuine 3d $\N=2^{\ast}$ gauge theories. The above analysis extends naturally to this case upon examining its co-Seifert fibration structure in more detail. As discussed in \cite{lambert2013}, $B_{4}^{(9)}$ can be realized as a fibration of the $B_{3}^{(2)}$ space over an interval $I_{1}$, with structure group $\mathbb{D}_{1}$. In turn, $B_{3}^{(2)}$ admits a co-Seifert fibration as a $T^{2}$ bundle over another interval $I_{2}$, also with structure group $\mathbb{D}_{1}$. By noticing that the two $\mathbb{D}_{1}$ groups are the same, the total space can be effectively described as
\begin{equation}
  T^{2}\,\hookrightarrow\,B_{4}^{(9)}\, \to\, I_{1}\,\times\,I_{2}\,,
\end{equation}
with $\mathbb{D}_{1}$ as the structure group.

In this setup, we have a holomorphic Higgs field supported on the 2-torus and acquire a nilpotent vacuum expectation value as before. The remaining Higgs field, $\Phi_{1}$, is supported on the base $I_{1}\times I_{2}$, and its vacuum expectation value is given as $(t_{1}+t_{2})$ times the semisimple element $h$ that correspond to the nilpotent vev. Here, $t_{1}$ and $t_{2}$ are the coordinates on $I_{1}$ and $I_{2}$, respectively. The rest of the analysis proceeds analogously to the previous setup and the trap point is now given at $z=0=t_{1}=t_{2}$.

\paragraph{Trapped matter and the Slodowy slice.}

Furthermore, for all cases, one finds that the trapped matter at are given exactly by elements of the corresponding Slodowy slice $\mathcal{S}_{e}$. This lead us to conjecture that:
\begin{equation}\label{eq:conjecture-II}
\text{Conjecture--II} \ :\ \parbox{10cm}{In the M-theoretic framework developed in this work, the massless trapped charged matter and the Higgs branch moduli are geometrically encoded in the Slodowy slice associated with the nilpotent Higgsing.}
\end{equation}
This is a refined version of the conjecture presented in \eqref{eq:conjecture-I}.

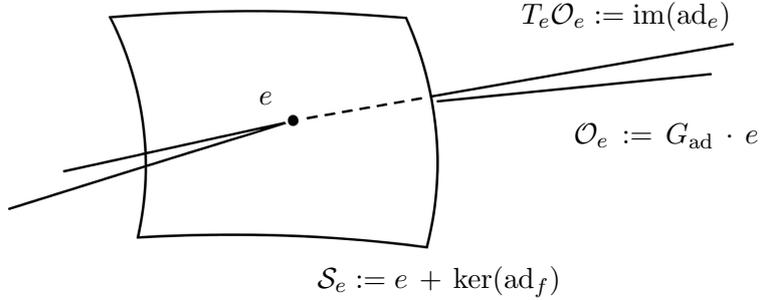
\begin{figure}[H]
\centering
\begin{tikzpicture}
\node[above right] (img) at (0,0) {\includegraphics[width=0.65\textwidth]{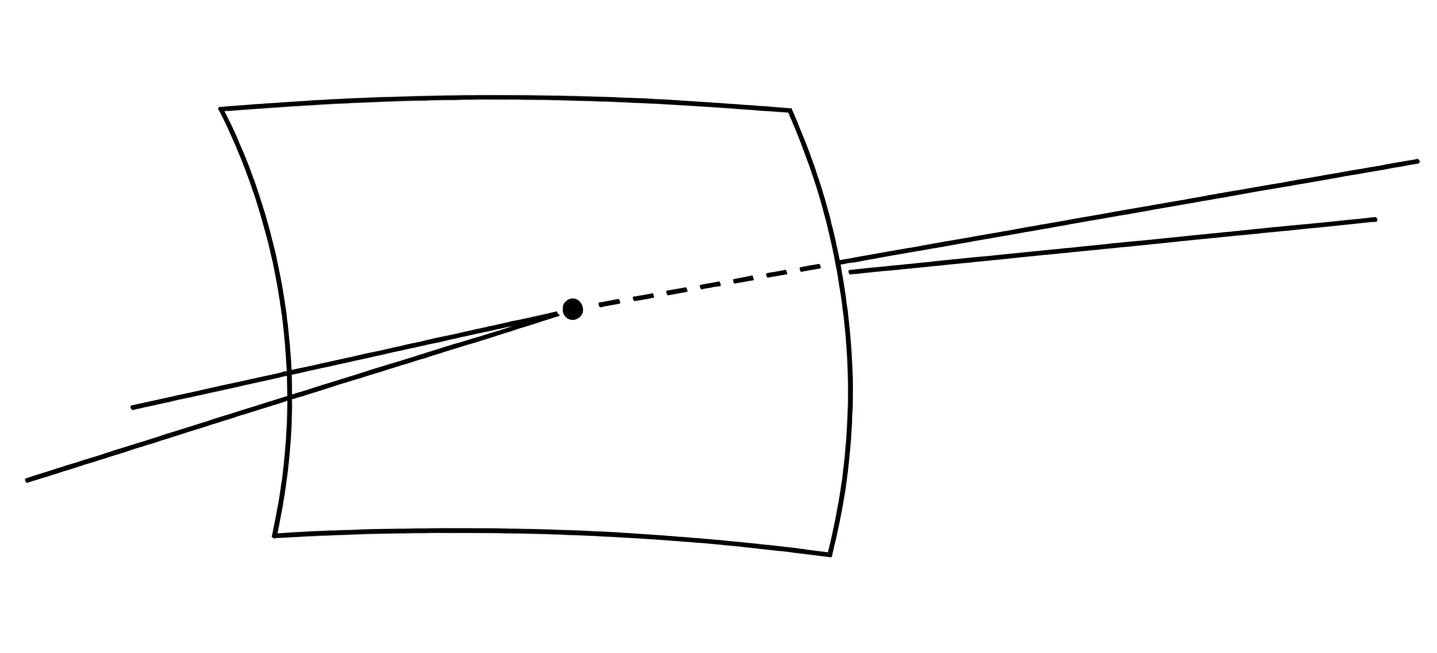}};

\node at (105pt,79pt) {{$e$}};

\node at (170pt,10pt) {{$\mathcal{S}_{e}:=e\,+\,\mathrm{ker}(\mathrm{ad}_{f})$}};

\node at (255pt,65pt) {{$\mathcal{O}_{e}\,:=\,G_{\mathrm{ad}}\,\cdot\,e$}};

\node at (240pt,110pt) {{$T_{e}\mathcal{O}_{e}:=\mathrm{im}(\mathrm{ad}_{e})$}};

\end{tikzpicture}
\caption{The figure illustrates a nilpotent element $e$ together with its orbit $\mathcal{O}_{e}$, generated by the adjoint action of the group on $e$. The tangent space at $e$, $T_{e}\mathcal{O}_{e} = \mathrm{im}(\mathrm{ad}_{e})$, admits a natural physical interpretation as the space of linearized gauge transformations. The transverse fluctuations at $e$ are captured by the Slodowy slice $\mathcal{S}_{e}$, which we interpret as the space of physical degrees of freedom. The figure is adapted from section 1.5 of \cite{slodowy1980four}.}
\label{Fig:Slodowy-slice}
\end{figure}

One way to intuitively realize the above conjecture is to consider the decomposition of the Lie algebra $\mk{g}$ under the $\mk{sl}(2,\C)$ triple $\{e,h,f\}$. In particular, $\mk{g}$ decomposes, see \cite{Collingwood1993,slodowy1980four,slodowy2006simple}, as 
\begin{equation}
  \mk{g}\,=\, \mathrm{im}\left(\mathrm{ad}_{e}\right)\,\oplus\,\mathrm{ker}\left(\mathrm{ad}_{f}\right)\,.
\end{equation}
Since our Higgs fields are adjoint-valued and the physical perturbations are defined modulo linearized gauge transformations, which belong to $\mathrm{im}\left(\mathrm{ad}_{e}\right)$, then we conclude that
\begin{equation}
   \delta\Phi_{\mathrm{phy}}\,\in\,\mathrm{ker}\left(\mathrm{ad}_{f}\right)\, =\, \mathcal{S}_{e}\,-\,e\,.
\end{equation}
Meaning that, at the trap point, the physical fluctuations belong to the transverse direction of the corresponding nilpotent orbit, i.e., the Slodowy slice. Whereas the tangent space of the nilpotent element at the trap point correspond to gauge transformations. The geometric description is depicted in Figure \ref{Fig:Slodowy-slice}.  

However, naively, these modes are seen to be massive given the expected physical potential in \eqref{eq:potential-eta}, with $\eta$ is replaced by $\delta\Phi_{\mathrm{phy}}$. It is only due to the above algebraic trap point framework, in which we one can construct $\delta\Phi_{\mathrm{phy}}$ satisfying \eqref{eq:annhilation-zm-delta-Phi-phys} such that it belongs to $\mathcal{S}_{e}$ and commute with the nilpotent element $e$ at the trap point.

%%%%%%%%%%%%%%%%%%%%%%%%%%%%%%%%%%%%%%%%%%%%%%%%%%%%%%%%%%%%%%%%%%%%%%%%%%%%%%%%%%%%%%%%%

\section{Conclusions and outlook}

The main novelty of our work lies in the following aspects:

\paragraph{Biberbach 4-manifolds in M-theory.}

We have considered Biberbach 4-manifolds $B_{4}$, which were classified in \cite{lambert2013}, within the geometric engineering framework in M-theory. These manifolds can be realized as finite quotients of a 4-torus, i.e., $T^{4}/H$. To obtain 3d gauge $ADE$ theories, we examined the fibration of $\R^{4}/\Gamma_{ADE}$ over $B_{4}$. For the 8d total space $X_{8}$, we analyzed the existence of a $Spin(7)$-structure on the resulting. Consistency of this structure requires that the group $H$ should act on the $Sp(1)$-structure of the $\R^{4}/\Gamma_{ADE}$ fibers. We further determined the possible holonomy groups of the spaces $X_{8}$.

We have found that the resulting 3d theories can be interpreted as mass deformations of 3d $\N=8$ theories. In particular, depending on the topology of the $B_{4}$ spaces, one obtains 3d $\N=2^{\ast}$ and 3d $\N=4^{\ast}$ theories. Furthermore, by exploiting the co-Seifert fibration structure of the $B_{4}$ spaces, we identified the cases in which the 3d theories can be viewed as $\bbS^{1}$ reductions of 4d $\N=1^{\ast}$ or $\N=2^{\ast}$ theories. Equivalently, the 4d theories associated with the 7d total space given as $\R^{4}/\Gamma_{ADE}$ fibered over $B_{3}$, which were considered earlier in \cite{Acharya:1998pm}, can now themselves be interpreted as mass deformations of $4$d $\N=4$ theories.

 \paragraph{Higgs branch of gauge theories.}

We have considered the action of a finite group\footnote{At this stage, we restrict to cyclic groups.} $H$ by permutations on the centres of $\R^{4}/\Gamma_{ADE}$, as discussed in \cite{wright2011quotientsgravitationalinstantons}. We found that such permutations can be interpreted as implementing nilpotent Higgsing of the corresponding 7d $\N=1$ $ADE$ gauge theory, at the expense of complexifying the gauge algebra $\mk{g}_{ADE}$. The corresponding Higgs fields arise from the triplet of scalar fields in the 7d theory. The Higgsing can also be seen from geometric perspective via the action on the harmonic 2-form\footnote{This was observed earlier in 4d theories in \cite{Acharya:2023xlx}.} $\{\widetilde{\mathrm{h}}_{2}^{a}\}$ of $\widetilde{\R^{4}/\Gamma_{ADE}}$, which are associated with the centres. However, the action of $H$ does not preserve the $Sp(1)$-structure on $\R^{4}/\Gamma_{ADE}$ and therefore breaks supersymmetry, rendering the resulting theory non-supersymmetric.

We argued that supersymmetry can be restored in lower-dimensional theories once we fiber $\R^{4}/\Gamma_{ADE}$ over an internal space $Y_{n}$ with $H$ acting on the total space. Here, the group $H$ is any finite group of the $SO(3)$ rotational group. We then focus on the case where the $Y_{n}$ space is given by $B_{4}$|the cases of $B_{3}$ were considered in \cite{Acharya:1998pm}. We also analyzed the possible $p$-forms on $X_{8}$ in terms of their behavior under the $H$ group.

We then turned to the Higgs branch of the 3d $\N=2^{\ast}$ and 4d $\N=1^{\ast}$ theories. These theories contain a triplet of adjoint-valued massive chiral multiplets $\Phi_{i}$, with $i=1,2,3$, together with a superpotential that enforces the $\mk{sl}(2,\C)$ commutation relations among them. Consistency then requires considering embeddings of $\mk{sl}(2,\C)$ into the complexified gauge algebra $\mk{g}$. Field configurations realizing such embeddings provide the necessary Higgs fields for the nilpotent Higgsing phenomenon. 

We analyzed the Higgs branch moduli from two complementary perspectives: geometric and field--theoretic.
\begin{itemize}
  \item From the geometric viewpoint, the Higgs branch moduli arise from $H$-invariant $p$-forms constructed as wedge products of $H$-twisted forms on the constituent spaces. Such $p$-forms give rise to massless scalar fields, corresponding to the bosonic components of chiral multiplets, which parametrize the Higgs branch.

  \item From the field-theoretic perspective, the Higgs branch moduli are described as fluctuations around a nilpotent vacuum configuration subject to the linearized gauge transformation. We conjectured that they should be identified with particular elements of the Slodowy slice $\mathcal{S}_{e}$ associated with the nilpotent vev $e$.
\end{itemize}
We find that the geometric and field-theoretic descriptions of the Higgs branch moduli agree. This provides a non-trivial consistency check of the proposed Higgs fields, the associated nilpotent Higgsing mechanism, and the conjecture. 

\paragraph{Charged matter.}

The correspondence between certain elements of the Slodowy slice and the geometric Higgs branch pushes us to consider other elements of $\mathcal{S}_{e}$. Some of these elements can be interpreted as non-chiral charged matter under the surviving gauge algebra. These fields are non-chiral in the sense that they come as pairs in representation $R$ and its anti-representation $\overline{R}$. We considered several examples which can be generalized to more complicated cases. 

These matter fields can be verified to be massless only when we link our work to existing literature on T-branes in \cite{Cecotti:2010bp,Barbosa:2019bgh}. In particular, we found that our charged matter can be identified with the trap matter fields developed in these references. Thus, we refined our previous conjecture to include the massless charged matter and HB moduli as elements of $\mathcal{S}_{e}$, with trapped-matter interpretation.

%%%%%%%%%%%%%%%%%%%%%%%%%%%%%%%%%%%%%%%%%%%%%%%
%%%%%%%%%%%%%%%%%%%%%%%%%%%%%%%%%%%%%%%%%%

\subsection*{Outlook}
Our systematic analysis suggests several promising avenues for future research:
\begin{itemize}
  \item Investigate the inner and outer automorphisms of gauge theories induced by the topology of Bieberbach spaces and their impact on the effective 3d theories. We will further explore the relationship between these automorphisms and the structure of Coulomb and Higgs branches. The discussion in \cite{Khlaif:2025jnx} can be used as a guide. 

    \item Analyze the decomposition of these spaces and the associated generalized symmetries,. In particular, examining different topological field theories coupled to the effective theories via the corresponding symmetry topological operators, following the framework of \cite{Najjar:2024vmm,Khlaif:2025jnx}; also see \cite{Najjar:2025htp}.

    \item  Examine non-orientable Bieberbach spaces, which do not admit $Spin$-structure, and their physical interpretation within geometric engineering, as well as in other string/M-theory frameworks.

    \item Extend our analysis beyond co-Seifert fibrations with a circle base. As noted in \cite{lambert2013}, orientable Bieberbach spaces may also admit co-Seifert fibrations with an interval base, leading to singular fibers. It would be interesting to study the resulting effective physics and generalized symmetries in such cases.

  \item It is known that the 7d theories engineered in $M$-theory via the $\R^{4}/\Gamma_{ADE}$ has type IIA string theory interpretation in terms of configurations of D6-branes. We have observed that when considering the nilpotent Higgsing interpretation of the permutation group $H$ action on the centres of $\R^{4}/\Gamma_{ADE}$ the theory become non-supersymmetric. Therefore, one should be careful when considering any configuration of D-/M-branes in nilpotent background. It is interesting to examine conditions on cases where such configurations remain supersymmetric without the need of wrapping the branes on compact internal spaces. Furthermore, if compactifying on such internal space is necessary, then one should examine what criteria these spaces should satisfy.

  \item As noted previously, it is a bit puzzling that the framework developed in this work do not sense the trap matter interpretation. It would be interesting to examine this puzzle further and provide more direct access to the trap matter framework.

  \item In this work, we have restricted our analysis to $\mk{su}(N)$ gauge theories. It would be important to extend this study to other $D$ and $E$ type gauge algebras, as well as to non-simply laced cases, and to investigate whether the observations and conjectures presented here continue to hold in these more general settings.

  \item In this work we focused on the Higgs branch of the engineered 3d $\N=2^{\ast}$ and 4d $\N=1^{\ast}$ theories. One could also consider and analyze the 3d $\N=4^{\ast}$ theories and their 4d partners.
\end{itemize}

%%%%%%%%%%%%%%%%%%%%%%%%%%%%%%%%%%%%%%%%%%%%%%%%%%%%%%%%%%%%%%%%%%%%%%%%%%%%%%%%%%%%%%%%%

\subsection*{Acknowledgments}

We would like to thank Sergio Cecotti, Kimyeong Lee, and Kotaro Kawai for fruitful discussions. Special thanks go to the author’s family for their unlimited support.

%%%%%%%%%%%%%%%%%%%%%%%%%%%%%%%%%%%%%%%%%%%%%%%%%%%%%%%%%%%%%%%%%%%%%%%%%%%%%%%%%%%%%%%%%

\appendix

\section{Bieberbach manifolds}\label{app:Bieberbach}

In this appendix, we review fundamental aspects of Bieberbach $B_{n}$ $n$-manifolds and their co-Seifert fibrations. The primary references for this material are: \cite{Auslander1960,auslander1965,Charlap-Vasquez-1973,brown1978,Charlap1986,Rossetti:1998qz,Pfaeffle2000Dirac,conway2003,Ratcliffe_2006,Ratcliffe_2010,putrycz2010,lutowski2011,ratcliffe2012,szczepanski2012geometry,lambert2013,Lutowski_2015,ocampo2019}.

\subsection{The general construction}

Given an Euclidean space $\R^{n}$, the map $\phi:\R^{n}\to\R^{n}$ is an isometry of $\R^{n}$, denoted by $\mathrm{Isom}(\R^{n})$, having that it can be written as
\begin{equation}
    \phi(x)\,=\,A(x)\,+\,v\,,
\end{equation}
with $x$ a point in $\R^{n}$, $A\in O(n)$ and $v\in\R^{n}$. In physical terms, such transformations are equivalent to Galilean transformations. The group operation, i.e., the composition law, is given as:
\begin{equation}\label{eq:composition-law}
    (A_{2},v_{2})\,\cdot\,(A_{1},v_{1}) \,=\, (A_{2}A_{1}\,,\,A_{2}(v_{1}) \,+\, v_{2})\,.
\end{equation}
In particular, the isometry group can be expressed as
\begin{equation}
   \mathrm{Isom}(\R^{n})\,=\, O(n) \,\rtimes\, \R^{n} \,. 
\end{equation}
Having that $A\in GL(n,\R)$, then one recovers the affinity of $\R^{n}$ denoted by $\mathrm{Aff}(\R^{n})$ that contains $\mathrm{Isom}(\R^{n})$ as a subgroup.

Compact $n$-dimensional flat Riemannian $B_{n}$ manifolds can be constructed by quotienting the Euclidean space $\R^{n}$ by some finite group $\Gamma\subset \mathrm{Isom}(\R^{n})$,
\begin{equation}
    B_{n} \,=\, \R^{n}/\Gamma\,.
\end{equation}
In general, the finite $n$-dimensional groups $\Gamma$ are known as the crystallographic groups (or $n$-space group). The torsion-free finite sub-groups|also denoted by $\Gamma$|are the so-called Bieberbach groups and accordingly the $B_{n}$ spaces are known as Bieberbach manifolds. Here, we are only be intrested in such torsion-free finite groups. 

For any group $G$, an element $g\in G$ is called a torsion element if it has finite order, i.e.,
\begin{equation}
    g^{k}\,=\, e,\,\quad k\in\Z_{>\,0} \,,
\end{equation}
with $e$ being the identity element in $G$. A group $G$ is called a torsion group if every element of $G$ is a torsion element.
Conversely, $G$ is said to be torsion-free if the only element of finite order is the identity.

Some remarks are in order:
\begin{itemize}
    \item If $\Gamma$ is a torsion-free crystallographic group, then the quotient space $B_{n}=\R^{n}/\Gamma$ is a closed flat $n$-manifold (or a space form). Conversely, every closed flat $n$-manifold arises as such a quotient by a torsion-free discrete group $\Gamma$.

    \item If $\Gamma$ contains torsion elements, then the quotient $B_{n}$ develops orbifold singularities | points fixed under non-trivial finite-order isometries. In contrast, if $\Gamma$  is torsion-free, the action is free, and $B_{n}$ is a smooth manifold.

\end{itemize}

The first Bieberbach’s theorem, see, e.g., \cite{szczepanski2012geometry}, states that the set of translations in $n$-space group $\Gamma$ forms a torsion-free, finitely generated, maximal abelian normal subgroup of rank $n$, which is isomorphic to $\Z^{n}\subset \R^{n}$. This allows us to represent $\Gamma$ via the short exact sequence:
\begin{equation}\label{eq:short-sequence-Gamma}
    1\,\to\,\Z^{n}\,\to\,\Gamma\,\to\,H\,\to\,1
    \,\,.
\end{equation} 
Here, the finite group $H\subset O(n)$ acts faithfully on $\Z^{n}$. The group $H$ is referred to as the holonomy group; this will be clarified below.

Let us present a simple example for a $B_{3}$ space. Let $e_{1},e_{2},e_{3}$ be the Cartesian coordinates on $\R^{3}$. We consider a 3-space group $\Gamma$ is generated by:
\begin{equation}
     \Gamma_{3}\,=\,  \langle \,t_{1},\,t_{2},\,t_{3}\,\,,\,A\,  \rangle\,; \qquad A\,=\, e_{3}/2\,+\,\mathrm{diag}(-1,-1,1)\,.
\end{equation}
The action of the generators $t_{i}$ define a 3-torus $T^{3}$, and the $B_{3}$ space can be expressed as: 
\begin{equation}
    B_{3}\,=\,T^{3}/H\,,\qquad \text{here, $H$ \,is generated by \, $A$ above}\,.
\end{equation}

In general, we represent a Bieberbach manifold $B_{n}$ as:
\begin{equation}\label{Bn=Tn-H}
    B_{n}\,=\, T^{n}/H\,.
\end{equation} 
With $T^{n}=\R^{n}/\Z^{n}$ is the $n$-torus defined by the action of the translation lattice $\langle t_{1},\cdots, t_{n}\rangle$. It is important to note that the action of $H$ may include additional translational components, as illustrated by the generator $A$ in the preceding example.

\paragraph{The fundamental and holonomy groups of $B_{n}$.}

By construction, the universal cover of the $B_{n}$ spaces is the Euclidean $\R^{n}$ space and the covering map is the quotient map itself:
\begin{equation}\label{eq:quotient-map}
    q\,:\,\R^{n}\,\to\, B_{n}\,.
\end{equation}
This sends a point $x \in \R^{n}$ to its orbit $[x] = \{ \gamma\cdot x  \,|\, \gamma \in \Gamma\}$ under the action of the group $\Gamma$.

To determine the fundamental group of $B_{n}$, i.e., $\pi_{1}(B_{n})$, we may use the fundamental theorem of covering spaces. Roughly speaking, the theorem says that the symmetries of the universal cover that respect the quotient map given in (\ref{eq:quotient-map}) is isomorphic to $\pi_{1}(B_{n})$. These symmetry transformations are known as the Deck transformations with the following consistency condition 
\begin{equation}
    q\,\circ\,D (x)\,=\, q(x)\,, \qquad \mathrm{for \,\,\, any}\  \  x \in\R^{n}\,.
\end{equation}
Let $D(x)=y$, then the above condition implies that $y$ should be in the orbit of $x$, i.e., $y\in[x]$. Therefore, the set of deck transformations should coincide with the finite group $\Gamma$. Thus, we conclude that 
\begin{equation}
    \pi_{1}(B_{n})\,=\,\Gamma\,.
\end{equation}

Moreover, one can compute the first homology group of $B_{n}$, i.e., $H_{1}(B_{n})$, through the abelianizations of the fundamental group, i.e., the quotient group $\Gamma$:
\begin{equation}
    H_{1}(B_{n})\,=\, \Gamma/[\Gamma,\Gamma]\,.
\end{equation}

For quotient spaces, like $B_{n}$, the holonomy groups can be defined in the following way. First, let us define charts $\alpha:U\to B_{n}$ on the $B_{n}$ spaces such that all points in the orbit $[x]$ belongs to $\alpha(U)$. Second, we define a loop based at a point $x\in U\subset B_{n}$, denoted by $\Delta_{x}$, on the $B_{n}$ space via the following order sequence of steps
\begin{equation}
    \Delta\,:=\,\{(0,\alpha_{0},A_{0});\, (\delta_{1},\alpha_{1},A_{1});\,(\delta_{2},\alpha_{2},A_{2});\,\cdots;(\delta_{m},\alpha_{m},A_{m})\}\,.
\end{equation}
Intuitively, the above can be understood as the following: before leaving the $\alpha=\alpha_{0}$ chart we apply a $A_{0}\in H$ transformation, then we move by an infinitesimal transformation $\delta_{1}$ to a new point belongs to the new chart $\alpha_{1}$ and again we apply a $A_{1}\in H$ transformation. We repeat this process $m$ times to get back to the original point $x$. Note that the $A_{\bullet}$ can be seen as a transition function that glue two different charts. The combined action of these transition functions is given by
\begin{equation}
  \widetilde{A}\,=\,  A_{0}\,\cdot\, A_{1}\,\cdot\,\,\cdots\,\,\cdot \, A_{m}\,\, \in\,\,  H,
\end{equation}
which does not depend on the infinitesimal transformations $\delta_{\bullet}$. Thus, $\widetilde{A}$ is associated to the  homotopy class of the loop $\Delta$, i.e., $\widetilde{A}_{[\Delta]}$. When considering the set of all possible loops, up to homotopy transformation, i.e., $\{\Delta\}$, one recovers the full group $H$ which defines the holonomy group of the $B_{n}$ space. 

The above discussion and the composition law in \eqref{eq:composition-law} implies that one can define the holonomy groups by the following projection operator \cite[chapter 13]{Ratcliffe_2006}
\begin{equation}
    \eta\ : \ \pi_{1}(\Gamma) \,\cong\, \Gamma \ \to\ H_{\mathrm{rot}}\,,
\end{equation}
with $ H_{\mathrm{rot}}$  denote the rotational part of the holonomy group $H$. Henceforth, we will omit the subscript and simply denote this group by $H$. That is, the holonomy group of the Bieberbach manifold is given as:
\begin{equation}
    \mathrm{Hol}(B_{n}) \,=\, \pi_{1}(B)/\Z^{n}\,\cong\, H\,.
\end{equation}

\paragraph{On the classification of Bieberbach spaces.}

Generically, a Bieberbach space $B_{n}$ is determined by its defining space group $\Gamma$. For instance, if $H$ is trivial, then $B_{n}$ reduces to the flat $n$-torus $T^{n}$. From this perspective, we have the following two important results:
\begin{itemize}
    \item Bieberbach second theorem, see, e.g., \cite{szczepanski2012geometry}, shows that there are only a finite number of isomorphism classes of $n$-space crystallographic groups $\Gamma_{n}$. Therefore, there are finitely many Bieberbach $B_{n}$ manifolds in any dimension $n$, up to equivalence. 

    \item Let $\Gamma_{1}$ and $\Gamma_{2}$ be two Bieberbach groups. If there exists $A \in \text{Aff}(n)$ such that,
    \begin{equation}
        \Gamma_{2} \,=\, A \, \Gamma_{1} \,.\,A^{-1}\,.
    \end{equation}
    Then $\Gamma_{1}$ and $\Gamma_{2}$ are isomorphic. This is the essence of  the third Bieberbach’s theorem.
\end{itemize}

Consequently, the corresponding flat manifolds $B_{n}^{(1)} = \R^{n}/\Gamma_{1}$ and $B_{n}^{(2)} = \R^{n}/\Gamma_{2}$ are affinely equivalent. That the affine transformation $A$ induces an affine transformation on $B_{n}^{(1)}$ and $B_{n}^{(2)}$ as: 
\begin{equation}
    \begin{split}
        \widetilde{A}\ &:\  B_{n}^{(1)}\ \to\ B_{n}^{(2)}\,
        \\
        &: \ [x] \ \mapsto \ [\widetilde{A}
        \,\cdot\, x]\,. 
    \end{split}
\end{equation}
%In other words, $B_{n}^{(1)}$ and $B_{n}^{(2)}$ are homeomorphic and diffeomorphic.

In $n$ dimensions, classifying isomorphism classes of Bieberbach groups is equivalent to classifying Bieberbach manifolds up to affine diffeomorphism. The complete classification of 4-dimensional Bieberbach spaces was accomplished in 2013 in \cite{lambert2013}.

\paragraph{Spin structure on Bieberbach $n$-manifolds.}

A natural question is whether the Bieberbach spaces $B_{n}$ admit spin structures, since their absence would invalidate the twisted reduction procedure in section \ref{sec:twisted-reduction}. Intuitively, this amounts to asking whether the $n$-space group $\Gamma$ can be embedded in $Spin(n)$, the double cover of $SO(n)$. However, as shown in \cite{Pfaeffle2000Dirac,putrycz2010,Lutowski_2015}, the complete answer is more subtle than this simple intuition suggests. In the following, we outline the key ideas underlying their construction. 

Let us consider the embedding of $\Gamma$ in $Spin(n)$ and $SO(n)$ via the following diagram:
\begin{equation}\label{eq:spin-diagram}
\begin{tikzpicture}[baseline=(current  bounding  box.center)]
\node (g) at (0,0) {$\Gamma$};
\node (sp) at (2,1.5) {$Spin(n)$};
\node (so) at (2,0) {$SO(n)$}; 
\draw[->] (g) --node[above left]{$S$} (sp);
\draw[->] (sp) -- node[right]{$\lambda$} (so);
\draw[->] (g) -- node[above]{$R$} (so);
\end{tikzpicture}
\end{equation}
Here, $R$ maps the rotational generators of $\Gamma$ to $SO(n)$, $R:\gamma  \mapsto  R(\gamma)$. $S$ a set of homomorphisms from $\Gamma$ to $Spin(n)$. The natural condition is that the above diagram should commute 
\begin{equation}
    \lambda(S(\gamma))\,=\,R(\gamma)\,,\qquad\forall \gamma\,\in\,\Gamma\,.
\end{equation}

In general, the group $\Gamma$ is given by the generators $\gamma_{\alpha}$ and the relations, denoted by $r_{i}$, among them. These relations can be written as
\begin{equation}
    \gamma_{1}\cdot\,\gamma_{2}\,\cdots\,\cdot\, \gamma_{k}\,=\,1\,,\quad \text{for finite $k$}\,.
\end{equation}
It turns out that $\Gamma$ can be embedded in $Spin(n)$ and the corresponding $B_{n}$ is a spin manifold wherever the $S$ maps respect the above relations: 
\begin{equation}
     S(\gamma_{1})\,\cdot\,S(\gamma_{2})\,\cdots\,\cdot\, S(\gamma_{k})\,=\,1\,,\quad \text{for finite $k$}\,.
\end{equation}
%Meaning the spin representation of each $\gamma_{\alpha}$ is constrained by the above relations.

The answer to which $B_{3}$ and $B_{4}$ spaces admit spin structures is provided in Tables \ref{Table:all-B3-spaces} and \ref{Table:all-B4-spaces}, respectively, following the results of \cite{lambert2013} and the references therein.

\subsection{Co-Seifert fibrations of \texorpdfstring{$B_{n}$}{Bn} spaces}\label{sec-co-Seifert-fibration}

In this subsection, we examine the geometric fibration structure of the Bieberbach manifolds in more details following \cite{Ratcliffe_2010,ratcliffe2012,lambert2013}. 

Let us consider a (complete) normal subgroup $N\trianglelefteq\Gamma$ defined as
\begin{equation}
    N\,=\,\{\, v\,+\,A \in\Gamma\,\,|\,\,v\in V \,,\quad W\,\subseteq\,\mathrm{Fix}(A)  \,\}\,.
\end{equation}
The translation-subset of $N$ is given as
\begin{equation}
    T_{N}\,=\,\{\, v\,\in\,\R^{n}\,\,|\,\, v\,+\,I  \,\in\, N  \,\}\,,
\end{equation}
with $I$ being the identity of $N$. This defines a vector subspace $V\subset \R^{n}$ given by the span of $T_{N}$. The vector subspace being invariant under the rotations of $N$, i.e., $W\subset \mathrm{Fix}(A)$, defines the orthogonal complement of $V$, $W:=V^{\perp}$, which can be defined as
\begin{equation}
    W\,=\, \{\, w\,\in \, \R^{n}\,|\,\, \ip{v}{w} \,=\,0\,,\quad \forall v\,\in\, V   \,\}\,.
\end{equation}
Therefore, the normal subgroup $N$ gives a natural splitting of the Euclidean space $\R^{n}$ as
\begin{equation}
    \R^{n}\, = \,\R^{n-m}\,\times\,\R^{m}\,:=\, V\,\times\,W\,.
\end{equation}
Here, we assume that $V$ is $(n-m)$-dimensional while $W$ is $m$-dimensional.

Let us consider the complement subgroup of $N$ denoted by $K$ and defined as
\begin{equation}
   K\,=\, \{\, b\,+\,B\,\in\,\Gamma\,\,|\,\, b\in W,\,\, V\,\subset\,\mathrm{Fix}(B)   \,\}\,.
\end{equation}
Given the splitting of $\R^{n}$ and think of $\Gamma$ as the direct product $N\times K$, then the Bieberbach manifold $B_{n}$ can be rewritten as
\begin{equation}
    B_{n}\,=\, \left(\frac{V}{N}\right)\,\times\, \left(\frac{W}{K}\right)\,.
\end{equation}
Here, we think of the $V/N$ as the fiber and the $W/K$ as the base of $B_{n}$.

However, the situation can be slightly more involved due to the fact that $\Gamma$ may not be given simply by the direct product $N\times K$.

\paragraph{Co-Seifert fibrations.}

One may regard $B_{n}$ as the total space of a bundle whose base is a closed flat $(n-1)$-manifold $B_{(n-1)}$ and whose typical fiber is a closed flat 1-manifold. Such a structure is referred to as a geometric Seifert fibration.

Alternatively, a co-Seifert fibration is obtained by interchanging the roles of the base and fiber in the Seifert fibration. In this case, $B_{n}$ is viewed as a fibration of $B_{(n-1)}$ over a closed flat 1-dimensional orbifold, where the base is either $\bbS^{1}$ or an interval $I$. In what follows, we will focus primarily on co-Seifert fibrations with $\bbS^{1}$ base following \cite{Hillman1995,Ratcliffe_2010,ratcliffe2012,lambert2013}.  

To construct a co-Seifert fibration for a closed flat $n$-manifold $B_{n}$ with space group $\Gamma_{n}$, we begin by choosing an $(n-1)$-dimensional space group $\Gamma_{n-1}$ as a normal subgroup of $\Gamma_{n}$. The group $\Gamma_{n-1}$ acts on $\R^{n-1}\subset \R^{n}$, producing the typical fiber $B_{n-1}$ of the co-Seifert fibration. The quotient $\Gamma_{n}/\Gamma_{n-1}$ defines a 1-dimensional space group $\Delta$, which is either the infinite cyclic group $\Z$ or the infinite dihedral group $\mathbb{D}$.

In this geometric picture, $\Gamma_{n}$ is a geometric fibration with $\Gamma_{n-1}$ being the typical fiber and $\Delta$ is the base,
\begin{equation}
    \Gamma_{n-1}\,\hookrightarrow\,\Gamma_{n}\,\to\, \Delta\,, \quad \text{with}\quad \Delta \, \cong\, \Gamma_{n}/\Gamma_{n-1}\,.
\end{equation}

An element $\delta\in \Delta$ acts on the fiber group $\Gamma_{n-1}$ via an outer automorphism:
\begin{equation}
    \mathrm{Mon}(\delta)\,:\, \Gamma_{n-1}\,\to\,\Gamma_{n-1}\,,\qquad \delta\,\in\,\Delta\,.
\end{equation}
We refer to this action as the monodromy action of $\delta$, denoted $\mathrm{Mon}(\delta)$, for reasons that will become clear shortly. Roughly, the set of all such monodromy actions $\{ \mathrm{Mon}(\delta)\}$ constitutes the group of the outer automorphism of $\Gamma_{n-1}$, $\mathrm{Out}(\Gamma_{n-1})$. The exact relation is given in theorem 23 of \cite{ratcliffe2012}, which is beyond the discussion we present here.

The fibration structure of $\Gamma_{n}$ reflects itself on the Biberbach $B_{n}=\R^{n}/\Gamma_{n}$ space as the following
\begin{equation}
   \frac{\R^{n-1}}{\Gamma_{n-1}}\,\hookrightarrow\, \frac{\R^{n}}{\Gamma_{n}}\,\to\,\frac{\R}{\Delta}\,.
\end{equation}
For the case, $\Delta=\Z$ we get $\bbS^{1}$ as the base manifold, while for $\Delta=\mathbb{D}$ we get an interval $I$. Taking the first case, the $B_{n}$ space then can be viewd as 
\begin{equation}\label{eq:Bn-fibresBn-1-baseS1}
    B_{n-1}\,\hookrightarrow\,B_{n}\,\to\, \bbS^{1}
\end{equation}

The action of $\Delta$ on $\Gamma_{n-1}$ extends naturally to a free action on the fibers $B_{n-1}$. Geometrically, as one traverses the base $\bbS^{1}$, the fiber $B_{n-1}$ returns to itself via an affine isomorphism. More precisely, there is a correspondence (not necessary $1-1$) between the outer automorphism group $\mathrm{Out}(\Gamma_{n-1})$ and the group of affine symmetries of $B_{n-1}$. This establishes a correspondence between the monodromy action and the affine isomorphism class of $B_{n-1}$, justifying the term ``monodromy'' in this context. Consequently, the total space $B_{n}$, described by the fibration in \eqref{eq:Bn-fibresBn-1-baseS1} may correspond to multiple distinct affine equivalence classes \cite{lambert2013}.

The distinction between different affine equivalence classes will not be central to our analysis. We work under the assumption that all affine equivalence classes associated with a given $B_{n}$ yield equivalent physical theories. Consequently, our focus rests on the distinct $B_{n}$ spaces themselves, which are characterized by: (i) The groups $\Gamma_{n}$. (ii) The fiber space $B_{n}$. (iii) The structure group $G$|to be defined shortly.

\paragraph{Revisiting the holonomy of $B_{n}$.}

In general, the above discussion enable us topologically to express the Bieberbach $B_{n}$ space as 
\begin{equation}
    B_{n}\,=\, \frac{B_{(n-1)}\,\times\,\bbS^{1}}{G}\,.
\end{equation}
Here, $G$ is the structure group of the above fibration and defied as $G:=\Gamma_{n}/\Gamma_{n-1}\Delta$. In other words, it is the subgroup consists of the elements that are neither in $\Gamma_{n-1}$ nor in $\Delta=\Z$.

Let us express $B_{n-1}$ as $T^{(n-1)}/\widetilde{H}$ using the general expression in \eqref{Bn=Tn-H}. Substituting this into the above equation, we have  
\begin{equation}\label{eq:Bn=TnmHtilde-TmG}
    B_{n}\,=\, \frac{\left(T^{(n-1)}/\widetilde{H}\right)\,\times\,\bbS^{1}}{G}\,.
\end{equation}
Here, $\widetilde{H}$ is the holonomy group of $B_{n-1}$ and it is related to $\Gamma_{n-1}$ through \eqref{eq:short-sequence-Gamma}. Comparing this expression with the standard form $B_{n}=T^{n}/H$, we deduce the following inclusion of holonomy groups:
\begin{equation}\label{eq:H-subgroup-Htilda-G}
    H\,\subseteq\, \widetilde{H}\,\rtimes\, G\,.
\end{equation}

%The framework of geometric co-Seifert fibrations, combined with our characterization of the holonomy group $H$, provides the key features for analysing the physics in section \ref{sec:G-structure-co-seifert}.

\paragraph{Example: The $B_{4}^{(9)}$ space.}

Let $(e_{1},\cdots, e_{4})$ be the Cartesian coordinates on $\R^{4}$. The 4-space $\Gamma_{4}$ that acts on $\R^{4}$ and correspond to the $B_{4}^{(9)}$ space is given by the usual 4 translations $t_{i}$ acting as $t_{i}:e_{i}\sim e_{i} +1 $ to produce the 4-torus $T^{4}$. Further, we have the subgroup $H$ generated by
\begin{equation}
    A\,=\, e_{4}/2  \, +\, \mathrm{diag}(-1,-1,1,1)\,,\qquad B\,=\, e_{2}/2\,+\,e_{4}/4\,+\, \mathrm{diag}(1,-1,-1,1)\,.
\end{equation}
We note that the rotational subgroup of $H$ leave $e_{4}$ invariant. Hence, $e_{4}\sim e_{4}+1$ is taken to the base circle of the co-Seifert fibration, on which $\Gamma_{4}$ acts by $e_{4}/2$. The 3-group $\Gamma_{3}$ acting on $e_{1},e_{2},e_{3}$ is given by 
\begin{equation}
    \Gamma_{3}\,=\,  \langle \,t_{1},\,t_{2},\,t_{3}\,,A\,  \rangle\,.
\end{equation}
Following the tables in \cite{brown1978} and the discussion in \cite{lambert2013}, one notices that $\Gamma_{3}$ is the same type as the 3-space group acting on $\R^{3}$ to produce the $B_{3}^{(2)}$ space. One can refer to as the $B_{3}^{(2)}$-type quotient group. According to the general discussion above, the co-Seifert fibre is the $B_{3}^{(2)}$ space. The structure group $G$ can be generated by the rotational part of the $AB$ element, i.e., 
\begin{equation}
 G\,:\, e_{2}/2\,+\,  \mathrm{diag}(-1,1,-1)\,.
\end{equation}
Further, $G$ acts on the base as $e_{4}\sim e_{4} +e_{4}/2$ as can be seen from the element $B$ above.

\paragraph{Example: The $B_{4}^{(25)}$ space.}

The 4-space $\Gamma_{4}$ that acts on $\R^{4}$ and correspond to the $B_{4}^{(25)}$ space is given by the usual 4 translations $t_{i}$ acting as $t_{i}:e_{i}\sim e_{i} +1 $ to produce the 4-torus $T^{4}$. The subgroup $H$ generated by:
\begin{equation}
  A \,=\,  e_{2}/6\,+\,  \begin{pmatrix}
        1&0&0&0\\
        0&1&0&0\\
        0&0&0&-1\\
        0&0&1&1
    \end{pmatrix}\,\,;
    \qquad  
   B\,=\, e_{1}/2\,+\, \begin{pmatrix}
        1&0&0&0\\
        0&-1&0&0\\
        0&0&0&1\\
        0&0&1&0
    \end{pmatrix}
\end{equation}
The co-Seifert $\bbS^{1}$ base is taken along $e_{1}\sim e_{1}+1$, which is acted on as $e_{1}\sim e_{1}+e_{1}/2$ as seen from the translation of the generator $B$. The 3-space group is given by neglecting the first row of the $A$ and $B$ generators and can be represented as:
    \begin{equation}\label{eq:AB-gen-B425-3space}
  A \,=\,  e_{2}/6\,+\,  \begin{pmatrix}
        1&0&0\\
        0&0&-1\\
        0&1&1
    \end{pmatrix}\,\,;
    \qquad  
   B\,=\,  \begin{pmatrix}
        -1&0&0\\
        0&0&1\\
        0&1&0
    \end{pmatrix}
\end{equation}
We can take different representative of the above 3-space group as:
\begin{equation}
    \widetilde{A}\,=\,A^{3}\,=\,e_{2}/2\,+\, \begin{pmatrix}
        1&0&0\\
        0&-1&0\\
        0&0&-1
    \end{pmatrix}\,,\qquad \widetilde{B}\,=\, A^{2}\widetilde{A}_{\mathrm{rot}}\,=\, e_{2}/3\,+\, \begin{pmatrix}
        1&0&0\\
        0&1&1\\
        0&-1&0
    \end{pmatrix}\,.
\end{equation}
Here, $\widetilde{A}_{\mathrm{rot}}$ means only the rotational part of the $\widetilde{A}$ generator. Comparing to the group $(6/1/1/4)$ of \cite{brown1978}, this is exactly the same type of 3-space group that give the $B_{3}^{(5)}$ Bieberbach space. Hence, it is identified with the typical fiber and the structure group $G$ is given by the generator $B$ in \eqref{eq:AB-gen-B425-3space} \cite{lambert2013}.

\begin{longtable}{|c|c|c|c|c|c|r|}
\caption{List of orientable, closed, flat Bieberbach 3-manifolds \cite{lambert2013}. The 3-space groups $\Gamma_{3}$ are classified according to the BBNWZ notation and comprehensively listed in \cite{brown1978}.}
\label{Table:all-B3-spaces} \\
\hline
$B_{3}^{(k)}$  & Structure group & $H_{1}(B_{3}^{(k)})$  & Holonomy group & Spin & BBNWZ & IT \\ \hline
\hline
\endfirsthead
\multicolumn{7}{c}{{\tablename\ \thetable{} -- continued from previous page}} \\
\hline
$B_{3}^{(k)}$  & Structure group & $H_{1}(B_{3}^{(k)})$  & Holonomy group & Spin & BBNWZ & IT \\ \hline
\hline
\endhead
\hline
%\multicolumn{7}{r}{{Continued on next page}} \\
\endfoot
\hline
\endlastfoot
$B_3^{(1)}$   & $\mathbb{Z}_{1}$ & 
	  	$\mathbb Z^3$ & $\mathbb{Z}_{1}$ & \phantom{\Big(}Y\phantom{\Big)} & 1/1/1/1 & 1 \\ \hline
$B_{3}^{(2)}$   & $\mathbb{Z}_{2}$ &
		$\mathbb Z \oplus (\mathbb Z_2)^{\oplus 2}$  & $\mathbb{Z}_{2}$ & \phantom{\Big(}Y\phantom{\Big)} & 2/1/1/2 & 4 \\ \hline
$B_{3}^{(3)}$  & $\mathbb{Z}_{3}$ &
 		$\mathbb Z \oplus \mathbb Z_3$  & $\mathbb{Z}_{3}$ & \phantom{\Big(}Y\phantom{\Big)} & 5/1/2/2 & 144 \\ \hline
$B_{3}^{(4)}$  & $\mathbb{Z}_{4}$ &
 		$\mathbb Z \oplus \mathbb Z_2$  & $\mathbb{Z}_{4}$ & \phantom{\Big(}Y\phantom{\Big)} & 4/1/1/2 &  76 \\ \hline
$B_{3}^{(5)}$   & $\mathbb{Z}_{6}$ &
 		$\mathbb Z$  & $\mathbb{Z}_{6}$ & \phantom{\Big(}Y\phantom{\Big)} & 6/1/1/4 & 169 \\ \hline
$B_{3}^{(6)}$    & $\mathbb{Z}_{1}$ &
 	$(\mathbb Z_4)^{\oplus 2}$  & $\mathbb{Z}_{2}\oplus\mathbb{Z}_{2}$ &  \phantom{\Big(}Y\phantom{\Big)}  & 3/1/1/4 & 19 \\
\end{longtable}

%\begin{table}[H]
%\centering{
%\begin{tabular}{|c|c|c|c|c|c|r|}
%\hline	
%	$B_{3}^{(k)}$  & Structure group & $H_{1}(B_{3}^{(k)})$  & Holonomy group & Spin & BBNWZ & IT \\ \hline	
%    \hline	
%	$B_3^{(1)}$   & $\Z_{1}$ & 
%	  	$\mathbb Z^3$ & $\Z_{1}$ & \phantom{\Big(}Y\phantom{\Big)} & 1/1/1/1 & 1 \\ \hline
%	$B_{3}^{(2)}$   & $\Z_{2}$ &
%		$\mathbb Z \oplus (\mathbb Z_2)^{\oplus 2}$  & $\Z_{2}$ & \phantom{\Big(}Y\phantom{\Big)} & 2/1/1/2 & 4 \\ \hline
%	$B_{3}^{(3)}$  & $\Z_{3}$ &
% 		$\mathbb Z \oplus \mathbb Z_3$  & $\Z_{3}$ & \phantom{\Big(}Y\phantom{\Big)} & 5/1/2/2 & 144 \\ \hline
%	$B_{3}^{(4)}$  & $\Z_{4}$ &
% 		$\mathbb Z \oplus \mathbb Z_2$  & $%\Z_{4}$ & \phantom{\Big(}Y\phantom{\Big)} & 4/1/1/2 &  76 \\ \hline
%	$B_{3}^{(5)}$   & $\Z_{6}$ &
% 		$\mathbb Z$  & $\Z_{6}$ & \phantom{\Big(}Y\phantom{\Big)} & 6/1/1/4 & 169 \\ \hline
% 	$B_{3}^{(6)}$    & $\Z_{1}$ &
% 	$(\mathbb Z_4)^{\oplus 2}$  & $\Z_{2}\oplus\Z_{2}$ &  \phantom{\Big(}Y\phantom{\Big)}  & 3/1/1/4 & 19 \\ \hline
%\end{tabular}
%}
%\caption{List of orientable, closed, flat Bieberbach 3-manifolds \cite{lambert2013}. The 3-space groups $\Gamma_{3}$ are classified according to the BBNWZ notation and comprehensively listed in \cite{brown1978}.}
%\label{Table:all-B3-spaces}
%\end{table}

\begin{longtable}{|c|c|c|c|c|l|}
\caption{List of orientable, closed, flat Bieberbach 4-manifolds \cite{lambert2013}. The 4-space groups $\Gamma_{4}$ are classified according to the BBNWZ notation and comprehensively listed in \cite{brown1978}. Here, the cyclic groups denoted by $\mathbb{Z}_{N}$, the dihedral groups denoted by $\mathbb{D}_{N}$ of order $2N$, and the polyhedral groups: tetrahedral denoted by $\mathbb{T}$ of order 12.}
\label{Table:all-B4-spaces} \\
\hline
$B_{4}^{(k)}$  & Structure group & $H_1(B_{4}^{(k)})$ & Holonomy group & Spin & BBNWZ \\ \hline
\hline
\endfirsthead
\multicolumn{6}{c}{{\tablename\ \thetable{} -- continued from previous page}} \\
\hline
$B_{4}^{(k)}$  & Structure group & $H_1(B_{4}^{(k)})$ & Holonomy group & Spin & BBNWZ \\ \hline
\hline
\endhead
\hline
%\multicolumn{6}{r}{{Continued on next page}} \\
\endfoot
\hline
\endlastfoot
$B_{4}^{(1)}$  &  $\mathbb{Z}_{1}$ &
 		$\mathbb Z^4$ & $\mathbb{Z}_{1}$ & \phantom{\Big(}Y\phantom{\Big)} & 1/1/1/1\\ \hline
$B_{4}^{(2)}$    & $\mathbb{Z}_{2}$ & 
		$\mathbb Z^2 \oplus (\mathbb Z_2)^{\oplus 2}$  & $\mathbb{Z}_{2}$ & \phantom{\Big(}Y\phantom{\Big)} &  3/1/1/2 \\ \hline
$B_{4}^{(3)}$   & $(\mathbb{Z}_{2})^2$ &
		$\mathbb Z^2 \oplus \mathbb Z_2$  & $\mathbb{Z}_{2}$ & \phantom{\Big(}Y\phantom{\Big)} & 3/1/2/2  \\ \hline
$B_{4}^{(4)}$   & $\mathbb{Z}_{3}$ & 
		$\mathbb Z^2 \oplus \mathbb Z_3$  & $\mathbb{Z}_{3}$ & \phantom{\Big(}Y\phantom{\Big)} & 8/1/2/2\\ \hline
$B_{4}^{(5)}$   & $(\mathbb{Z}_{3})^2$ & 
		$\mathbb Z^2$ & $\mathbb{Z}_{3}$ & \phantom{\Big(}Y\phantom{\Big)} & 8/1/1/2 \\ \hline
$B_{4}^{(6)}$  & $\mathbb{Z}_{4}$ & 
		$\mathbb Z^2 \oplus \mathbb Z_2$  & $\mathbb{Z}_{4}$ & \phantom{\Big(}Y\phantom{\Big)} & 7/2/1/2 \\ \hline
$B_{4}^{(7)}$  & $\mathbb{Z}_{4}\times \mathbb{Z}_{2}$ & 
			$\mathbb Z^2$ & $\mathbb{Z}_{4}$ & \phantom{\Big(}Y\phantom{\Big)} & 7/2/2/2 \\ \hline
$B_{4}^{(8)}$   & $\mathbb{Z}_{6}$ & 
			$\mathbb Z^2$  & $\mathbb{Z}_{6}$ & \phantom{\Big(}Y\phantom{\Big)} & 9/1/1/2 \\ \hline
$B_{4}^{(9)}$  & $\mathbb{Z}_{2}$ & 
 		$\mathbb Z \oplus (\mathbb Z_2)^{\oplus 3}$  & $(\mathbb{Z}_{2})^2$ & 
 		\phantom{\Big(}Y\phantom{\Big)} & 5/1/2/7\\ \hline
$B_{4}^{(10)}$   & $\mathbb{Z}_{2}$ &
 		$\mathbb Z \oplus \mathbb Z_2 \oplus \mathbb Z_4$ & $(\mathbb{Z}_{2})^2$ & 
 		\phantom{\Big(}Y\phantom{\Big)} & 5/1/2/8 \\ \hline
 	$B_{4}^{(11)}$   & $\mathbb{Z}_{2}$ &
 		$\mathbb Z \oplus \mathbb Z_2 \oplus \mathbb Z_4$ &  
		$(\mathbb{Z}_{2})^2$ & \phantom{\Big(}Y\phantom{\Big)} & 5/1/2/10 \\ \hline
	$B_{4}^{(12)}$   & $\mathbb{Z}_{2}$ & 
 		$\mathbb Z \oplus (\mathbb Z_2)^{\oplus 2}$ & 
			$(\mathbb{Z}_{2})^2$ & \phantom{\Big(}Y\phantom{\Big)} & 5/1/3/6 \\ \hline
	$B_{4}^{(13)}$ & $\mathbb{Z}_{4}$ &
 		$\mathbb Z \oplus \mathbb Z_4$ & 
			$(\mathbb{Z}_{2})^2$ & \phantom{\Big(}Y\phantom{\Big)} & 5/1/10/4\\ \hline	
	$B_4^{(14)}$  & $\mathbb{Z}_{1}$ &
 		$\mathbb Z \oplus (\mathbb Z_4)^{\oplus 2}$ & 
			$(\mathbb{Z}_{2})^2$ & \phantom{\Big(}Y\phantom{\Big)} & 5/1/2/9 \\ \hline
	$B_4^{(15)}$   & $\mathbb{Z}_{2}$ &
 		$\mathbb Z \oplus (\mathbb Z_2)^{\oplus 2}$ &  
			$(\mathbb{Z}_{2})^2$ & \phantom{\Big(}Y\phantom{\Big)} & 5/1/7/4 \\ \hline
	$B^4_{16}$   & $\mathbb{Z}_{2}$ &
 		$\mathbb Z \oplus (\mathbb Z_2)^{\oplus 2}$ & 
			$(\mathbb{Z}_{2})^2$ & \phantom{\Big(}N\phantom{\Big)} & 5/1/6/6 \\ \hline
	$B^4_{17}$   & $\mathbb{Z}_{2}$ &	
 		$\mathbb Z \oplus \mathbb Z_2 \oplus \mathbb Z_4$ &  
			$(\mathbb{Z}_{2})^2$ & \phantom{\Big(}N\phantom{\Big)} & 5/1/4/6 \\ \hline	
	$B^4_{18}$   &$\mathbb{Z}_{2}$ &
 		$\mathbb Z \oplus \mathbb Z_6$ & 
			$\mathbb{D}_{3}$ & \phantom{\Big(}Y\phantom{\Big)} & 14/3/5/4 \\ \hline
	$B^4_{19}$   & $\mathbb{Z}_{2}$ & 
 		$\mathbb Z \oplus \mathbb Z_2$ &  
 			 $\mathbb{D}_{3}$ & \phantom{\Big(}Y\phantom{\Big)} & 14/3/6/4 \\ \hline
	$B^4_{20}$  & $\mathbb{Z}_{6}$ &
 		$\mathbb Z \oplus \mathbb Z_2$ & 
 			$\mathbb{D}_{3}$ & \phantom{\Big(}Y\phantom{\Big)} & 14/3/1/4 \\ \hline
	$B^4_{21}$  & $\mathbb{Z}_{2}$ &
 		$\mathbb Z \oplus (\mathbb Z_2)^{\oplus 2}$ &  
 			$\mathbb{D}_{4}$ & \phantom{\Big(}Y\phantom{\Big)} & 13/4/1/14 \\ \hline
	$B^4_{22}$    & $\mathbb{Z}_{2}$ & 
 		$\mathbb Z \oplus \mathbb Z_4$ &  
 			$\mathbb{D}_{4}$ & \phantom{\Big(}Y\phantom{\Big)} & 13/4/1/20 \\ \hline
	$B^4_{23}$   & $\mathbb{Z}_{2}$ & 
		$\mathbb Z \oplus \mathbb Z_4$ &  
			$\mathbb{D}_{4}$ & \phantom{\Big(}Y\phantom{\Big)} & 13/4/1/23 \\ \hline
	$B^4_{24}$   & $\mathbb{Z}_{4}$ &
		$\mathbb Z \oplus \mathbb Z_2$ &  
		$\mathbb{D}_{4}$ & \phantom{\Big(}N\phantom{\Big)} & 13/4/4/11 \\ \hline
	$B^4_{25}$  & $\mathbb{Z}_{2}$ &
 		$\mathbb Z \oplus \mathbb Z_2$ &  
 		$\mathbb{D}_{6}$ & \phantom{\Big(}Y\phantom{\Big)} & 15/4/1/10 \\ \hline
	$B^4_{26}$   & $\mathbb{Z}_{3}$ &
 		$\mathbb Z$ &  
 		$\mathbb{T}$ & \phantom{\Big(}Y\phantom{\Big)} & 24/1/2/4 \\ \hline
	$B^4_{27}$   & $\mathbb{Z}_{6}$ &
 		$\mathbb Z$ &  
 		$\mathbb{T}$ & \phantom{\Big(}Y\phantom{\Big)} & 24/1/4/4\\ \hline
\end{longtable}

%%%%%%%%%%%%%%%%%%%%%%%%%%%%%%%%%%%%%%%%%%%%%%%%%%%%%%%%%%%%%%%%%%%%%%%%%%%%%%%%%%%%%

\section{Metrics and harmonic 2-forms on \texorpdfstring{$\R^{4}/\Z_{N}$}{R4/ZN}}\label{sec:harmonic-2-forms}

The space $\widetilde{\R^{4}/\Z_{N}}$ admit a geometric description in terms of a degenerate $U(1)$-bundle fibering over a base $\R^{3}$. The metric on the total space is explicitly known and referred to as the  multi-centred gravitational instanton. Here, we review the metric along with its $Sp(1)$-structure and  $L^{2}$-normalizable harmonic 2-forms following \cite{Eguchi1979,GIBBONS1978430,Ruback1986TheMO, Hausel:2002xg, Franchetti:2014lza} and chapter 2 of \cite{Najjar:2022eci}. In M-theory geometric engineering such geometry leads to enhanced $\mk{su}(N)$ gauge symmetries in M-theory, e.g., see \cite{Sen:1997js,Sen:1997kz}.

\paragraph{Metric and self-dual 2-forms.}

The metric on the total space is captured by the Gibbons–Hawking (GH) ansatz, which can be expressed as:
\begin{equation}\label{eq:GH-metric}
    \dd s^{2} \,=\, V(r)\, \dd\vec{x}\cdot \dd\vec{x} \,+\, V^{-1}(r)\,(\dd\psi \,+\, \mathcal{A})^{2}\, ,\, \qquad   \text{with}\,\,\,r \,=\, \sqrt{\vec{x}\cdot\vec{x}}\,.
\end{equation}
Here, $\mathcal{A}=\mathcal{A}_{i}(x)\,\dd x^{i}$ is a magnetic gauge field on $\R^{3}$, and $V(r)$ is an electric potential over $\R^{3}$. This is precisely the type of metric given on the $\widetilde{\R^{4}/\Z_{N}}$ subspace of the 8-dimensional space $X_{8}$ given in \eqref{eq:general-our-X8-topology}.

From the metric, one can write down the frame-fields, or tetrad, for this geometry as:
\begin{equation}\label{eq:form-fields}
    e^{4} \,=\, V^{-1/2}(r)\,(\dd\psi\,+\,\mathcal{A})\,,\qquad e^{i} \,=\, V^{1/2}(r)\,\dd x^{i}\,,\qquad i = 1,2,3.
\end{equation}
We can construct 2-forms by taking the exterior derivative of the frame-fields as
\begin{equation}\label{eq:SD-ASD-2-forms}
   \begin{split}
      \omega^{i}_{\pm} \,=\, &\, (\dd\psi\,+\,\mathcal{A})\,\wedge\,\dd x^{i}\,\pm\,\frac{1}{2}\,V(r)\,\varepsilon^{ijk}\,\dd x^{j}\,\wedge\,\dd x^{k}\,
      \\
\,=\, & \, e^{4}\,\wedge\, e^{i} \,\pm\, \frac{1}{2}\,\epsilon^{ijk}\,e^{j}\,\wedge\, e^{k}\,.
   \end{split}
\end{equation}
Here, the set $\{\omega^{i}_{+}\}$ is the self-dual 2-forms and $\{\omega^{i}_{-}\}$ defines the set of anti-self-dual 2-forms on the geometry. The closure conditions on $\{\omega^{i}_{+}\}$ implies a relation between the gauge field $\mathcal{A}$ and the potential given as:  
\begin{equation}
  \dd \mathcal{A} \,=\,  \ast_{3}\, \dd V(r)\,.
\end{equation}
Here, $\ast_{3}$ is the Hodge stare operator on $\R^{3}$.

Taking the exterior derivative on the above relation implies that $V(r)$ solves the Laplace equations on $\R^{3}$. In particular, the solutions may be expressed as:
\begin{equation}\label{eq:V(r)}
    \begin{split}
     V(r)\, = \,\left\{
   \begin{aligned}
      &\sum_{a=1}^{N}\, V_{a} \,=\, \sum_{a=1}^{N} \, \frac{1}{|\vec{x}-\vec{x}_{a}|}\,, \ \ \ \ \ \ \ \, && \text{ALE} \,,
     \\
     & 1 \,+\, \sum_{a=1}^{N}\, V_{a} \,=\, 1 \,+\, \sum_{a=1}^{N} \, \frac{1}{|\vec{x}-\vec{x}_{a}|}\ ,\ \ \ && \text{ALF}\,.
   \end{aligned}
   \right.
    \end{split}
\end{equation}
Here, $a$ runs over $1,\cdots, N$ with $N$ fixed by the $\Z_{N}$ quotient in $\widetilde{\R^{4}/\Z_{N}}$ and $\vec{x}_{a}$ referred to as the centres of the GH metric. ALE stands for asymptotically locally Euclidean and  ALF for asymptotically locally flat. At $\vec{x}=\vec{x}_{a}$, we have coordinate singularities of the metric. The two-centred solution is considered in \cite{Eguchi1979}.

The generic 1-form gauge field is then given as 
\begin{equation}
    \mathcal{A} \,=\, \sum_{a=1}^{n} \,\mathcal{A}_{a} \,=\,  \ \sum_{a=1}^{n}\, \cos{(\theta_{a})}\,\dd\phi_{a}\,, \quad \text{locally around each centre}\,.
\end{equation}

\paragraph{$L^{2}$-normalizable harmonic 2-forms.}

We would like to consider the space of $L^{2}$-normalizable harmonic two-forms, which we denote by $L^{2}\mathcal{H}^{2}_{(N-1)}$, as discussed in \cite{Ruback1986TheMO, Hausel:2002xg, Franchetti:2014lza} and reviewed in \cite{Najjar:2022eci}. 

First, we start with the following generic 2-form  
\begin{equation}\label{eq:h2-harmonic-correspond-centres}
    \mathrm{h}_{2} \,=\, \partial_{i}F(r) \, e^{4} \,\wedge\, e^{i}\,+\,\partial_{i}G(r)\,\frac{1}{2}\,\epsilon^{ijk}\,e^{j}\,\wedge \,e^{k}\,,
\end{equation}
for generic functions $F(r)$ and $G(r)$ defined only over the base $\R^{3}$ and depend on the radial directions, i.e., spherical symmetric functions. 

Second, we impose a closure condition and anti-self-duality on $\mathrm{h}_{2}$, i.e., $\dd \mathrm{h}_{2}$, which give constraints on the functions $F(r)$ and $G(r)$ as: 
\begin{equation}
    \partial_{i}F(r)\,=\,-\,\partial_{i}G(r)\,,\quad \text{and}\quad \,\, F(r)=f(r)/V(r)\,,
\end{equation}
Here, $f(r)$ is a spherical symmetric harmonic function on $\R^{3}$. Generically, we could take it as: $f(r) = \sum_{a=1}^{n} c_{a}\,V_{a}(r)$, with $c_{a}$ are constants and $V_{a}(r)$ are given in \eqref{eq:V(r)}. As a result, the harmonic 2-form is given as
\begin{equation}
    \mathrm{h}_{2} \,=\, \partial_{i}\left(\frac{f(r)}{V(r)} \right)\,\left(  \, e^{4} \,\wedge\, e^{i}\,- \,\frac{1}{2}\,\epsilon^{ijk}\,e^{j}\,\wedge\, e^{k}\,\right)\,.
\end{equation}
Here, the 2-form in the parentheses is the anti-self-dual 2-form given in \eqref{eq:SD-ASD-2-forms}. In the generic form of the function $f(r)$, given as a linear combination of $N$ linearly independent harmonic 2-forms, we then write:
\begin{equation}\label{eq:h2-harmonic-correspond-centres-Va-V}
        \mathrm{h}_{2}^{a} \,=\, \partial_{i}\left(\,\frac{V_{a}(r)}{V(r)} \,\right)\,\left(\,  e^{4} \wedge e^{i}\,- \,\frac{1}{2}\,\epsilon^{ijk}\,e^{j}\wedge e^{k}\,\right)\,.
\end{equation}

By examining the large $r$ limit of the above harmonic 2-forms, i.e.,  
\begin{equation}
  \displaystyle{\lim_{r \to \infty}} \,\,\left(\,\partial_{i} ( V_{a}(r)/V(r)\,) \,\right)^{2}\,\,\sim\,\, \displaystyle{\lim_{r \to \infty}}\,\,\frac{1}{r}\,\,=\,\,0\,,
\end{equation}
we observe that they are $L^{2}$-normalizable.

On these space, however, their exist  $(N-1)$ independent $L^{2}$-normalizable harmonic 2-forms, i.e., basis of the $L^{2}\mathcal{H}_{(N-1)}^{2}$ space. To overcome the above redundancy, the basis of $L^{2}\mathcal{H}_{(N-1)}^{2}$ can be defined as, see, e.g., \cite{Sen:1997kz,Sen:1997js}, 
\begin{equation}\label{eq:tilde-h2}
    \widetilde{\mathrm{h}}_{2}^{a}\,=\,\mathrm{h}_{2}^{a}\,-\,\mathrm{h}_{2}^{a+1}\,,\qquad a\,=\,1\,\cdots\,N-1\,.
\end{equation}

\paragraph{Poincar\'e duality and McKay correspondence.}

Poincar\'e duality implies that the set of harmonic 2-forms $\{\widetilde{\mathrm{h}}_{2}^{a}\}$ is dual to the collection of vanishing 2-cycles $\{\bbS^{2}_{a}\}$ arising in the crepant resolution of the $\R^{4}/\Z_{N}$ singularity. In particular, the forms $\widetilde{\mathrm{h}}_{2}^{a}$ parametrize the K\"ahler moduli of $\widetilde{\R^{4}/\Gamma_{ADE}}$, and satisfy
\begin{equation}
    \int_{\bbS^{2}_{b}}\,\widetilde{\mathrm{h}}_{2}^{a}
    \, =\, \delta^{a}_{\,b}\,\,\mathrm{vol}(\bbS^{2}_{b})\,.
\end{equation}

The McKay correspondence further relates the harmonic forms $\{\widetilde{\mathrm{h}}_{2}^{a}\}$ to the set of simple roots $\{\alpha^{a}\}$ of the $\mk{su}(N)$ algebra. Recall that the simple roots can be expressed in terms of an orthonormal basis $\{e^{a}\}$, with $a=1,\dots,N$, as
\begin{equation}
    \alpha^{a} \,=\, e^{a}\, - \, e^{a+1}\,, 
    \qquad a\,=\,1,\,\dots,\,N-1\,.
\end{equation}
From this perspective, the McKay correspondence can be viewed as identifying the centres, or the $\mathrm{h}_{2}^{a}$ 2-forms, of the resolved geometry with the basis vectors $\{e^{a}\}$.

Moreover, the intersection pairing of the 2-cycles is encoded in the wedge product of the corresponding harmonic forms, and is given by (minus) the Cartan matrix of $\mk{su}(N)$:
\begin{equation}
    \int_{\widetilde{\R^{4}/\Z_{N}}}\,
    \widetilde{\mathrm{h}}_{2}^{a}\,\wedge\,
    \widetilde{\mathrm{h}}_{2}^{b}
    \,=\, -\,C_{ab}\,,
\end{equation}
where $C_{ab}$ denotes the $\mk{su}(N)$ Cartan matrix, which coincides with the intersection matrix
\begin{equation}
    C_{ab} \,=\, -\,\bbS^{2}_{a}\,\cdot\, \bbS^{2}_{b}\,.
\end{equation}

%%%%%%%%%%%%%%%%%%%%%%%%%%%%%%%%%%%%%%%%%%%%%%%%%%%%%%%%%%%%%%%%%%%%%%%%%%%%%%%%%%%%%

\section{Holonomy and covering maps}\label{app:holonomy}

The following argument draws upon the theory of holonomy groups as presented, e.g., in \cite{kobayashi1963I,kobayashi1969II,Besse1987,Berger2003,clarke2012,RudolphSchmidt2017}. This approach is general and dimension-independent, extending beyond the specific compactification scenario considered in this paper. Consequently, we will omit explicit reference to the dimensions of $X_{8}$ and $\widetilde{X}_{8}$ in the subsequent discussion.

Let us take the covering map $\pi:\widetilde{X}\to X $ that correspond to the group $H$. We observe that all points $\wp\in \widetilde{X}$ related by $H$ transformations correspond to a point in $p\in X$, i.e., $\rho(h)\cdot\wp\sim \wp$ and $\pi(\rho(h)\cdot\wp)=\pi(\wp)=p$ for all $h\in H$. Two immediate consequences follow:
\begin{itemize}
    \item All loops $\widetilde{\gamma}$ based at $\wp\in\widetilde{X}$ descend to loops $\gamma$ based at $p\in X$.

    Therefore, all parallel transports $P_{\widetilde{\gamma}}$ along $\widetilde{\gamma}$ descend to $P_{\gamma}$ on $X$. Since the holonomy group based at $\wp$, denoted by $\mathrm{Hol}_{\wp}(\widetilde{X})$ (or at $p$ and denoted by $\mathrm{Hol}_{p}(X)$) are defined via $P_{\widetilde{\gamma}}$ (or by $P_{\gamma}$), then regardless of the base point(s) the holonomy of $X$ can not be smaller than that of $\widetilde{X}$ and \eqref{eq:HoltildX-inside-X} follow.  

     \item All paths $\sigma$ start at $\wp$ and end at $\rho(h)\cdot \wp$, for each $h\neq\mathds{1}\in H$, descend to loops $\sigma(h)$ based at $p\in X$.

     This implies that the holonomy groups based at $p\in X$ receives new element due to the parallel transport along the $\widetilde{\sigma}(h)$ loops that are not present in $\mathrm{Hol}_{\wp}(\widetilde{X})$. Further, it implies that the paths starting at $\wp$, goes around $\widetilde{\gamma}$, and end at $\rho(h)\cdot \wp$ are also new loops on $X$. These case are seen as $\gamma \sigma(h)$ loops on $X$. This means that $\mathrm{Hol}_{p}(X)$ has a reacher structure and can be written as  
     \begin{equation}
         \mathrm{Hol}_{p}(X)\,=\, \{\, P_{\gamma\sigma(h)} \ \,  |\ \, \gamma\sigma(h) \ \  \text{loops at $p$ correspond to all $h\in H$ }  \,\}\,.
     \end{equation}
     On the covering space $\widetilde{X}$, we have 
     \begin{equation}
         P_{\widetilde{\gamma}\widetilde{\sigma}(h)}\,:\, T_{\wp}\widetilde{X}\,\to\, T_{\rho(h)\cdot\wp}\widetilde{X}\,,\quad  \text{and}\quad \dd\pi_{h}\,:\, T_{\rho(h)\cdot\wp}\widetilde{X}\,\to\, T_{\wp}\widetilde{X}\,,
     \end{equation}
     with $P_{\widetilde{\gamma}\widetilde{\sigma}(h)} = P_{\widetilde{\gamma}}P_{\widetilde{\sigma}(h)}$. 
     Then parallel transports along $\gamma\sigma(h)$ loops in $X$ are defined as
     \begin{equation}
    P_{\gamma\sigma(h)}\,=\,\dd\pi_{h}\,\circ\, P_{\widetilde{\gamma}\widetilde{\sigma}(h)}\,, \quad \forall h \, \in \, H\,.
\end{equation}

     For a finite group $H$, of order $N$, and elements given as  $\{\mathds{1},h_{1},h_{2},\cdots, h_{N-1}\}$, the holonomy group decomposes as
     \begin{equation}\label{eq:hol-p-X-decomposition}
         \mathrm{Hol}_{p}(X)\,=\,\{\,P_{\gamma\sigma(\mathds{1})}\,\}\,\cup\, \{\,P_{\gamma\sigma(h_{1})}\,\}\,\cdots\, \{\,P_{\gamma\sigma(h_{N-1})}\,\} \,.
     \end{equation}
From the perspective of the quotient group $H$, $\mathrm{Hol}_{p}(X)$ has components that are not connected to the identity. The connected component, $\mathrm{Hol}^{\mathrm{con.}}_{p}(X)$, matches exactly $\mathrm{Hol}_{\wp}(\widetilde{X})$ as $\{P_{\gamma\sigma(\mathds{1})}\}=\{P_{\gamma}\}\cong\{\dd\pi_{\mathds{1}}\circ P_{\widetilde{\gamma}}\}$. We observe that the decomposition structure of $\mathrm{Hol}_{p}(X)$ can be described by a homomorphism map,
     \begin{equation}
         \vartheta\,:\, H\, \to\, \mathrm{Hol}_{p}(X)/\mathrm{Hol}^{\mathrm{con.}}_{p}(X)
     \end{equation}
     Therefore, for any point $p$, the holonomy $\mathrm{Hol}(X)$ is a splittable extension of $\mathrm{Hol}^{\mathrm{con.}}(X)$ by $H$, i.e., 
     \begin{equation}\label{eq:Hol-rtimes-H}
        \mathrm{Hol}(X)\,=\, \mathrm{Hol}(\widetilde{X}) \rtimes\, H\,.
     \end{equation}

\end{itemize}

\paragraph{Holonomy and co-Seifert fibrations.}

In this paragraph, we clarify the relation between the co-Seifert fibrations discussed in appendix \ref{sec-co-Seifert-fibration} and section \ref{sec:G-structure-co-seifert}, and the general strategy for determining the holonomy of covering spaces.

Suppose that a $d$-dimensional manifold $X$ admits a co-Seifert fibration over a base $\bbS^{1}$ with typical fiber a $(d-1)$–dimensional manifold $Y$. At the level of topology, one may express this as
\begin{equation}
    X\,=\, \frac{Y\,\times\, \bbS^{1}}{G}\,,
\end{equation}
where $G$ is the structure group of the fibration. The group $G$ acts as a monodromy: transporting the fiber $Y$ once around the base $\bbS^{1}$ returns it to itself up to a transformation by $G$. Interpreting $Y \times \bbS^{1}$ as the covering space of $X$, the holonomy group of $X$ fits naturally into a semidirect product:
\begin{equation}\label{eq:hol-X=Hol-Y-rtimes-G-monodromy}
      \mathrm{Hol}(X)\,=\,\mathrm{Hol}(Y)\,\rtimes \,G\,.
\end{equation}
according to the discussion of the previous paragraph. In this way, $\mathrm{Hol}(X)$ may be viewed as an extension of $\mathrm{Hol}(Y)$ by the monodromy group $G$.

%%%%%%%%%%%%%%%%%%%%%%%%%%%%%%%%%%%%%%%%%%%%%%%%%%%%%%%%%%%%%%%%%%%%%%%%%%%

\bibliographystyle{JHEP}
\bibliography{M-ref.bib}

@Article{Closset:2020scj,
  author        = {Closset, Cyril and Schafer-Nameki, Sakura and Wang, Yi-Nan},
  journal       = {JHEP},
  title         = {{Coulomb and Higgs Branches from Canonical Singularities: Part 0}},
  year          = {2021},
  pages         = {003},
  volume        = {02},
  archiveprefix = {arXiv},
  doi           = {10.1007/JHEP02(2021)003},
  eprint        = {2007.15600},
  primaryclass  = {hep-th},
}

@article{Apruzzi:2019opn,
      author         = "Apruzzi, Fabio and Lawrie, Craig and Lin, Ling and
                        Schafer-Nameki, Sakura and Wang, Yi-Nan",
      title          = "{Fibers add Flavor, Part I: Classification of 5d SCFTs,
                        Flavor Symmetries and BPS States}",
      journal        = "JHEP",
      volume         = "11",
      year           = "2019",
      pages          = "068",
      doi            = "10.1007/JHEP11(2019)068",
      eprint         = "1907.05404",
      archivePrefix  = "arXiv",
      primaryClass   = "hep-th",
      SLACcitation   = "%%CITATION = ARXIV:1907.05404;%%"
}

@article{Heckman:2013pva,
      author         = "Heckman, Jonathan J. and Morrison, David R. and Vafa,
                        Cumrun",
      title          = "{On the Classification of 6D SCFTs and Generalized ADE
                        Orbifolds}",
      journal        = "JHEP",
      volume         = "05",
      year           = "2014",
      pages          = "028",
      doi            = "10.1007/JHEP06(2015)017, 10.1007/JHEP05(2014)028",
      note           = "[Erratum: JHEP06,017(2015)]",
      eprint         = "1312.5746",
      archivePrefix  = "arXiv",
      primaryClass   = "hep-th",
      SLACcitation   = "%%CITATION = ARXIV:1312.5746;%%"
}

@Article{Morrison:1996xf,
  author =        {Morrison, David R. and Seiberg, Nathan},
  title =         {{Extremal transitions and five-dimensional supersymmetric field theories}},
  journal =       {Nucl. Phys.},
  year =          {1997},
  volume =        {B483},
  pages =         {229-247},
  archiveprefix = {arXiv},
  doi =           {10.1016/S0550-3213(96)00592-5},
  eprint =        {hep-th/9609070},
  primaryclass =  {hep-th},
  reportnumber =  {DUKE-TH-96-130, RU-96-80},
  slaccitation =  {%%CITATION = HEP-TH/9609070;%%}
}

@article{Bershadsky:1996nh,
      author         = "Bershadsky, M. and Intriligator, Kenneth A. and Kachru,
                        S. and Morrison, David R. and Sadov, V. and Vafa, Cumrun",
      title          = "{Geometric singularities and enhanced gauge symmetries}",
      journal        = "Nucl. Phys.",
      volume         = "B481",
      year           = "1996",
      pages          = "215-252",
      doi            = "10.1016/S0550-3213(96)90131-5",
      eprint         = "hep-th/9605200",
      archivePrefix  = "arXiv",
      primaryClass   = "hep-th",
      reportNumber   = "HUTP-96-A017, IASSNS-HEP-96-49, RU-96-40",
      SLACcitation   = "%%CITATION = HEP-TH/9605200;%%"
}

@Article{Jefferson:2018irk,
  author        = {Jefferson, Patrick and Katz, Sheldon and Kim, Hee-Cheol and Vafa, Cumrun},
  title         = {{On Geometric Classification of 5d SCFTs}},
  journal       = {JHEP},
  year          = {2018},
  volume        = {04},
  pages         = {103},
  archiveprefix = {arXiv},
  doi           = {10.1007/JHEP04(2018)103},
  eprint        = {1801.04036},
  primaryclass  = {hep-th},
  slaccitation  = {%%CITATION = ARXIV:1801.04036;%%},
}

@Article{Intriligator:1997pq,
  author        = {Intriligator, Kenneth A. and Morrison, David R. and Seiberg, Nathan},
  title         = {{Five-dimensional supersymmetric gauge theories and degenerations of Calabi-Yau spaces}},
  journal       = {Nucl. Phys.},
  year          = {1997},
  volume        = {B497},
  pages         = {56-100},
  archiveprefix = {arXiv},
  doi           = {10.1016/S0550-3213(97)00279-4},
  eprint        = {hep-th/9702198},
  primaryclass  = {hep-th},
  reportnumber  = {RU-96-99, IASSNS-HEP-96-112},
  slaccitation  = {%%CITATION = HEP-TH/9702198;%%},
}

@Article{Witten:1996qb,
  author        = {Witten, Edward},
  title         = {{Phase transitions in M theory and F theory}},
  journal       = {Nucl. Phys.},
  year          = {1996},
  volume        = {B471},
  pages         = {195-216},
  archiveprefix = {arXiv},
  doi           = {10.1016/0550-3213(96)00212-X},
  eprint        = {hep-th/9603150},
  primaryclass  = {hep-th},
  reportnumber  = {IASSNS-HEP-96-26},
  slaccitation  = {%%CITATION = HEP-TH/9603150;%%},
}

@Article{Seiberg:1996bd,
  author        = {Seiberg, Nathan},
  title         = {{Five-dimensional SUSY field theories, nontrivial fixed points and string dynamics}},
  journal       = {Phys. Lett.},
  year          = {1996},
  volume        = {B388},
  pages         = {753-760},
  archiveprefix = {arXiv},
  doi           = {10.1016/S0370-2693(96)01215-4},
  eprint        = {hep-th/9608111},
  primaryclass  = {hep-th},
  reportnumber  = {RU-96-69},
  slaccitation  = {%%CITATION = HEP-TH/9608111;%%},
}

@Article{Xie:2015rpa,
  author        = {Xie, Dan and Yau, Shing-Tung},
  title         = {{4d N=2 SCFT and singularity theory Part I: Classification}},
  year          = {2015},
  month         = {10},
  archiveprefix = {arXiv},
  eprint        = {1510.01324},
  primaryclass  = {hep-th},
}

@Article{Shapere:1999xr,
  author        = {Shapere, Alfred D. and Vafa, Cumrun},
  title         = {{BPS structure of Argyres-Douglas superconformal theories}},
  year          = {1999},
  month         = {10},
  archiveprefix = {arXiv},
  eprint        = {hep-th/9910182},
  reportnumber  = {HUTP-99-A057, UKHEP-99-15},
}

@article{Klemm:1996bj,
    author = "Klemm, Albrecht and Lerche, Wolfgang and Mayr, Peter and Vafa, Cumrun and Warner, Nicholas P.",
    title = "{Selfdual strings and N=2 supersymmetric field theory}",
    eprint = "hep-th/9604034",
    archivePrefix = "arXiv",
    reportNumber = "CERN-TH-96-95, HUTP-96-A014, USC-96-008",
    doi = "10.1016/0550-3213(96)00353-7",
    journal = "Nucl. Phys. B",
    volume = "477",
    pages = "746--766",
    year = "1996"
}

@Article{Beasley:2008dc,
  author        = {Beasley, Chris and Heckman, Jonathan J. and Vafa, Cumrun},
  journal       = {JHEP},
  title         = {{GUTs and Exceptional Branes in F-theory - I}},
  year          = {2009},
  pages         = {058},
  volume        = {01},
  archiveprefix = {arXiv},
  doi           = {10.1088/1126-6708/2009/01/058},
  eprint        = {0802.3391},
  primaryclass  = {hep-th},
}

@Article{Donagi:2008ca,
  author        = {Donagi, Ron and Wijnholt, Martijn},
  journal       = {Adv. Theor. Math. Phys.},
  title         = {{Model Building with F-Theory}},
  year          = {2011},
  number        = {5},
  pages         = {1237--1317},
  volume        = {15},
  archiveprefix = {arXiv},
  doi           = {10.4310/ATMP.2011.v15.n5.a2},
  eprint        = {0802.2969},
  primaryclass  = {hep-th},
  reportnumber  = {AEI-2007-174},
}

@article{Katz:1996fh,
    author = "Katz, Sheldon H. and Klemm, Albrecht and Vafa, Cumrun",
    title = "{Geometric engineering of quantum field theories}",
    eprint = "hep-th/9609239",
    archivePrefix = "arXiv",
    reportNumber = "EFI-96-37, HUTP-96-A046, OSU-M-96-24",
    doi = "10.1016/S0550-3213(97)00282-4",
    journal = "Nucl. Phys. B",
    volume = "497",
    pages = "173--195",
    year = "1997"
}

@article{Closset:2021lwy,
    author = {Closset, Cyril and Sch\"afer-Nameki, Sakura and Wang, Yi-Nan},
    title = "{Coulomb and Higgs branches from canonical singularities. Part I. Hypersurfaces with smooth Calabi-Yau resolutions}",
    eprint = "2111.13564",
    archivePrefix = "arXiv",
    primaryClass = "hep-th",
    doi = "10.1007/JHEP04(2022)061",
    journal = "JHEP",
    volume = "04",
    pages = "061",
    year = "2022"
}

@article{Bourget:2023wlb,
    author = "Bourget, Antoine and Collinucci, Andr\'es and Schafer-Nameki, Sakura",
    title = "{Generalized Toric Polygons, T-branes, and 5d SCFTs}",
    eprint = "2301.05239",
    archivePrefix = "arXiv",
    primaryClass = "hep-th",
    month = "1",
    year = "2023"
}

@article{Acharya_2004,
   title={M theory and singularities of exceptional holonomy manifolds},
   volume={392},
   ISSN={0370-1573},
   url={http://dx.doi.org/10.1016/j.physrep.2003.10.017},
   DOI={10.1016/j.physrep.2003.10.017},
   number={3},
   journal={Physics Reports},
   publisher={Elsevier BV},
   author={Acharya, Bobby S. and Gukov, Sergei},
   year={2004},
   month=mar, pages={121–189} }

@article{Sen:1997kz,
    author = "Sen, Ashoke",
    title = "{A Note on enhanced gauge symmetries in M and string theory}",
    eprint = "hep-th/9707123",
    archivePrefix = "arXiv",
    reportNumber = "MRI-PHY-P970717",
    doi = "10.1088/1126-6708/1997/09/001",
    journal = "JHEP",
    volume = "09",
    pages = "001",
    year = "1997"
}

@article{Najjar:2023hee,
    author = "Najjar, Marwan and Tian, Jiahua and Wang, Yi-Nan",
    title = "{3d $ \mathcal{N} $ = 2 theories from M-theory on CY4 and IIB brane box}",
    eprint = "2312.17082",
    archivePrefix = "arXiv",
    primaryClass = "hep-th",
    doi = "10.1007/JHEP05(2024)038",
    journal = "JHEP",
    volume = "05",
    pages = "038",
    year = "2024"
}

@article{Oh:1998qi,
    author = "Oh, Kyungho and Tatar, Radu",
    title = "{Three-dimensional SCFT from M2-branes at conifold singularities}",
    eprint = "hep-th/9810244",
    archivePrefix = "arXiv",
    reportNumber = "BROWN-HET-1148",
    doi = "10.1088/1126-6708/1999/02/025",
    journal = "JHEP",
    volume = "02",
    pages = "025",
    year = "1999"
}

@article{Acharya:2020vmg,
    author = "Acharya, Bobby Samir and Foscolo, Lorenzo and Najjar, Marwan and Svanes, Eirik Eik",
    title = "{New G$_{2}$-conifolds in M-theory and their field theory interpretation}",
    eprint = "2011.06998",
    archivePrefix = "arXiv",
    primaryClass = "hep-th",
    doi = "10.1007/JHEP05(2021)250",
    journal = "JHEP",
    volume = "05",
    pages = "250",
    year = "2021"
}

@article{Apruzzi:2021nmk,
    author = "Apruzzi, Fabio and Bonetti, Federico and Garc\'\i{}a Etxebarria, I\~naki and Hosseini, Saghar S. and Schafer-Nameki, Sakura",
    title = "{Symmetry TFTs from String Theory}",
    eprint = "2112.02092",
    archivePrefix = "arXiv",
    primaryClass = "hep-th",
    doi = "10.1007/s00220-023-04737-2",
    journal = "Commun. Math. Phys.",
    volume = "402",
    number = "1",
    pages = "895--949",
    year = "2023"
}

@article{Herzog:2000rz,
    author = "Herzog, Christopher P. and Klebanov, Igor R.",
    title = "{Gravity duals of fractional branes in various dimensions}",
    eprint = "hep-th/0101020",
    archivePrefix = "arXiv",
    reportNumber = "PUPT-1974",
    doi = "10.1103/PhysRevD.63.126005",
    journal = "Phys. Rev. D",
    volume = "63",
    pages = "126005",
    year = "2001"
}

@phdthesis{Najjar:2022eci,
    author = "Najjar, Marwan A. M.",
    title = "{Field Theory Dynamics from M-theory on Special Holonomy Manifolds}",
    school = "King's Coll. London",
    year = "2022"
}

@article{Acharya:2021jsp,
    author = "Acharya, Bobby and Lambert, Neil and Najjar, Marwan and Svanes, Eirik Eik and Tian, Jiahua",
    title = "{Gauging discrete symmetries of T$_{N}$-theories in five dimensions}",
    eprint = "2110.14441",
    archivePrefix = "arXiv",
    primaryClass = "hep-th",
    reportNumber = "KCL-PH-TH/2021-78",
    doi = "10.1007/JHEP04(2022)114",
    journal = "JHEP",
    volume = "04",
    pages = "114",
    year = "2022"
}

@article{Acharya:2001hq,
    author = "Acharya, Bobby Samir",
    title = "{Confining strings from G(2) holonomy space-times}",
    eprint = "hep-th/0101206",
    archivePrefix = "arXiv",
    reportNumber = "RUNHETC-2001-03",
    month = "1",
    year = "2001"
}

@article{Bryant1989OnTC,
  title={On the construction of some complete metrics with exceptional holonomy},
  author={Robert L. Bryant and Simon Salamon},
  journal={Duke Mathematical Journal},
  year={1989},
  volume={58},
  pages={829-850},
  url={https://api.semanticscholar.org/CorpusID:117940964}
}

@article{Najjar:2024vmm,
    author = "Najjar, Marwan and Santilli, Leonardo and Wang, Yi-Nan",
    title = "{(\ensuremath{-}1)-form symmetries from M-theory and SymTFTs}",
    eprint = "2411.19683",
    archivePrefix = "arXiv",
    primaryClass = "hep-th",
    reportNumber = "USTC-ICTS/PCFT-24-53",
    doi = "10.1007/JHEP03(2025)134",
    journal = "JHEP",
    volume = "03",
    pages = "134",
    year = "2025"
}

@article{mckay,
    author = "McKay, John",
    title = "{Graphs, singularities and finite groups}",
    doi = "10.1090/pspum/037/604577",
    journal = "Proc. Symp. Pure Math., Proceedings of Symposia in Pure Mathematics, Amer. Math. Soc.",
    volume = "37",
    pages = "183–186",
    year = "1980"
}

@article{Najjar:2025htp,
    author = "Najjar, Marwan",
    title = "{Modified instanton sum and 4-group structure in 4d $\mathcal{N}=1$$SU(M)$ SYM from holography}",
    eprint = "2503.17108",
    archivePrefix = "arXiv",
    primaryClass = "hep-th",
    month = "3",
    year = "2025"
}

@article{Najjar:2025rgt,
    author = "Najjar, Marwan and Wang, Yi-Nan",
    title = "{Confinement of 3d $\mathcal{N}=2$ Gauge Theories from M-theory on CY4}",
    eprint = "2501.07116",
    archivePrefix = "arXiv",
    primaryClass = "hep-th",
    month = "1",
    year = "2025"
}

@article{Hausel:2002xg,
    author = "Hausel, Tamas and Hunsicker, Eugenie and Mazzeo, Rafe",
    title = "{\hyperlink{https://arxiv.org/pdf/math/0207169.pdf}{Hodge cohomology of gravitational instantons}}",
    eprint = "math/0207169",
    archivePrefix = "arXiv",
    month = "7",
    year = "2002"
}

@article{Gibbons:1989er,
    author = "Gibbons, G. W. and Page, Don N. and Pope, C. N.",
    title = "{Einstein Metrics on S**3 R**3 and R**4 Bundles}",
    reportNumber = "CTP-TAMU-08-89",
    doi = "10.1007/BF02104500",
    journal = "Commun. Math. Phys.",
    volume = "127",
    pages = "529",
    year = "1990"
}

@article{Katz:1996th,
    author = "Katz, Sheldon H. and Vafa, Cumrun",
    title = "{Geometric engineering of N=1 quantum field theories}",
    eprint = "hep-th/9611090",
    archivePrefix = "arXiv",
    reportNumber = "HUTP-96-A051, OSU-M-96-25, OSU-MATH-1996-25",
    doi = "10.1016/S0550-3213(97)00283-6",
    journal = "Nucl. Phys. B",
    volume = "497",
    pages = "196--204",
    year = "1997"
}

@article{Witten:1995ex,
    author = "Witten, Edward",
    title = "{String theory dynamics in various dimensions}",
    eprint = "hep-th/9503124",
    archivePrefix = "arXiv",
    reportNumber = "IASSNS-HEP-95-18",
    doi = "10.1201/9781482268737-32",
    journal = "Nucl. Phys. B",
    volume = "443",
    pages = "85--126",
    year = "1995"
}

@article{Ooguri:1997ih,
    author = "Ooguri, Hirosi and Vafa, Cumrun",
    title = "{Geometry of N=1 dualities in four-dimensions}",
    eprint = "hep-th/9702180",
    archivePrefix = "arXiv",
    reportNumber = "HUPT-97-A010, UCB-PTH-97-11, LBL-40032, LBNL-40032",
    doi = "10.1016/S0550-3213(97)00304-0",
    journal = "Nucl. Phys. B",
    volume = "500",
    pages = "62--74",
    year = "1997"
}

@article{Acharya:1998pm,
    author = "Acharya, Bobby Samir",
    title = "{M theory, Joyce orbifolds and superYang-Mills}",
    eprint = "hep-th/9812205",
    archivePrefix = "arXiv",
    reportNumber = "QMW-PH-98-42",
    doi = "10.4310/ATMP.1999.v3.n2.a3",
    journal = "Adv. Theor. Math. Phys.",
    volume = "3",
    pages = "227--248",
    year = "1999"
}

@book{szczepanski2012geometry,
  title={Geometry of Crystallographic Groups},
  author={Szczepa{\'n}ski, A.},
  isbn={9789814412254},
  lccn={2012016310},
  series={Algebra and discrete mathematics},
  url={https://books.google.com.hk/books?id=HG_6s-RP-tsC},
  year={2012},
  publisher={World Scientific}
}

@book{brown1978,
  title={Crystallographic Groups of Four-Dimensional Space},
  author={Brown, H. and Bülow, R. and Neubüser, J. and Wondratschek, H. and Zassenhaus, H.},
  year={1978},
  publisher={Wiley},
  address={New York},
  series={Wiley Monographs in Crystallography},
  isbn={978-0471030959},
  pages={443}
}

@misc{lambert2013,
      title={Closed flat Riemannian 4-manifolds}, 
      author={Thomas P. Lambert and John G. Ratcliffe and Steven T. Tschantz},
      year={2013},
      eprint={1306.6613},
      archivePrefix={arXiv},
      primaryClass={math.GT},
      url={https://arxiv.org/abs/1306.6613}, 
}

@article{auslander1965,
  title={{An account of the theory of crystallographic groups}},
  author={Auslander, Louis},
  journal={The American Mathematical Monthly},
  volume={72},
  number={11},
  pages={1231--1235},
  year={1965},
  publisher={JSTOR}
}

@book{Ratcliffe_2006,
    author = {Ratcliffe, John G.},
    title = {Foundations of Hyperbolic Manifolds},
    series = {Graduate Texts in Mathematics},
    volume = {149},
    edition = {2nd},
    publisher = {Springer},
    address = {New York, NY},
    year = {2006},
    isbn = {0-387-33197-2}
}

@book{charlap1986,
  title={{Bieberbach Groups and Flat Manifolds}},
  author={Charlap, Leonard S.},
  year={1986},
  publisher={Springer-Verlag},
  address={New York},
  series={Universitext},
  isbn={978-0-387-96395-2}
}

@article{Lutowski_2015,
   title={Spin structures on flat manifolds},
   volume={436},
   ISSN={0021-8693},
   url={http://dx.doi.org/10.1016/j.jalgebra.2015.03.037},
   DOI={10.1016/j.jalgebra.2015.03.037},
   journal={Journal of Algebra},
   publisher={Elsevier BV},
   author={Lutowski, Rafał and Putrycz, Bartosz},
   year={2015},
   month=aug, pages={277–291} }

@article{putrycz2010,
  title={{Existence of spin structures on flat four-manifolds}},
  author={Putrycz, Bartosz and Szczepa{\'n}ski, Andrzej},
  journal={Advances in Geometry},
  volume={10},
  number={2},
  pages={323--332},
  year={2010},
  publisher={Walter de Gruyter GmbH}
}

@article{Ratcliffe_2010,
   title={Fibered orbifolds and crystallographic groups},
   volume={10},
   ISSN={1472-2747},
   url={http://dx.doi.org/10.2140/agt.2010.10.1627},
   DOI={10.2140/agt.2010.10.1627},
   number={3},
   journal={Algebraic \&amp; Geometric Topology},
   publisher={Mathematical Sciences Publishers},
   author={Ratcliffe, John G and Tschantz, Steven T},
   year={2010},
   month=jul, pages={1627–1664} }

@article{Rossetti:1998qz,
    author = "Rossetti, J. P. and Tirao, Paulo",
    title = "{Five-dimensional Bieberbach groups with holonomy group Z(2) + Z(2)}",
    reportNumber = "IC-98-79",
    month = "7",
    year = "1998"
}

@book{kobayashi1963I,
  title={Foundations of Differential Geometry},
  author={Kobayashi, S. and Nomizu, K.},
  number={v. 1},
  isbn={9780470496473},
  lccn={68019209},
  series={Foundations of Differential Geometry [by] Shoshichi Kobayashi and Katsumi Nomizu},
  url={https://books.google.com.sg/books?id=iLANAQAAIAAJ},
  year={1963},
  publisher={Interscience Publishers}
}

@book{kobayashi1969II,
  title={Foundations of Differential Geometry, Volume 2},
  author={Kobayashi, S. and Nomizu, K.},
  isbn={9780471157328},
  lccn={96135215},
  series={Wiley Classics Library},
  url={https://books.google.com.sg/books?id=zM7bEAAAQBAJ},
  year={1969},
  publisher={Wiley}
}

@misc{ratcliffe2012,
      title={Fibered orbifolds and crystallographic groups, II}, 
      author={John G. Ratcliffe and Steven T. Tschantz},
      year={2012},
      eprint={1112.3981},
      archivePrefix={arXiv},
      primaryClass={math.GT},
      url={https://arxiv.org/abs/1112.3981}, 
}

@article{Charlap-Vasquez-1973,
 ISSN = {00029327, 10806377},
 URL = {http://www.jstor.org/stable/2373726},
 author = {L. S. Charlap and A. T. Vasquez},
 journal = {American Journal of Mathematics},
 number = {3},
 pages = {471--494},
 publisher = {The Johns Hopkins University Press},
 title = {Compact Flat Riemannian Manifolds III: The Group of Affinities},
 urldate = {2025-10-25},
 volume = {95},
 year = {1973}
}

@article{Auslander1960,
 ISSN = {0003486X, 19398980},
 URL = {http://www.jstor.org/stable/1969945},
 author = {Louis Auslander},
 journal = {Annals of Mathematics},
 number = {3},
 pages = {579--590},
 publisher = {[Annals of Mathematics, Trustees of Princeton University on Behalf of the Annals of Mathematics, Mathematics Department, Princeton University]},
 title = {Bieberbach's Theorems on Space Groups and Discrete Uniform Subgroups of Lie Groups},
 urldate = {2025-10-25},
 volume = {71},
 year = {1960}
}

@misc{conway2003,
      title={Describing the platycosms}, 
      author={John Horton Conway and Juan Pablo Rossetti},
      year={2003},
      eprint={math/0311476},
      archivePrefix={arXiv},
      primaryClass={math.DG},
      url={https://arxiv.org/abs/math/0311476}, 
}

@misc{lutowski2011,
      title={Seven dimensional flat manifolds with cyclic holonomy}, 
      author={Rafał Lutowski},
      year={2011},
      eprint={1101.2633},
      archivePrefix={arXiv},
      primaryClass={math.GR},
      url={https://arxiv.org/abs/1101.2633}, 
}

@misc{ocampo2019,
      title={Bieberbach groups and flat manifolds with finite abelian holonomy from Artin braid groups}, 
      author={Oscar Ocampo},
      year={2019},
      eprint={1905.05123},
      archivePrefix={arXiv},
      primaryClass={math.GT},
      url={https://arxiv.org/abs/1905.05123}, 
}

@article{Hillman1995,
  author = {Hillman, Jonathan A.},
  title = {Flat 4-manifold groups},
  journal = {New Zealand Journal of Mathematics},
  volume = {24},
  pages = {29--40},
  year = {1995},
  publisher = {New Zealand Mathematical Society}
}

@misc{clarke2012,
      title={Holonomy Groups in Riemannian Geometry}, 
      author={Andrew Clarke and Bianca Santoro},
      year={2012},
      eprint={1206.3170},
      archivePrefix={arXiv},
      primaryClass={math.DG},
      url={https://arxiv.org/abs/1206.3170}, 
}

@book{Besse1987,
  author = {Besse, Arthur L.},
  title = {Einstein Manifolds},
  series = {Classics in Mathematics},
  publisher = {Springer-Verlag},
  address = {Berlin, Heidelberg, New York},
  year = {1987},
  edition = {Reprint of the 1987 edition, published in the Classics in Mathematics series},
  isbn = {978-3-540-74120-6},
  doi = {10.1007/978-3-540-74311-8},
  note = {Originally published as Vol. 10 of the Ergebnisse der Mathematik und ihrer Grenzgebiete, 3rd series},
}

@book{Berger2003,
    author = {Berger, Marcel},
    title = {A Panoramic View of Riemannian Geometry},
    year = {2003},
    publisher = {Springer},
    address = {Berlin, Heidelberg},
    isbn = {978-3-540-65317-2}
}

@book{RudolphSchmidt2017,
  author    = {Rudolph, Gerd and Schmidt, Matthias},
  title     = {Differential Geometry and Mathematical Physics: Part II. Fibre Bundles, Topology and Gauge Fields},
  series    = {Theoretical and Mathematical Physics},
  publisher = {Springer},
  year      = {2017},
  doi       = {10.1007/978-94-024-1451-6},
  isbn      = {978-94-024-1449-3}
}

@article{Ruback1986TheMO,
  title={The motion of Kaluza-Klein monopoles},
  author={Peter J. Ruback},
  journal={Communications in Mathematical Physics},
  year={1986},
  volume={107},
  pages={93-102}
}

@article{Franchetti:2014lza,
    author = "Franchetti, Guido",
    title = "Harmonic forms on ALF gravitational instantons",
    eprint = "arXiv:1410.2864",
    archivePrefix = "arXiv",
    primaryClass = "hep-th",
    doi = "10.1007/JHEP12(2014)075",
    journal = "JHEP",
    volume = "12",
    pages = "075",
    year = "2014"
}

@article{GIBBONS1978430,
title = {Gravitational multi-instantons},
journal = {Physics Letters B},
volume = {78},
number = {4},
pages = {430-432},
year = {1978},
issn = {0370-2693},
doi = {https://doi.org/10.1016/0370-2693(78)90478-1},
url = {https://www.sciencedirect.com/science/article/pii/0370269378904781},
author = {G.W. Gibbons and S.W. Hawking}
}

@article{Eguchi1979,
  author = {Eguchi, Tohru and Hanson, Andrew J.},
  title = {Selfdual solutions to {E}uclidean gravity},
  journal = {Annals of Physics},
  volume = {120},
  number = {1},
  pages = {82--105},
  year = {1979},
  doi = {10.1016/0003-4916(79)90282-3},
  bibcode = {1979AnPhy.120...82E},
  osti = {1447072}
}

@article{Sen:1997js,
    author = "Sen, Ashoke",
    title = "{Dynamics of multiple Kaluza-Klein monopoles in M and string theory}",
    eprint = "hep-th/9707042",
    archivePrefix = "arXiv",
    reportNumber = "MRI-PHY-P970716",
    doi = "10.4310/ATMP.1997.v1.n1.a3",
    journal = "Adv. Theor. Math. Phys.",
    volume = "1",
    pages = "115--126",
    year = "1998"
}

@misc{wright2011quotientsgravitationalinstantons,
      title={Quotients of gravitational instantons}, 
      author={Evan P. Wright},
      year={2011},
      eprint={1102.2442},
      archivePrefix={arXiv},
      primaryClass={math.DG},
      url={https://arxiv.org/abs/1102.2442}, 
}

@article{Acharya:2023xlx,
    author = "Acharya, Bobby Samir and Baldwin, Daniel Andrew",
    title = "{Coulomb and Higgs phases of G$_{2}$-manifolds}",
    eprint = "2309.12869",
    archivePrefix = "arXiv",
    primaryClass = "hep-th",
    doi = "10.1007/JHEP01(2024)147",
    journal = "JHEP",
    volume = "01",
    pages = "147",
    year = "2024"
}

@article{Pfaeffle2000Dirac,
  title={The Dirac spectrum of Bieberbach manifolds},
  author={Pfäffle, Frank},
  journal={Journal of Geometry and Physics},
  volume={35},
  number={4},
  pages={367--385},
  year={2000},
  publisher={Elsevier},
  doi={10.1016/S0393-0440(00)00005-X}
}

@article{Cecotti:2010bp,
    author = "Cecotti, Sergio and Cordova, Clay and Heckman, Jonathan J. and Vafa, Cumrun",
    title = "{T-Branes and Monodromy}",
    eprint = "1010.5780",
    archivePrefix = "arXiv",
    primaryClass = "hep-th",
    doi = "10.1007/JHEP07(2011)030",
    journal = "JHEP",
    volume = "07",
    pages = "030",
    year = "2011"
}

@article{Donagi:2003hh,
    author = "Donagi, Ron and Katz, S. and Sharpe, E.",
    title = "{Spectra of D-branes with higgs vevs}",
    eprint = "hep-th/0309270",
    archivePrefix = "arXiv",
    reportNumber = "ILL-TH-03-09, ILL-(TH)-03-09",
    doi = "10.4310/ATMP.2004.v8.n5.a3",
    journal = "Adv. Theor. Math. Phys.",
    volume = "8",
    number = "5",
    pages = "813--859",
    year = "2004"
}

@article{Collinucci:2014qfa,
    author = "Collinucci, Andres and Savelli, Raffaele",
    title = "{T-branes as branes within branes}",
    eprint = "1410.4178",
    archivePrefix = "arXiv",
    primaryClass = "hep-th",
    doi = "10.1007/JHEP09(2015)161",
    journal = "JHEP",
    volume = "09",
    pages = "161",
    year = "2015"
}

@article{Donagi:2011jy,
    author = "Donagi, Ron and Wijnholt, Martijn",
    title = "{Gluing Branes, I}",
    eprint = "1104.2610",
    archivePrefix = "arXiv",
    primaryClass = "hep-th",
    doi = "10.1007/JHEP05(2013)068",
    journal = "JHEP",
    volume = "05",
    pages = "068",
    year = "2013"
}

@book{Collingwood1993,
    author = {Collingwood, David H. and McGovern, William M.},
    title = {Nilpotent Orbits in Semisimple Lie Algebras: An Introduction},
    year = {1993},
    publisher = {CRC Press},
    isbn = {978-0534188344},
    url = {},
    }

@article{Collinucci:2014taa,
    author = "Collinucci, Andres and Savelli, Raffaele",
    title = "{F-theory on singular spaces}",
    eprint = "1410.4867",
    archivePrefix = "arXiv",
    primaryClass = "hep-th",
    doi = "10.1007/JHEP09(2015)100",
    journal = "JHEP",
    volume = "09",
    pages = "100",
    year = "2015"
}

@book{slodowy2006simple,
  title={Simple Singularities and Simple Algebraic Groups},
  author={Slodowy, P.},
  isbn={9783540381914},
  series={Lecture Notes in Mathematics},
  url={https://books.google.ps/books?id=-u96CwAAQBAJ},
  year={2006},
  publisher={Springer Berlin Heidelberg}
}

@article{Donagi:1995cf,
    author = "Donagi, Ron and Witten, Edward",
    title = "{Supersymmetric Yang-Mills theory and integrable systems}",
    eprint = "hep-th/9510101",
    archivePrefix = "arXiv",
    reportNumber = "IASSNS-HEP-95-78",
    doi = "10.1016/0550-3213(95)00609-5",
    journal = "Nucl. Phys. B",
    volume = "460",
    pages = "299--334",
    year = "1996"
}

@misc{henderson2015,
      title={Singularities of nilpotent orbit closures}, 
      author={Anthony Henderson},
      year={2015},
      eprint={1408.3888},
      archivePrefix={arXiv},
      primaryClass={math.RT},
      url={https://arxiv.org/abs/1408.3888}, 
}

@article{Dorey:1999sj,
    author = "Dorey, N.",
    title = "{An Elliptic superpotential for softly broken N=4 supersymmetric Yang-Mills theory}",
    eprint = "hep-th/9906011",
    archivePrefix = "arXiv",
    reportNumber = "UW-PT-99-10",
    doi = "10.1088/1126-6708/1999/07/021",
    journal = "JHEP",
    volume = "07",
    pages = "021",
    year = "1999"
}

@article{Polchinski:2000uf,
    author = "Polchinski, Joseph and Strassler, Matthew J.",
    title = "{The String dual of a confining four-dimensional gauge theory}",
    eprint = "hep-th/0003136",
    archivePrefix = "arXiv",
    reportNumber = "IAS-TH-00-18, NSF-ITP-00-16",
    month = "3",
    year = "2000"
}

@article{Barbosa:2019bgh,
    author = "Barbosa, Rodrigo and Cveti{\v{c}}, Mirjam and Heckman, Jonathan J. and Lawrie, Craig and Torres, Ethan and Zoccarato, Gianluca",
    title = "{T-branes and $G_2$ backgrounds}",
    eprint = "1906.02212",
    archivePrefix = "arXiv",
    primaryClass = "hep-th",
    reportNumber = "UPR-1298-T",
    doi = "10.1103/PhysRevD.101.026015",
    journal = "Phys. Rev. D",
    volume = "101",
    number = "2",
    pages = "026015",
    year = "2020"
}

@article{Cecotti:2009zf,
    author = "Cecotti, Sergio and Cheng, Miranda C. N. and Heckman, Jonathan J. and Vafa, Cumrun",
    title = "{Yukawa Couplings in F-theory and Non-Commutative Geometry}",
    eprint = "0910.0477",
    archivePrefix = "arXiv",
    primaryClass = "hep-th",
    month = "10",
    year = "2009"
}

@article{Yamron:1988qc,
    author = "Yamron, Jonathan P.",
    title = "{Topological Actions From Twisted Supersymmetric Theories}",
    reportNumber = "IASSNS-HEP-88/12",
    doi = "10.1016/0370-2693(88)91769-8",
    journal = "Phys. Lett. B",
    volume = "213",
    pages = "325--330",
    year = "1988"
}

@article{Vafa:1994tf,
    author = "Vafa, Cumrun and Witten, Edward",
    title = "{A Strong coupling test of S duality}",
    eprint = "hep-th/9408074",
    archivePrefix = "arXiv",
    reportNumber = "HUTP-94-A017, IASSNS-HEP-94-54",
    doi = "10.1016/0550-3213(94)90097-3",
    journal = "Nucl. Phys. B",
    volume = "431",
    pages = "3--77",
    year = "1994"
}

@article{Witten:1988ze,
    author = "Witten, Edward",
    title = "{Topological Quantum Field Theory}",
    reportNumber = "IASSNS-HEP-87-72",
    doi = "10.1007/BF01223371",
    journal = "Commun. Math. Phys.",
    volume = "117",
    pages = "353",
    year = "1988"
}

@article{Anderson:2013rka,
    author = "Anderson, Lara B. and Heckman, Jonathan J. and Katz, Sheldon",
    title = "{T-Branes and Geometry}",
    eprint = "1310.1931",
    archivePrefix = "arXiv",
    primaryClass = "hep-th",
    doi = "10.1007/JHEP05(2014)080",
    journal = "JHEP",
    volume = "05",
    pages = "080",
    year = "2014"
}

@article{Collinucci:2016hpz,
    author = "Collinucci, Andres and Giacomelli, Simone and Savelli, Raffaele and Valandro, Roberto",
    title = "{T-branes through 3d mirror symmetry}",
    eprint = "1603.00062",
    archivePrefix = "arXiv",
    primaryClass = "hep-th",
    doi = "10.1007/JHEP07(2016)093",
    journal = "JHEP",
    volume = "07",
    pages = "093",
    year = "2016"
}

@article{Bena:2016oqr,
    author = {Bena, Iosif and Bl{\r{a}}b{\"a}ck, Johan and Minasian, Ruben and Savelli, Raffaele},
    title = "{There and back again: A T-brane's tale}",
    eprint = "1608.01221",
    archivePrefix = "arXiv",
    primaryClass = "hep-th",
    reportNumber = "IPHT-T16-071",
    doi = "10.1007/JHEP11(2016)179",
    journal = "JHEP",
    volume = "11",
    pages = "179",
    year = "2016"
}

@article{Marchesano:2016cqg,
    author = "Marchesano, Fernando and Schwieger, Sebastian",
    title = "{T-branes and $\alpha'$-corrections}",
    eprint = "1609.02799",
    archivePrefix = "arXiv",
    primaryClass = "hep-th",
    reportNumber = "IFT-UAM-CSIC-16-082",
    doi = "10.1007/JHEP11(2016)123",
    journal = "JHEP",
    volume = "11",
    pages = "123",
    year = "2016"
}

@article{Anderson:2017rpr,
    author = "Anderson, Lara B. and Heckman, Jonathan J. and Katz, Sheldon and Schaposnik, Laura P.",
    title = "{T-Branes at the Limits of Geometry}",
    eprint = "1702.06137",
    archivePrefix = "arXiv",
    primaryClass = "hep-th",
    doi = "10.1007/JHEP10(2017)058",
    journal = "JHEP",
    volume = "10",
    pages = "058",
    year = "2017"
}

@article{Collinucci:2017bwv,
    author = "Collinucci, Andres and Giacomelli, Simone and Valandro, Roberto",
    title = "{T-branes, monopoles and S-duality}",
    eprint = "1703.09238",
    archivePrefix = "arXiv",
    primaryClass = "hep-th",
    doi = "10.1007/JHEP10(2017)113",
    journal = "JHEP",
    volume = "10",
    pages = "113",
    year = "2017"
}

@article{Cicoli:2017shd,
    author = "Cicoli, Michele and Garc{\`\i}a-Etxebarria, I{\~n}aki and Mayrhofer, Christoph and Quevedo, Fernando and Shukla, Pramod and Valandro, Roberto",
    title = "{Global Orientifolded Quivers with Inflation}",
    eprint = "1706.06128",
    archivePrefix = "arXiv",
    primaryClass = "hep-th",
    doi = "10.1007/JHEP11(2017)134",
    journal = "JHEP",
    volume = "11",
    pages = "134",
    year = "2017"
}

@article{Cvetic:2018xaq,
    author = "Cveti{\v{c}}, Mirjam and Heckman, Jonathan J. and Lin, Ling",
    title = "{Towards Exotic Matter and Discrete Non-Abelian Symmetries in F-theory}",
    eprint = "1806.10594",
    archivePrefix = "arXiv",
    primaryClass = "hep-th",
    doi = "10.1007/JHEP11(2018)001",
    journal = "JHEP",
    volume = "11",
    pages = "001",
    year = "2018"
}

@article{Apruzzi:2018xkw,
    author = "Apruzzi, Fabio and Hassler, Falk and Heckman, Jonathan J. and Rochais, Thomas B.",
    title = "{Nilpotent Networks and 4D RG Flows}",
    eprint = "1808.10439",
    archivePrefix = "arXiv",
    primaryClass = "hep-th",
    doi = "10.1007/JHEP05(2019)074",
    journal = "JHEP",
    volume = "05",
    pages = "074",
    year = "2019"
}

@article{Heckman:2018pqx,
    author = "Heckman, Jonathan J. and Rudelius, Tom and Tomasiello, Alessandro",
    title = "{Fission, Fusion, and 6D RG Flows}",
    eprint = "1807.10274",
    archivePrefix = "arXiv",
    primaryClass = "hep-th",
    doi = "10.1007/JHEP02(2019)167",
    journal = "JHEP",
    volume = "02",
    pages = "167",
    year = "2019"
}

@article{Carta:2018qke,
    author = "Carta, Federico and Giacomelli, Simone and Savelli, Raffaele",
    title = "{SUSY enhancement from T-branes}",
    eprint = "1809.04906",
    archivePrefix = "arXiv",
    primaryClass = "hep-th",
    reportNumber = "ROM2F-2018-05, IFT-UAM-CSIC-18-93",
    doi = "10.1007/JHEP12(2018)127",
    journal = "JHEP",
    volume = "12",
    pages = "127",
    year = "2018"
}

@article{Cicoli:2015ylx,
    author = "Cicoli, Michele and Quevedo, Fernando and Valandro, Roberto",
    title = "{De Sitter from T-branes}",
    eprint = "1512.04558",
    archivePrefix = "arXiv",
    primaryClass = "hep-th",
    doi = "10.1007/JHEP03(2016)141",
    journal = "JHEP",
    volume = "03",
    pages = "141",
    year = "2016"
}

@article{Heckman:2016ssk,
    author = "Heckman, Jonathan J. and Rudelius, Tom and Tomasiello, Alessandro",
    title = "{6D RG Flows and Nilpotent Hierarchies}",
    eprint = "1601.04078",
    archivePrefix = "arXiv",
    primaryClass = "hep-th",
    doi = "10.1007/JHEP07(2016)082",
    journal = "JHEP",
    volume = "07",
    pages = "082",
    year = "2016"
}

@article{Marchesano:2019azf,
    author = "Marchesano, Fernando and Savelli, Raffaele and Schwieger, Sebastian",
    title = "{T-branes and defects}",
    eprint = "1902.04108",
    archivePrefix = "arXiv",
    primaryClass = "hep-th",
    reportNumber = "IFT-UAM/CSIC-19-015; ROM2F/2019/01",
    doi = "10.1007/JHEP04(2019)110",
    journal = "JHEP",
    volume = "04",
    pages = "110",
    year = "2019"
}

@article{Myers:1999ps,
    author = "Myers, Robert C.",
    title = "{Dielectric branes}",
    eprint = "hep-th/9910053",
    archivePrefix = "arXiv",
    reportNumber = "MCGILL-99-27, NSF-ITP-99-113",
    doi = "10.1088/1126-6708/1999/12/022",
    journal = "JHEP",
    volume = "12",
    pages = "022",
    year = "1999"
}

@article{Naculich:2001us,
    author = "Naculich, Stephen G. and Schnitzer, Howard J. and Wyllard, Niclas",
    title = "{Vacuum states of N=1* mass deformations of N=4 and N=2 conformal gauge theories and their brane interpretations}",
    eprint = "hep-th/0103047",
    archivePrefix = "arXiv",
    reportNumber = "BRX-TH-484, HUTP-00-A052, BOW-PH-121",
    doi = "10.1016/S0550-3213(01)00291-7",
    journal = "Nucl. Phys. B",
    volume = "609",
    pages = "283--312",
    year = "2001"
}

@article{Beasley:2008kw,
    author = "Beasley, Chris and Heckman, Jonathan J. and Vafa, Cumrun",
    title = "{GUTs and Exceptional Branes in F-theory - II: Experimental Predictions}",
    eprint = "0806.0102",
    archivePrefix = "arXiv",
    primaryClass = "hep-th",
    doi = "10.1088/1126-6708/2009/01/059",
    journal = "JHEP",
    volume = "01",
    pages = "059",
    year = "2009"
}

@article{Gaiotto:2008sa,
    author = "Gaiotto, Davide and Witten, Edward",
    title = "{Supersymmetric Boundary Conditions in N=4 Super Yang-Mills Theory}",
    eprint = "0804.2902",
    archivePrefix = "arXiv",
    primaryClass = "hep-th",
    doi = "10.1007/s10955-009-9687-3",
    journal = "J. Statist. Phys.",
    volume = "135",
    pages = "789--855",
    year = "2009"
}

@article{Caorsi:2018zsq,
    author = "Caorsi, Matteo and Cecotti, Sergio",
    title = "{Geometric classification of 4d $\mathcal{N}=2$ SCFTs}",
    eprint = "1801.04542",
    archivePrefix = "arXiv",
    primaryClass = "hep-th",
    doi = "10.1007/JHEP07(2018)138",
    journal = "JHEP",
    volume = "07",
    pages = "138",
    year = "2018"
}

@book{slodowy1980four,
  author    = {Slodowy, Peter},
  title     = {Four lectures on simple groups and singularities},
  series    = {Communications of the Mathematical Institute, Rijksuniversiteit Utrecht},
  volume    = {11},
  publisher = {Mathematical Institute, Rijksuniversiteit Utrecht},
  address   = {Utrecht},
  year      = {1980},
  note      = {Lectures delivered February--March 1979},
  oclc      = {6917278}
}

@article{Khlaif:2025jnx,
    author = "Khlaif, Osama and Najjar, Marwan",
    title = "{Aspects of 4d $ \mathcal{N}=1 $ADE gauge theories from M-theory: decomposition, automorphisms, and generalised symmetries}",
    eprint = "2508.00564",
    archivePrefix = "arXiv",
    primaryClass = "hep-th",
    doi = "10.1007/JHEP02(2026)067",
    journal = "JHEP",
    volume = "02",
    pages = "067",
    year = "2026"
}

\end{document}